\documentclass[12pt, letterpaper]{article}
\usepackage{geometry}
\usepackage{amsmath,amssymb,amsfonts}
\usepackage[onehalfspacing]{setspace}
\usepackage{abstract}
\usepackage{natbib,bibmods}
\usepackage{enumerate,enumitem}
\usepackage{url}
\usepackage{bbm}
\usepackage{libertine}
\usepackage[T1]{fontenc}
\usepackage{soul}
\usepackage[usenames,dvipsnames,svgnames]{xcolor}
\usepackage{xcolor,colortbl}
\usepackage{threeparttable,multirow}
\usepackage[colorlinks,urlcolor=DarkBlue,citecolor=DarkBlue, linkcolor=DarkBlue]{hyperref}

\usepackage{longtable}
\usepackage{supertabular}
\usepackage{caption}
\usepackage{subcaption}
\usepackage{ushort}
\usepackage{xfrac}
\usepackage{threeparttable}
\usepackage{lscape}
\usepackage{rotating}
\usepackage{pgf,tikz}
\usetikzlibrary{arrows,decorations.markings}
\usepackage{tabularx,ragged2e,booktabs}
\usepackage{graphicx} 
\usepackage{eurosym}
\usepackage{siunitx}
\usepackage{ntheorem}
\usepackage{float}
\usepackage{changepage}
\usepackage{bbm}
\begin{document}

\title{Welfare Effects of Labor Income Tax Changes\\on Married Couples: A Sufficient Statistics Approach}
\author{Egor Malkov\footnote{~University of Minnesota and the Federal Reserve Bank of Minneapolis. E-mail: \href{mailto:malko017@umn.edu}{malko017@umn.edu}. I thank Fatih Guvenen, Larry Jones, and Fabrizio Perri for their valuable advice and constant support. I thank Carlos E. da Costa, Diane Lim, and Xincheng Qiu for insightful discussions. I thank Nezih Guner, Andreas Haufler, Julie Hotchkiss, Jonas Loebbing, Andreas Peichl, Dominik Sachs, Minchul Yum, Daniel Weishaar, and the participants of the European Economic Association Congress 2020, VMACS Junior Conference, National Tax Association Annual Conference 2020, Midwest Economics Association 2021 Annual Meeting, SEHO 2021 Meeting, ESPE 2021 Conference, COMPIE 2021 Conference, Public Economics Seminar at the LMU Munich, NIPFP-IIPF International Conference on Public Finance, 2021 ECINEQ Meeting, China Meeting of the Econometric Society, and Brazilian Meeting in Family and Gender Economics for fruitful comments. The views expressed herein are those of the author and not necessarily those of the Federal Reserve Bank of Minneapolis or the Federal Reserve System.}}
\date{This version: July 2021}

\maketitle

\begin{abstract}
\begin{spacing}{1}
\noindent This paper develops a framework for assessing the welfare effects of labor income tax changes on married couples. I build a static model of couples' labor supply that features both intensive and extensive margins and derive a tractable expression that delivers a transparent understanding of how labor supply responses, policy parameters, and income distribution affect the reform-induced welfare gains. Using this formula, I conduct a comparative welfare analysis of four tax reforms implemented in the United States over the last four decades, namely the Tax Reform Act of 1986, the Omnibus Budget Reconciliation Act of 1993, the Economic Growth and Tax Relief Reconciliation Act of 2001, and the Tax Cuts and Jobs Act of 2017. I find that these reforms created welfare gains ranging from -0.16 to 0.62 percent of aggregate labor income. A sizable part of the gains is generated by the labor force participation responses of women. Despite three reforms resulted in aggregate welfare gains, I show that each reform created both winners and losers. Furthermore, I uncover two patterns in the relationship between welfare gains and couples' labor income. In particular, the reforms of 1986 and 2017 display a monotonically increasing relationship, while the other two reforms demonstrate a U-shaped pattern. Finally, I characterize the bias in welfare gains resulting from the assumption about a linear tax function. I consider a reform that changes tax progressivity and show that the linearization bias is given by the ratio between the tax progressivity parameter and the inverse elasticity of taxable income. Quantitatively, it means that linearization overestimates the welfare effects of the U.S. tax reforms by 3.6-18.1\%.
\medskip

\smallskip

\end{spacing}
\end{abstract}

\newpage

\section{Introduction}

\begin{figure}[t!]
\centering
\includegraphics[width=0.75\linewidth]{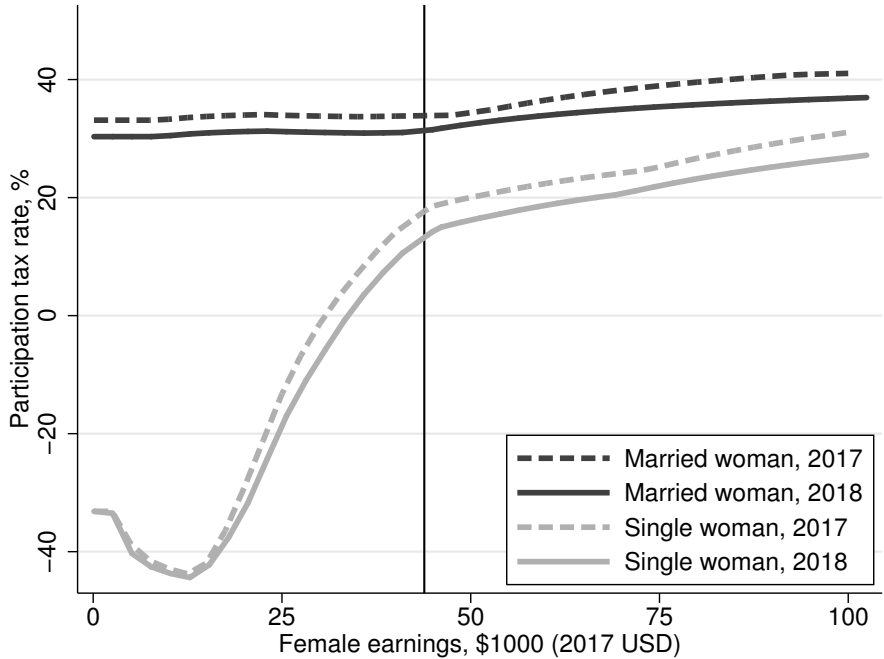}
\caption{Participation tax rates of married and single women in the United States}
\label{fig: mtr1_2018}
\justify\footnotesize{\textsc{Notes:} Participation tax rate is defined as the change in household's tax liability divided by female earnings when she starts working. The tax rates are calculated using NBER TAXSIM and include federal, state, and FICA tax rates. Both women aged 40, live in Minnesota, and have two children under 19. A married woman husband's annual earnings are fixed at the median level in Minnesota. Individuals do not have any non-labor income. Married couple is assumed to file jointly. Vertical line corresponds to median annual female earnings in Minnesota (I multiply median weekly female earnings in Minnesota by median annual working hours of women in the United States divided by 40).}
\end{figure}

What are the welfare effects of tax reforms on married couples? How are the welfare gains and losses distributed among them? The answers to these questions are of crucial importance for both academic economists and policymakers for several reasons. First, the scope is significant. Married couples constitute a sizable share of the population (e.g., they account for almost a half of all the U.S. households in 2019) and taxpayers (e.g., the number of tax returns of married couples filing jointly constitutes more than a third of all tax returns in the United States). Second, positive assortative mating, when similarly educated individuals tend to marry each other, is considered as one of the driving forces of between-household inequality \citep{dupuy2021marriage}. Therefore, it is crucial to know who benefits and who loses from the redistributive policies such as income taxation. Finally, the tax and transfer systems that feature jointness, such as in Germany and the United States, create substantial disincentive effects for the married women's labor supply \citep{bick2017quantifying, holter2019tax}. Under joint taxation of spousal incomes, the marginal tax rate on the first dollar earned by the secondary (lower income) earner is equal to the marginal tax rate on the last dollar earned by the primary (higher income) earner. Figure \ref{fig: mtr1_2018} illustrates the last point by showing the participation tax rates (change in household's tax liability divided by woman's earnings when she starts working) for married and single women in the United States. Except for the marital status, these two women are otherwise identical. Clearly, a married woman faces a significantly higher tax rate, when she starts working, than a single one. A tax reform that results in lower participation tax rates, as shown by the difference between dashed and solid lines, can enhance employment of the married women. Overall, studying the positive aspects of income taxation of couples is critically important because it is closely related to a vast array of topics including between- and within-household inequality, female labor supply, marriage decisions, and so on.

In this paper, I develop a framework for studying the welfare effects of income tax changes on married couples. First, I build a static model of couples' labor supply that features both intensive and extensive margins. Using this model, I derive a tractable expression for welfare gains, resulting from any arbitrary small tax policy reform, as a function of several sufficient statistics: labor supply elasticities (elasticities of hours, cross-elasticities of spousal hours, and participation elasticity), policy parameters (pre-reform marginal and participation tax rates and their reform-induced changes), and labor income shares. In a transparent way, it allows decomposing the changes in aggregate efficiency gains into the effects that operate through labor supply responses. I use this sufficient statistics formula to quantify the welfare effects of four tax reforms, implemented in the United States over the last four decades. The reforms include the Tax Reform Act of 1986 (TRA 1986), the Omnibus Budget Reconciliation Act of 1993 (OBRA 1993), the Economic Growth and Tax Relief Reconciliation Act of 2001 (EGTRRA 2001), and the Tax Cuts and Jobs Act of 2017 (TCJA 2017). To map my expression for welfare gains to the data, I use the Current Population Survey combined with the NBER TAXSIM calculator.

Under the baseline parameterization, I estimate the efficiency gains from four tax reforms to range from -0.16 (OBRA 1993) to 0.62 (TCJA 2017) percent of aggregate labor income. The welfare gains per dollar spent range from 0.63 USD (OBRA 1993) to 1.10 USD (TCJA 2017). Overall, three reforms, the TRA 1986, the EGTRRA 2001, and the TCJA 2017, created aggregate welfare gains for married couples. A substantial part of these gains comes from the labor force participation responses of women. Furthermore, I also emphasize that the spousal cross-effects of working hours (change in working hours of one spouse resulting from the change in the net-of-tax rate of the other spouse) are quantitatively important. Abstracting from them can lead to substantial overestimation of efficiency gains. For example, if I abstract from the cross-effects in the case of the TRA 1986 reform, I overestimate the welfare gains by 34.6\%. The sensitivity analysis, where I consider the range of reasonable values of elasticities, confirms this argument.

Next, despite three out of four considered reforms generate aggregate welfare gains, I find that each reform created winners and losers. For example, the TRA 1986 left 12.3 percent of couples with welfare losses. Furthermore, I study how the welfare gains vary by income and uncover two general patterns. First, the TRA1986, the OBRA 1993 (excluding the bottom 10 percent), and the TCJA 2017 reforms display a monotonically increasing relationship between welfare gains and income. In other words, richer taxpayers benefited more than poorer taxpayers. Second, the OBRA 1993 and the EGTRRA 2001 reforms demonstrate a U-shaped pattern in the welfare gains. In the case of these reforms, the main winners are located at the lower and upper ends of the income distribution. Overall, my general takeaway from this part of the analysis is that the aggregate welfare measures mask significant heterogeneity in welfare gains.

After that, I discuss the robustness of my findings. I start with considering alternative parameterizations of elasticities. Under the scenario that delivers an upper bound on the efficiency gains, I find that they range from -0.07 to 1.15 percent of aggregate labor income. Under the ``lower bound'' scenario, the range from -0.25 to 0.12 percent of the aggregate labor income. To address the concern that between the 1970s and 2000s there was a sizable reduction in own- and cross-elasticities of married women's labor supply \citep{blau2007changes, heim2007incredible}, I also consider the parameterization where all elasticities have reasonably high (and low) values. Next, to address the concerns about the role of the initial income distribution and pre-reform tax rates, I conduct two sets of counterfactual reforms. I begin with the exercises where I apply the actual tax reforms to the counterfactual income distributions. For example, I show that if the TRA 1986 reform were to be applied to the 2017 income distribution, aggregate welfare gains would be 1.32 percent of aggregate labor income, or 1.14 USD per dollar spent. This exceeds the actual welfare gains per dollar spent from the TCJA by 5.48\%. Overall, I find that counterfactual welfare gains per dollar spent do not differ by more than 7.54\% from the actual ones. In another set of exercises, I fix the income distribution and tax law in a given pre-reform year and calculate the welfare effects of moving to the other post-reform's tax law. For example, I show that moving from the pre-TRA 1986 economy to the post-EGTRRA 2001 and post-TCJA 2017 economies creates higher efficiency gains, 0.88 and 1.19 percent of aggregate labor income, than the actual TRA 1986 (0.55 percent). However, when I make the efficiency gains comparable, the actual TRA 1986 generates more welfare gain per dollar than these alternative counterfactual reforms.

Finally, I address the concern that the assumption about linearity of the tax function, commonly used in the sufficient statistics literature, may deliver biased estimates of efficiency gains, because real tax codes feature nonlinearities. To do so, I characterize the linearization bias, defined as the percentage difference between reform-induced efficiency loss under true and linearized tax and transfer functions. I assume quasilinear preferences and use the log-linear specification $T (y) = y - \lambda y^{1 - \theta}$ that provides a good approximation of the actual tax and transfer system in the United States \citep{heathcote2017optimal}. More precisely, I consider joint, i.e. $T \left( y_m, y_f \right) = T \left( y_m + y_f \right)$, and separate, i.e. $T \left( y_m, y_f \right) = T \left( y_m \right) + T \left( y_m \right)$, taxation of spouses. Assuming a small reform that changes progressivity of the tax system, $d \theta \approx 0$, I show that the linearization bias is given by the ratio between the tax progressivity parameter (tax function curvature) and the inverse elasticity of taxable income (utility curvature). Using the estimates of these objects from the literature, I conclude that linearization biases upward the welfare effects of the U.S. tax reforms in the range from 3.6\% to 18.1\%.

My paper is related to several strands of literature. First, I contribute to the literature studying the welfare effects of tax and transfer reforms initiated by the classic paper \cite{harberger1964taxation}, and further developed by \cite{dahlby1998progressive}, \cite{feldstein1999tax}, \cite{kleven2006marginal}, and \cite{blomquist2019marginal}, among many others. On a related note, \cite{finkelstein2020welfare} and \cite{hendren2020unified} emphasize the attractiveness of the marginal value of public funds (MVPF) as a tool that allows comparing the effects of various policies on social welfare. The paper most closely related to mine is \cite{eissa2008evaluation}. They quantify the welfare effects of the U.S. tax reforms on single mothers. Importantly, the results for single individuals may be quite different from ones for married couples. First, the interactions between spouses, captured by the cross-elasticities of working hours, are naturally absent in a framework with singles. Second, single mothers more likely represent the lower part of the household income distribution than married couples, and any given tax reform may differently affect households that belong to different income groups. Another related work by \cite{immervoll2009evaluation} studies the welfare effects of tax policy changes on married couples, but the authors consider hypothetical reforms. Instead, I evaluate the welfare gains of the actual U.S. tax reforms, including the most recent one, the Tax Cuts and Jobs Act of 2017. Furthermore, \cite{bar2009work} investigate the impact of the U.S. tax reforms on married couples' participation but do not allow for the intensive margin of labor supply. Next, \cite{hotchkiss2012assessing} and \cite{hotchkiss2021impact} evaluate the welfare effects of the U.S. tax reforms on households, but using a different approach than mine. Beyond that, two other related strands of literature study the macroeconomic effects of tax reforms \citep{barro2011macroeconomic, mertens2012empirical, mertens2013dynamic, barro2018macroeconomic} and heterogeneity in the effects of economic policy \citep{domeij2004distributional, bitler2006mean, zidar2019tax}.

Next, my paper is related to the literature that studies the taxation of couples and its effects on the female labor supply. \cite{eissa2004taxes} find that the Earned Income Tax Credit (EITC) expansions between 1984 and 1996 reduced the total family labor supply of couples mainly through lowering the labor force participation of married women. Next, \cite{guner2012taxation}, using a rich general equilibrium life-cycle model calibrated to the U.S. economy, show that the reform replacing joint taxation to separate taxation would substantially increase the labor supply of married women. Similarly, \cite{bick2017quantifying} emphasize that joint taxation creates significant disincentive effects for the labor supply of married women in the United States and Europe. In the same vein, \cite{borella2021marriage} show that eliminating marriage-related taxes and old age Social Security benefits in the United States would significantly increase the participation of married women over their entire life cycle.

The rest of the paper is organized as follows. The model is presented in Section \ref{Model}. The data and sample selection are discussed in Section \ref{Data}. The U.S. tax reforms and construction of reform-induced tax changes using NBER TAXSIM are described in Section \ref{The U.S. Tax Reforms}. The quantitative findings along with the sensitivity analysis, evidence on welfare gains distribution, and the results of the counterfactual reforms are reported in Section \ref{Quantitative Results}. The linearization bias is discussed in Section \ref{Efficiency Loss and Nonlinear Taxation of Couples}. Section \ref{Conclusion} concludes.

\newpage
\section{Model}\label{Model}

\noindent \textbf{Economic Environment.} To study the welfare effects of the changes in labor income taxes, I build a static model of married couples along the lines of \cite{kaygusuz2010taxes} and \cite{bick2017taxation}.\footnote{~Here and thereafter, I use ``labor income taxes'' and ``taxes'' interchangeably.} Consider an economy populated by $N$ married couples. In each couple, spouses choose joint private consumption, $c$, males choose how much to work (intensive margin), $h^m$, and females choose whether to work (extensive margin) and, conditional on participation, how much to work (intensive margin), $h^f$. I abstract from modeling the extensive margin for males because in the data their participation rates are traditionally high and demonstrate little variation over time. Hence the households in the model are either single-earner or dual-earner couples. To model the extensive margin of labor supply for women, I assume that each couple draw fixed utility cost of work $q_i$ from a distribution $F_i \left( q_i \right)$. This cost is incurred when a wife enters the labor market. I interpret it as a utility loss that is related to inconvenience of scheduling joint work for both spouses or childcare responsibilities \citep{cho1988family}. Modeling the extensive margin with the fixed cost of work allows generating the distribution of hours that is consistent with the data. In particular, as Figure \ref{fig:hoursf} reports, the empirical distribution of married women's annual working hours has a little mass at low number of hours.

The wages of a male and a female in couple $i$ are denoted by $w^m_i$ and $w^f_i$ correspondingly. The tax and transfer system is introduced by function $T \left( w^m_i h^m, w^f_i h^f, \theta \right)$. It embodies all labor income taxes and transfers and may feature non-separabilities between the arguments. I assume that $T (\cdot)$ is piecewise linear, so that the spouses face locally constant marginal tax rates. The policy reform is modeled in a flexible way by allowing $T (\cdot)$ to be a function of a treatment parameter $\theta$. Changes in $\theta$ capture any arbitrary tax reform. In what follows, I focus on small reforms ($d \theta \approx 0$). Furthermore, I assume that there no other externalities than those operating through the government budget.
\bigskip

\noindent \textbf{Household Optimization.} The utility maximization problem of couple $i$ is given by
\begin{equation}
\max_{c, h^m, h^f} U_i \left( c, h^m, h^f \right) = v_i \left( c, h^m, h^f \right) - q_i \cdot \mathbbm{1} \{ h^f > 0 \}
\label{eq: utility}
\end{equation}
\begin{equation}
\text{s.t.}~~~~~c = w^m_i h^m + w^f_i h^f - T \left( w^m_i h^m, w^f_i h^f, \theta \right)
\label{eq: budget constraint}
\end{equation}
where $v_i (\cdot)$ is a well-behaved utility function, and $\mathbbm{1} \{ h^f > 0 \}$ takes the value one if the wife works and zero otherwise. Similarly to \cite{eissa2008evaluation}, my formulation accounts for the presence of income effects. This problem can be solved in two stages. First, conditional on wife's participation, the couple choose the hours of work. Second, the wife makes a participation decision at the optimal level of working hours. Consider these stages in turn.

\newpage
First, given that both spouses work, $h^m > 0$ and $h^f > 0$, the optimal allocation for couple $i$ is characterized by the following first-order conditions:
\begin{equation}
(1 - \tau^j_i (\theta)) w^j_i \frac{\partial v_i \left( c^2_i, h^{m,2}_i, h^f_i \right)}{\partial c} = - \frac{\partial v_i \left( c^2_i, h^{m,2}_i, h^f_i \right)}{\partial h^j},~~~~~~~j = m, f
\label{eq: foc_h_umax}
\end{equation}
where $c^2_i$ denotes the optimal consumption in a dual-earner couple, $h^{m,2}_i$ and $h^f_i$ denote the optimal male's and female's hours of work in a dual-earner couple, and the effective marginal tax rates on male's and female's earnings are denoted by $\tau^j_i (\theta) \equiv \partial T \left( w^m_i h^m_i, w^f_i h^f_i, \theta \right) / \partial \left( w^j_i h^j_i \right)$, with $j = m,f$. Note that these marginal tax rates include the marginal claw-back on special provisions of the tax code such as deductions or tax credits. To simplify notation, I omit explicit dependence of marginal tax rates on $\theta$.

Next, at the second stage, the wife makes a decision whether to enter the labor market or not. There exists a threshold level of fixed cost of work, $\bar{q}_i$, such that the wife chooses to work if the utility of the dual-earner couple is greater than or equal to the utility of the single-earner couple. This threshold is given by
\begin{equation}
\bar{q}_i = v_i \left( c^2_i, h^{m,2}_i, h^f_i \right) - v_i \left( c^1_i, h^{m,1}_i, 0 \right)
\label{eq: q_threshold}
\end{equation}
where $c^1_i \equiv w^m_i h^{m,1}_i - T \left( w^m_i h^{m,1}_i, 0, \theta \right)$ denotes the optimal consumption of the single-earner couple. Consumption allocations of the dual-earner and single-earner couples are connected through the following equation:
\begin{equation}
c^2_i = w^m_i h^{m,2}_i + w^f_i h^f_i - T \left( w^m_i h^{m,2}_i, w^f_i h^f_i, \theta \right) = c^1_i + (1 - a_i (\theta)) \left[ w^m_i \left( h^{m,2}_i - h^{m,1}_i \right) + w^f_i h^f_i  \right]
\label{eq: c_2 and c_1}
\end{equation}
where $a_i (\theta) \equiv \left[ T \left( w^m_i h^{m,2}_i, w^f_i h^f_i, \theta \right) - T \left( w^m_i h^{m,1}_i, 0, \theta \right) \right] / \left( w^m_i \left( h^{m,2}_i - h^{m,1}_i \right) + w^f_i h^f_i \right)$ is a participation tax rate of the couple \citep{prescott2004americans}. It captures the change in tax liability as a share of the change in earnings following the wife's decision to participate. In the quantitative part of this paper, I assume that the husband's earnings do not change if the wife enters the labor market, and hence $a_i$ is effectively the women's participation tax rate. To simplify notation, I omit explicit dependence of participation tax rates on $\theta$.

To evaluate the welfare effects of a tax policy reform, I obtain the compensated (Hicksian) consumption and labor supply by solving the expenditure minimization problem. This dual problem is given by
\begin{equation}
\min_{c, h^m, h^f} c - w^m_i h^m - w^f_i h^f + T \left( w^m_i h^m, w^f_i h^f, \theta \right)
\label{eq: dual_problem}
\end{equation}
\begin{equation}
\text{s.t.}~~~~~v_i \left( c, h^m, h^f \right) - q_i \cdot \mathbbm{1} \{ h^f > 0 \} \geq \bar{U}_i
\label{eq: dual_problem_constr}
\end{equation}
where $\bar{U}_i$ is some fixed level of utility.

\newpage
Similarly to problem \eqref{eq: utility}-\eqref{eq: budget constraint}, I solve it in two stages. First, given that both spouses work, $h^m > 0$ and $h^f > 0$, the solution is characterized by the following first-order conditions:
\begin{equation}
(1 - \tau^j_i) w^j_i \frac{\partial v_i \left( \tilde{c}^2_i, \tilde{h}^{m,2}_i, \tilde{h}^f_i \right)}{\partial c} = - \frac{\partial v_i \left( \tilde{c}^2_i, \tilde{h}^{m,2}_i, \tilde{h}^f_i \right)}{\partial h^j},~~~~~~~j = m, f
\label{eq: foc_h_dual_two_earner}
\end{equation}
\begin{equation}
v_i \left( \tilde{c}^2_i, \tilde{h}^{m,2}_i, \tilde{h}^f_i \right) = \bar{U}_i + q_i
\label{eq: foc_bc_dual_two_earner}
\end{equation}

From \eqref{eq: foc_h_dual_two_earner}-\eqref{eq: foc_bc_dual_two_earner}, I get compensated consumption, $\tilde{c}^2_i = \tilde{c}^2_i \left( \cdot \right)$, and labor supply, $\tilde{h}^{m,2}_i = \tilde{h}^{m,2}_i \left( \cdot \right)$ and $\tilde{h}^f_i = \tilde{h}^f_i \left( \cdot \right)$, for dual-earner couples. Using these compensated functions, I write down the expenditure function for a dual-earner couple:
\begin{equation}
E^2_i \left( \bar{U}_i + q_i, \theta \right) = \tilde{c}^2_i (\cdot) - w^m_i \tilde{h}^{m,2}_i (\cdot) - w^f_i \tilde{h}^f_i (\cdot) + T \left( w^m_i \tilde{h}^{m,2}_i (\cdot), w^f_i \tilde{h}^f_i (\cdot), \theta \right)
\label{eq: exp_func_two_earner}
\end{equation}
This expenditure function is evaluated at the current (pre-reform) tax and transfer system and depends on the treatment parameter $\theta$ both directly and through compensated consumption and labor supply.

For a single-earner couple, the solution to the dual problem is characterized by
\begin{equation}
(1 - \tau^m_i) w^m_i \frac{\partial v_i \left( \tilde{c}^1_i, \tilde{h}^{m,1}_i, 0 \right)}{\partial c} = - \frac{\partial v_i \left( \tilde{c}^1_i, \tilde{h}^{m,1}_i, 0 \right)}{\partial h^m}
\label{eq: foc_hm_dual_one_earner}
\end{equation}
\begin{equation}
v_i \left( \tilde{c}^1_i, \tilde{h}^{m,1}_i, 0 \right) = \bar{U}_i
\label{eq: foc_bc_dual_one_earner}
\end{equation}
and it delivers compensated consumption $\tilde{c}^1_i = \tilde{c}^1_i \left( \cdot \right)$ and male's labor supply, $\tilde{h}^{m,1}_i = \tilde{h}^{m,1}_i \left( \cdot \right)$.

The expenditure function of a single-earner couple is given by
\begin{equation}
E^1_i \left( \bar{U}_i , \theta \right) = \tilde{c}^1_i (\cdot) - w^m_i \tilde{h}^{m,1}_i (\cdot) + T \left( w^m_i \tilde{h}^{m,1}_i (\cdot), 0, \theta \right)
\label{eq: exp_func_one_earner}
\end{equation}

Given utility $\bar{U}_i$, the wife chooses to enter the labor market if $E^2_i \left( \bar{U}_i + q_i, \theta \right) \leq E^1_i \left( \bar{U}_i , \theta \right)$, and not participate otherwise. Therefore, I write the expenditure function in the following way:
\begin{equation}
E_i \left( \bar{U}_i, q_i, \theta \right) = \min \Big\{ E^1_i \left( \bar{U}_i, \theta \right), E^2_i \left( \bar{U}_i + q_i, \theta \right) \Big\}
\label{eq: exp_function_full}
\end{equation}

Defining a compensated threshold of the fixed cost of work, $\tilde{q}_i$, such that $E^2_i \left( \bar{U}_i + \tilde{q}_i, \theta \right) = E^1_i \left( \bar{U}_i , \theta \right)$, and plugging \eqref{eq: exp_func_two_earner} and \eqref{eq: exp_func_one_earner} into this definition, I obtain the equation connecting compensated consumption in dual-earner and single-earner couples:
\begin{equation}
\tilde{c}^2_i = \tilde{c}^1_i + (1 - a_i) \left[ w^m_i \left( \tilde{h}^{m,2}_i - \tilde{h}^{m,1}_i \right) + w^f_i \tilde{h}^f_i \right]
\label{eq: c_tilde}
\end{equation}

Furthermore, evaluating \eqref{eq: foc_bc_dual_two_earner} and \eqref{eq: foc_bc_dual_one_earner} at $\tilde{q}_i$, I get
\begin{equation}
\tilde{q}_i = v_i \left( \tilde{c}^2_i, \tilde{h}^{m,2}_i, \tilde{h}^f_i \right) - v_i \left( \tilde{c}^1_i, \tilde{h}^{m,1}_i, 0 \right)
\label{eq: q_tilde_main}
\end{equation}

Setting $\bar{U}_i$ to be equal to the indirect utility obtained from problem \eqref{eq: utility}-\eqref{eq: budget constraint}, I guarantee that the solution to the expenditure minimization problem is consistent with the solution to the utility maximization problem.
\bigskip

\noindent \textbf{Aggregate Labor Supply.} Since the focus of the paper is in aggregate welfare effects of tax reforms, it is natural to ask how does a tax policy affect the aggregate labor supply of couples. Turning from individual optimization to the aggregates, I write down aggregate compensated labor supply:
\begin{multline}
\tilde{L} = \sum_{i = 1}^{N} \Bigg[ \underbrace{\int_0^{\tilde{q}_i} \tilde{h}^{m,2}_i \left( (1 - \tau^m_i) w^m_i, (1 - \tau^f_i) w^f_i, \bar{U}_i + q_i \right) dF_i \left( q_i \right)}_\text{compensated labor supply of males in dual-earner couples} +\\
\underbrace{\int_0^{\tilde{q}_i} \tilde{h}^f_i \left( (1 - \tau^m_i) w^m_i, (1 - \tau^f_i) w^f_i, \bar{U}_i + q_i \right) dF_i \left( q_i \right)}_\text{compensated labor supply of females in dual-earner couples} +\\
\underbrace{\int_{\tilde{q}_i}^\infty \tilde{h}^{m,1}_i \left( (1 - \tau^m_i) w^m_i, 0, \bar{U}_i \right) dF_i \left( q_i \right)}_\text{compensated labor supply of males in single-earner couples} \Bigg]
\label{eq: agg_labor_supply_I}
\end{multline}

The assumption about separability of $q_i$ implies that the utility net of fixed cost of work, i.e. $v_i = \bar{U}_i + q_i$, is independent of the realization of $q_i$. Therefore, I rewrite \eqref{eq: agg_labor_supply_I} as
\begin{equation}
\tilde{L} = \sum_{i = 1}^{N} \Bigg[ \underbrace{F_i \left( \tilde{q}_i \right)}_\text{affected by $a_i (\theta)$} \underbrace{\left( \tilde{h}^{m,2}_i + \tilde{h}^f_i \right)}_\text{affected by $\tau^m_i(\theta)$ and $\tau^f_i(\theta)$} + \underbrace{\left(1 - F_i \left( \tilde{q}_i \right) \right)}_\text{affected by $a_i (\theta)$} \underbrace{\tilde{h}^{m,1}_i}_\text{affected by $\tau^m_i(\theta)$} \Bigg]
\label{eq: agg_labor_supply_II}
\end{equation}
where $F_i \left( \tilde{q}_i \right)$ denotes the probability of being a dual-earner couple or, alternatively, the individual probability of the woman's participation. This term can be also interpreted as the married women's participation rate. Aggregate compensated labor supply depends on extensive and intensive margins of spousal labor supply. The former affects $\tilde{L}$ through the woman's participation rate $F_i(\tilde{q}_i)$ that is driven by the participation tax rate $a_i$. The latter affects $\tilde{L}$ through the working hours that are driven by the marginal tax rates $\tau^m_i$ and $\tau^f_i$. These labor supply behavioral responses are the key factors in assessing the efficiency gains of tax policy reforms.
\bigskip

\noindent \textbf{Elasticities.} To evaluate the welfare effects of a tax reform, I reformulate my results in terms of the compensated elasticities of labor supply. First, define the compensated participation elasticity as the percentage change in the woman's individual participation rate resulting from a one percentage change in the participation net-of-tax rate:
\begin{equation}
\eta_i \equiv \frac{\partial F_i \left( \tilde{q}_i \right)}{\partial \left( 1 - a_i \right)} \cdot \frac{1 - a_i}{F_i \left( \tilde{q}_i \right)}
\label{eq: eta_definition}
\end{equation}

Next, define the compensated elasticity of working hours for males and females as the percentage change in working hours resulting from a one percentage change in the effective marginal net-of-tax rate:
\begin{equation}
\varepsilon^{m,\iota}_i \equiv \frac{\partial \tilde{h}^{m,\iota}_i}{\partial \left( 1 - \tau^m_i \right)} \cdot \frac{1 - \tau^m_i}{\tilde{h}^{m,\iota}_i},~~~~~~~\iota = 1,2
\label{eq: epsilon_m_definition}
\end{equation}
\begin{equation}
\varepsilon^f_i \equiv \frac{\partial \tilde{h}^f_i}{\partial \left( 1 - \tau^f_i \right)} \cdot \frac{1 - \tau^f_i}{\tilde{h}^f_i}
\label{eq: epsilon_f_definition}
\end{equation}

Finally, for dual-earner couples, define the cross-elasticities of working hours as the percentage change in individual's working hours resulting from a one percentage change in the effective marginal net-of-tax rate of his/her spouse. These elasticities are absent in the framework with singles.
\begin{equation}
\varepsilon^{mf}_i \equiv \frac{\partial \tilde{h}^{m,2}_i}{\partial \left( 1 - \tau^f_i \right)} \cdot \frac{1 - \tau^f_i}{\tilde{h}^{m,2}_i}
\label{eq: epsilon_mf_definition}
\end{equation}
\begin{equation}
\varepsilon^{fm}_i \equiv \frac{\partial \tilde{h}^f_i}{\partial \left( 1 - \tau^m_i \right)} \cdot \frac{1 - \tau^m_i}{\tilde{h}^f_i}
\label{eq: epsilon_fm_definition}
\end{equation}


\noindent \textbf{Efficiency Loss.} To study the effects of a tax reform on individual welfare, I define the measure of excess burden using the equivalent variation. Under this definition, the excess burden from the current tax and transfer system $\theta$ is the difference between the sum of money that the couple is willing to pay to move to the economy without distortionary taxes and transfers and collected tax revenue \citep{auerbach1985theory}:
\begin{equation}
D_i \left( \bar{U}_i, q_i, \theta \right) = E_i \left( \bar{U}_i, q_i, \theta \right) - E_i \left( \bar{U}_i, q_i, 0 \right) - R \left( \bar{U}_i, q_i, \theta \right)
\label{eq: excess_burden_i}
\end{equation}
where $R \left( \bar{U}_i, q_i, \theta \right)$ is given by
\begin{equation}
R \left( \bar{U}_i, q_i, \theta \right) =
\begin{cases}
T \left( w^m_i \tilde{h}^{m,2}_i (\cdot), w^f_i \tilde{h}^f_i (\cdot), \theta \right),~~~~~\text{if}~~q_i < \tilde{q}_i\\
T \left( w^m_i \tilde{h}^{m,1}_i (\cdot), 0, \theta \right),~~~~~~~~~~~~~~~~\text{otherwise}
\end{cases}
\label{eq: R_definition}
\end{equation}

Aggregate excess burden under a tax and transfer system $\theta$ is defined as the sum of excess burdens over all couples:
\begin{equation}
D = \sum_{i = 1}^N \int_0^\infty D_i \left( \bar{U}_i, q_i, \theta \right) dF_i \left( q_i \right)
\label{eq: agg_excess_burden}
\end{equation}
Aggregate efficiency loss $D$ measures additional revenue that can be collected, keeping couples at their initial utility levels $\bar{U}_i$, if the tax and transfer system $\theta$ were to be replaced by a lump-sum tax system. With heterogeneous agents, aggregate excess burden depends on the initial income distribution except under very strong conditions on preferences \citep{auerbach1985theory, auerbach2002taxation}. In Section \ref{Counterfactual Tax Reforms}, I discuss the sensitivity of my results to alternative initial income distributions.

Plugging \eqref{eq: exp_function_full} and \eqref{eq: R_definition} into \eqref{eq: excess_burden_i}, I rewrite aggregate excess burden \eqref{eq: agg_excess_burden} as
\begin{multline}
D = \sum_{i = 1}^N \Bigg[ \int_0^{\tilde{q}_i} \left( E^2_i \left( \bar{U}_i + q_i, \theta \right) - T \left( w^m_i \tilde{h}^{m,2}_i (\cdot), w^f_i \tilde{h}^f_i (\cdot), \theta \right) \right) dF_i \left( q_i \right) +\\
\int_{\tilde{q}_i}^\infty \left( E^1_i \left( \bar{U}_i, \theta \right) - T \left( w^m_i \tilde{h}^{m,1}_i (\cdot), 0, \theta \right) \right) dF_i \left( q_i \right) -
\int_0^\infty E_i \left( \bar{U}_i, q_i, 0 \right) dF_i \left( q_i \right) \Bigg]
\label{eq: D_detailed}
\end{multline}

I focus on a small tax reform ($d \theta \approx 0$), and to capture its welfare effects, I study how aggregate efficiency loss $D$ changes with $\theta$. At this step, I refer to the envelope theorem and the assumption that there are no other externalities beyond those operating through the government budget, and show that any arbitrary small reform affects the expenditure function only through mechanical revenue effect, i.e. $d E^2_i \left( \bar{U}_i + q_i, \theta \right) / d \theta = \partial T \left( w^m_i \tilde{h}^{m,2}_i (\cdot), w^f_i \tilde{h}^f_i (\cdot), \theta \right) / \partial \theta$ for dual-earner couples and $d E^1_i \left( \bar{U}_i , \theta \right) / d \theta = \partial T \left( w^m_i \tilde{h}^{m,1}_i (\cdot), 0, \theta \right) / \partial \theta$ for single-earner couples. Since the spouses optimize and there are no non-tax or non-transfer externalities, a small tax reform does not have the first-order effects on the expenditure functions and utility. In turn, the first-order effects come from externalities that operate through the government budget. In particular, when the spouses adjust their working hours or labor force participation, they create fiscal externality on all the other households. Having stated this, it follows from definition \eqref{eq: excess_burden_i} that the effect of any arbitrary small tax reform on economic efficiency is captured by the behavioral revenue effect (``fiscal externality'') or the difference between mechanical revenue effect, $\partial T_i / \partial \theta$, and total revenue effect, $d T_i / d \theta$.

Differentiating \eqref{eq: D_detailed} and using the result from the previous paragraph, I obtain
\begin{multline}
\frac{d D}{d \theta} = - \sum_{i = 1}^N \Bigg[ \tau^m_i w^m_i \frac{\partial \tilde{h}^{m,2}_i}{\partial \theta} F_i \left( \tilde{q}_i \right) + \tau^m_i w^m_i \frac{\partial \tilde{h}^{m,1}_i}{\partial \theta} \left( 1 - F_i \left( \tilde{q}_i \right) \right) +\\ \tau^f_i w^f_i \frac{\partial \tilde{h}^f_i}{\partial \theta} F_i \left( \tilde{q}_i \right) + a_i \left[ w^m_i \left( \tilde{h}^{m,2}_i - \tilde{h}^{m,1}_i \right) + w^f_i \tilde{h}^f_i \right] \frac{\partial F_i \left( \tilde{q}_i \right)}{\partial \theta} \Bigg]
\label{eq: dD/d_theta}
\end{multline}
The effect of a small tax reform on economic efficiency is driven by behavioral responses along intensive and extensive margins of labor supply. The first three terms in \eqref{eq: dD/d_theta} stand for reform-induced changes in the working hours. The last term captures the effect of reform-induced changes in female labor force participation.

Denote aggregate labor income by
\begin{equation}
W \equiv \sum_{i = 1}^N \left( w^m_i \tilde{h}^{m,2}_i + w^f_i \tilde{h}^f_i \right) F_i \left( \tilde{q}_i \right) + w^m_i \tilde{h}^{m,1}_i \left( 1 - F_i \left( \tilde{q}_i \right) \right)
\label{eq: agg_income}
\end{equation}
so that the expected labor income shares are given by $s^{m,2}_i \equiv w^m_i \tilde{h}^{m,2}_i F_i \left( \tilde{q}_i \right) / W$ for males in dual-earner couples, $s^f_i \equiv w^f_i \tilde{h}^f_i F_i \left( \tilde{q}_i \right) / W$ for females, and $s^{m,1}_i \equiv w^m_i \tilde{h}^{m,1}_i \left( 1 - F_i \left( \tilde{q}_i \right) \right) / W$ for males in single-earner couples. Finally, in Proposition 1, I state the main formula that expresses the reform-induced change in economic efficiency in terms of the empirically estimable objects.
\bigskip

\noindent\textbf{Proposition 1 (Reform-Induced Change in Economic Efficiency).} \textit{The effect of any arbitrary small tax reform $d \theta \approx 0$ on economic efficiency, captured by marginal excess burden as a fraction of aggregate labor income, is given by}
\begin{multline}
\frac{d D/ d \theta}{W} = \sum_{i = 1}^N \Bigg[ \left( \frac{\tau^m_i}{1 - \tau^m_i} \cdot \frac{d \tau^m_i}{d \theta} \varepsilon^{m,2}_i + \frac{\tau^m_i}{1 - \tau^f_i} \cdot \frac{d \tau^f_i}{d \theta} \varepsilon^{mf}_i \right) s^{m,2}_i + \frac{\tau^m_i}{1 - \tau^m_i} \cdot \frac{d \tau^m_i}{d \theta} \varepsilon^{m,1}_i s^{m,1}_i +\\
\left( \frac{\tau^f_i}{1 - \tau^f_i} \cdot \frac{d \tau^f_i}{d \theta} \varepsilon^f_i + \frac{\tau^f_i}{1 - \tau^m_i} \cdot \frac{d \tau^m_i}{d \theta} \varepsilon^{fm}_i \right) s^f_i + \frac{a_i}{1 - a_i} \cdot \frac{d a_i}{d \theta} \eta_i \left( s^{m,2}_i + s^f_i - \frac{F_i \left( \tilde{q}_i \right)}{1 - F_i \left( \tilde{q}_i \right)} s^{m,1}_i \right) \Bigg]
\label{eq: main_formula}
\end{multline}
\noindent\textbf{Proof.} \textit{See Appendix.}
\bigskip

One of the advantages of using the sufficient statistics approach is the transparency of the results. In particular, using equation \eqref{eq: main_formula}, I can decompose the aggregate effect of a tax reform into behavioral effects that operate through the men's working hours (the first and the third terms), the women's working hours (the fourth term), the spousal cross-effects of working hours (the second and the fifth terms), and, finally, the women's participation margin (the last term). It is useful to emphasize the difference between equation \eqref{eq: main_formula} and one that is obtained when households are modeled as single individuals. The first difference comes from the cross-elasticities, $\varepsilon^{mf}_i$ and $\varepsilon^{fm}_i$, that capture the changes in individual's working hours induced by the changes in spousal net-of-tax rate. The existing estimates of these elasticities are different from zero \citep{blau2007changes}, even though the empirical evidence is quite limited. In Section \ref{Quantitative Results}, I show that these terms matter for the overall effect. Second, my framework also accounts for the changes in husband's working hours following the wife's decision to join the labor force. Neither of these terms are present in the setting without couples.

In what follows, I construct the effective marginal and participation tax rates, reform-induced changes in the effective tax rates, and the expected labor income shares using the Current Population Survey data and the Internet NBER TAXSIM tax calculator. Furthermore, I take the estimates of the elasticities from the literature and, to study the sensitivity of the results, consider different ranges of values. Next, using these empirical estimates in equation \eqref{eq: main_formula}, I quantify the changes in economic efficiency caused by the tax reforms implemented in the United States over the last four decades.

\section{Data}\label{Data}

I use the data from the Annual Social and Economic Supplement (ASEC) of the Current Population Survey (CPS), or the ``March CPS''.\footnote{~The CPS data is extracted from IPUMS at \href{https://cps.ipums.org/cps}{https://cps.ipums.org/cps}. See \cite{flood2020cps}.} The March CPS contains data on annual earnings corresponding to the previous year. I define earnings as the sum of wage income and self-employment income. To be consistent with the model from Section \ref{Model}, I focus on different-sex married couples with working husbands in which both a husband and a wife are aged 25-54. Since I do not model education or retirement decisions, I do not include younger or older individuals. Furthermore, I exclude the couples where husbands do not have a reasonably strong labor market attachment. In particular, I drop the households where a husband earns less than a time-varying minimum threshold defined as one-half of the federal minimum wage times 520 hours (13 weeks at 40 hours per week) which amounts to annual earnings of \$1885 in 2012 USD \citep{guvenen2014nature}. Wives in my sample either work or not. To be consistent with the idea of reasonable labor market attachment of workers, I drop the couples where wives work but have annual earnings less than the time-varying minimum threshold described above. Using an alternative threshold of \$100 in 2012 USD does not change the results.

It is well known that the household survey data is subject to several caveats. First, there are two types of non-response: non-interview and item non-response. In the March CPS, non-response is accounted for by imputing missing values. As pointed out by \cite{meyer2015household}, the non-response rates in the major U.S. household surveys, including CPS, are growing over time. While imputation may cause serious problems for studying the trends over time, it works well if the object of interest is the cross-sectional distribution of individuals.\footnote{~For this reason, it is quite common in the literature, studying the trends, to drop the observations with imputed data \citep{ziliak2011earnings}.} I consider each tax reform separately, and use the cross-sectional distribution of couples in each pre-reform year. By this reason, I do not exclude the observations with imputed values. Second, earnings in the survey data are subject to bottom- or top-coding. Until 1995, the CPS used the traditional top-coding when the respondents, who reported income over the maximum allowed value, were assigned this maximum value. I drop all the top-coded and bottom-coded observations. In 1996-2010, the CPS used a replacement value system. The main difference with the traditional top-coding is that incomes above the maximum threshold are replaced by mean income of the other high-income individuals with similar demographic characteristics. Since 2011, the CPS has been using the rank proximity swapping procedure that preserves the distribution of values above the threshold. I do not drop the observations that are imputed using these procedures.

The summary statistics for the pre-reform years---1986, 1992, 2000, and 2017---is shown in Tables \ref{tab: cps_summary_1986_1992} and \ref{tab: cps_summary_2000_2017}. Several things worth emphasizing. First, mean and median annual hours of males have barely changed since the 1980s. In turn, mean and median annual hours of females have significantly increased. For example, the median went from 1400 hours in 1986 to 1872 hours in 2017, a 34 percent increase. Second, the employment rates among married women do not display such a significant increase. This observation echoes the discussion about stagnating female labor force participation in the United States \citep{blau2013female}. Third, in the 2010s the U.S. women are ahead of men in college education, even though in 1986 they were significantly behind \citep{goldin2014grand}. Finally, although not reported, the share of families receiving welfare benefits, such as the Aid to Families with Dependent Children (AFDC) and Temporary Assistance for Needy Families (TANF) or the Supplemental Nutrition Assistance Program (SNAP), is small in my sample. Therefore, in the simulations, I do not account for welfare benefits that are lost when a wife enters the labor market.

\section{The U.S. Tax Reforms}\label{The U.S. Tax Reforms}

\subsection{Background}\label{Background}

My goal is to evaluate the welfare gains of the labor income tax changes on married couples induced by four reforms in the United States: the Tax Reform Act of 1986, the Omnibus Budget Reconciliation Act of 1993, the Economic Growth and Tax Relief Reconciliation Act of 2001, and the Tax Cuts and Jobs Act of 2017. While they affected various parts of the tax code, I focus exclusively on labor income taxes. In what follows, I describe the main reform-induced changes in the tax schedule for married couples filing jointly.

The top-left panel of Figure \ref{fig:tax_changes_m} shows that the Tax Reform Act of 1986 significantly decreased the number of tax brackets. Despite the marginal tax rates were reduced for almost all the range of taxable income, they went up at the bottom of the income distribution and for the interval between \$60000 and \$69000 in 2012 USD. The top tax rate was decreased from 50 to 38.5 percent for tax year 1987, and then down to 28 percent in tax year 1988.\footnote{~In 1988-1990, the marginal tax rate structure included a 5 percent surtax within some range of taxable income.} Next, as reported in Tables \ref{tab:eitc_parameters} and \ref{tab: tax_parameters}, there was an expansion in the EITC, standard deductions, and personal exemptions.

Next, the top-right panel of Figure \ref{fig:tax_changes_m} reports that the Omnibus Budget Reconciliation Act of 1993 increased the top tax rate from 31 to 39.6 percent. Following the reform, the couples faced higher marginal tax rates for the taxable income above \$222000 in 2012 USD. However, on the other hand, the OBRA 1993 significantly expanded the EITC, thus benefiting low-income households \citep{kleven2020eitc}. As a result, this reform could potentially have different effects on married women with working husbands than on the other groups sensitive to the changes in the tax and transfer system, such as single women \citep{eissa1996labor, eissa2008evaluation}. The reason is that, in general, they belong to different parts of the income distribution, and the former are more likely affected by higher marginal tax rates rather than the EITC expansion.

The Economic Growth and Tax Relief Reconciliation Act of 2001 resulted into lower marginal tax rates for most tax brackets including the top income tax rate that went down from 39.6 to 35 percent. Moreover, the standard deduction for married couples filing jointly was increased relative to single filers. In particular, in 2000, the standard deduction for a married couple filing jointly was 67 percent higher than for a single filer, while in 2003 it became 100 percent higher.

Finally, the Tax Cuts and Jobs Act of 2017 featured changes in the federal income tax brackets and reduction in the marginal tax rates over almost the whole range of taxable income. The top income tax rate was decreased from 39.6 to 37 percent. Next, the standard deductions were increased, while personal exemptions were eliminated and itemized deductions were reduced. The individual income tax changes under the TCJA 2017 are effective for tax years 2018-2025. Moreover, there was a permanent shift from the Consumer Price Index (CPI) to the U.S. Chained Consumer Price Index (C-CPI-U) for indexing the tax brackets over time.

\subsection{Reform-Induced Changes in Tax Rates}\label{Reform-Induced Changes in Tax Rates}

I calculate tax liabilities and reform-induced changes in tax rates for each couple in my sample using NBER TAXSIM calculator.\footnote{~See \cite{feenberg1993introduction} for introduction to TAXSIM. Further details are available at \href{https://www.nber.org/taxsim/}{https://www.nber.org/taxsim/}.} This software provides accurate representation of the U.S. tax code and allows capturing the heterogeneous effects of tax reforms on households. In related papers, \cite{bick2017taxation} and \cite{bick2019long} emphasize the importance of accounting for nonlinearities of the labor income tax code for studying the effects of tax and transfer system on labor supply of married couples.

For each spouse in my sample, TAXSIM returns the federal, state, and the Federal Income Contributions Act (FICA) tax liabilities as well as corresponding effective marginal tax rates. In Online Appendix, I provide the full list of input variables and describe how I fill each field. To be consistent with the model from Section \ref{Model}, I abstract from all non-labor income. Next, because I do not explicitly model the children and childcare expenses, I set the number of children to two for each couple (this a median value for all the years in Tables \ref{tab: cps_summary_1986_1992} and \ref{tab: cps_summary_2000_2017}). I also assume that all couples choose joint filing.\footnote{~Married taxpayers in the United States pretty rarely file separate returns. According to IRS Income Statistics, in 2017 tax year about 95 percent of married couples filed jointly.} Finally, I assume that all the couples live in Michigan, a ``typical area'' in terms of state income taxation. Thus, heterogeneity in tax liabilities is solely driven by heterogeneity in couples' earnings. When I allow for variation in the factors that remain fixed in my analysis, the results do not significantly change.

To construct the participation and marginal tax rates, I need to know the potential earnings of all spouses, including non-working women. If a woman works, then I set her potential earnings to be equal to the actual earnings. If a woman does not work, her potential earnings are equal to her income in the case of entering the labor market. Since they are not observable, I apply a two-stage Heckman procedure to impute these earnings. I use the exclusion restrictions that the husband's earnings and the number of children under 6 do not directly influence the woman's earnings \citep{mulligan2008selection, bick2017taxation}.  Next, to obtain the expected labor income shares, I use the predicted probability of labor force participation as an empirical analogue of $F_i \left( \tilde{q}_i \right)$. Finally, I assume that workers bear the full incidence of employer payroll taxes. In this case, the proper measure of pre-tax labor income is equal to earnings plus the employer's share (50\%) of the FICA tax. Hence, when I construct the tax rates, I divide all of them by the factor of $(1 + 0.5 \cdot \text{FICA})$.

For each woman in the sample, I construct an effective participation tax rate. In particular, for a woman in couple $i$ it is given by
\begin{equation}
a_{it} = \frac{T_t \left( y^m_{it}, \hat{y}^f_{it}, Dem_{it} \right) - T_t \left( y^m_{it}, 0, Dem_{it} \right)}{\hat{y}^f_{it}}
\label{eq: participation_tax_data}
\end{equation}
where $y^m_{it}$ denotes the husband's taxable income in year $t$, $\hat{y}^f_{it}$ denotes the wive's taxable income in year $t$, $Dem_{it}$ denotes other TAXSIM inputs. I assume that the husband's earnings do not change when the wife enters the labor market.

The effective marginal tax rate in TAXSIM is calculated as the additional tax liabilities resulting from changing the taxable income by 10 cents. For example, for women it is given by
\begin{equation}
\tau^f_{it} = \frac{T_t \left( y^m_{it}, \hat{y}^f_{it} + \$ 0.1, Dem_{it} \right) - T_t \left( y^m_{it}, \hat{y}^f_{it}, Dem_{it} \right)}{\$ 0.1}
\label{eq: marginal_tax_data}
\end{equation}

\begin{figure}[t!]
\centering
\begin{subfigure}{.5\textwidth}
  \centering
  \includegraphics[width=\linewidth]{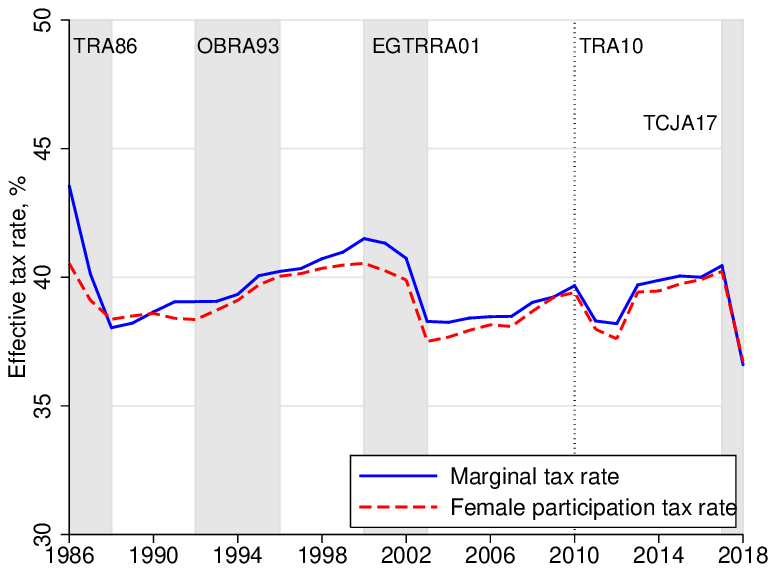}
  \label{fig: mtax_atax}
\end{subfigure}%
\begin{subfigure}{.5\textwidth}
  \centering
  \includegraphics[width=\linewidth]{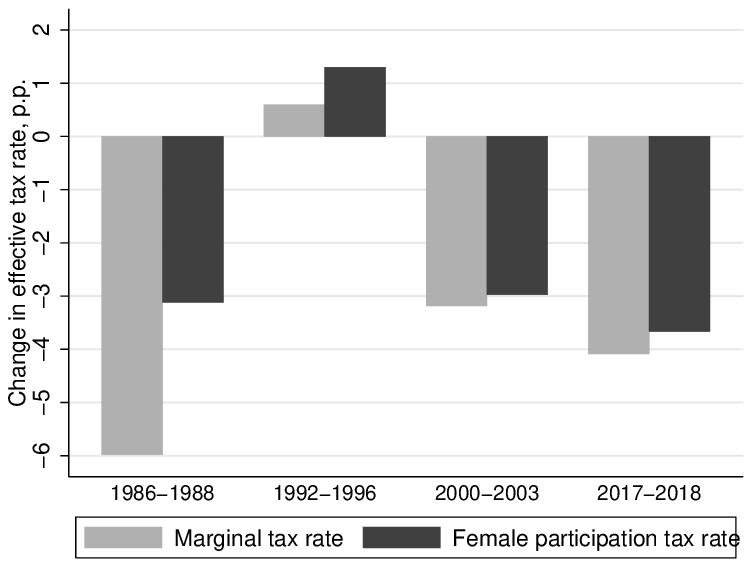}
  \label{fig: dmtax_datax}
\end{subfigure}
\caption{Left panel --- Mean effective marginal and female participation tax rates. Right panel --- Reform-induced changes in the mean effective marginal and female participation tax rates}
\label{fig:mt_at_tax}
\singlespacing\justify\footnotesize{\textsc{Left Panel Notes:} Marginal (solid blue) and participation (dashed red) tax rates include federal, state, and the FICA tax rates. Marginal tax rate series represents the mean marginal tax rate for males and females. Shaded areas indicate reform years. \textsc{Right Panel Notes:} To construct the changes, I apply the post-reform federal tax rules to the pre-reform taxable income and impute the post-reform federal income tax liabilities. The bars show the mean changes in the effective tax rates induced by the federal tax reforms. Changes in the marginal tax rates are calculated jointly for males and females.}
\end{figure}

The left panel of Figure \ref{fig:mt_at_tax} reports the income-weighted mean effective marginal and participation tax rates for my sample. The tax rates include federal, state, and the FICA tax rates. Grey shaded areas represent the periods of reforms when the changes in taxes came into effect. On the one hand, the TRA 1986, the EGTRRA 2001, and the TCJA 2017 resulted in a decrease in the mean effective tax rates for the married couples. On the contrary, the OBRA 1993 led to an increase in the effective tax rates.\footnote{~This is different from its effect on single women. In particular, their effective marginal and participation tax rates dropped, mainly due to the EITC expansion \citep{eissa2008evaluation}} The drop in the tax rates resulting from the Tax Relief, Unemployment Insurance Reauthorization, and Job Creation Act of 2010, that I do not analyze, was driven by a temporary reduction in the FICA tax.

The time series in the left panel of Figure \ref{fig:mt_at_tax} are driven by various factors, such as macroeconomic effects and behavioral responses. Furthermore, they capture the joint changes in the federal, state, and FICA tax rates. To isolate the changes in the effective tax rates separately induced by each federal tax reforms, I use the following procedure. For each spouse in pre-reform year $t$, I apply the federal tax rules of post-reform year $t+x$ to their year-$t$ real taxable income, keeping the state and FICA tax rules at year-$t$ level. Then I use the actual pre-reform and imputed post-reform federal tax liabilities to calculate the changes in the effective marginal and women's participation tax rates solely driven by the federal tax reforms. In this way, I find the empirical analogues of $d \tau^j_i / d \theta$ and $d a_i / d \theta$. The mean changes in the tax rates are reported in the right panel of Figure \ref{fig:mt_at_tax}.

\section{Quantitative Results}\label{Quantitative Results}

\subsection{Baseline Parameterization}\label{Baseline Parameterization}

To quantify the welfare effects of each tax reform separately, I take the pre-reform effective marginal and participation tax rates \eqref{eq: participation_tax_data} and \eqref{eq: marginal_tax_data}, the reform-induced changes in tax rates, the expected labor income shares, and the estimates of elasticities from the literature, and plug them into expression \eqref{eq: main_formula}. Because I assume that husband's earnings do not change when his wife starts working, the last bracket in \eqref{eq: main_formula} simplifies to $s^f_i$. The welfare gains are defined as \eqref{eq: main_formula} taken with the negative sign.

The rich existing empirical evidence suggests that women respond more along the participation margin rather than working hours margin when face tax and transfer changes \citep{bargain2014comparing}. As for men's elasticities, many studies suggest that their elasticity of working hours is very low and can almost be ignored for welfare purposes \citep{meghir2010labour}. The evidence on the cross-wage labor supply effects, that is crucial for my framework, is quite limited. In a related work, \cite{gayle2019optimal}, using the tax schedule corresponding to 2006, obtain the model-simulated working hours cross-elasticity is -0.08 for married men and -0.17 for married women. These numbers are, in general, consistent with existing empirical estimates \citep{blau2007changes}. Among other factors, the hours cross-elasticities may depend on the number of young children \citep{blundell2018children}. It is also worth noting that women's labor supply elasticities feature significant heterogeneity at the micro level, and any aggregate elasticity depends on the particular economic environment \citep{attanasio2018aggregating}.

Under the baseline parameterization, I set the elasticity of male hours to 0.05, the elasticity of female hours to 0.1, women's participation elasticity to 0.6, the hours cross-elasticity of males to -0.05, and the hours cross-elasticity of females to -0.1. In Section \ref{Sensitivity Analysis}, I conduct sensitivity analysis by varying the magnitudes of elasticities.

Beyond evaluating the actual welfare effects of tax reforms, it is also instructive to consider a benchmark case corresponding to a representative couple. This environment features no heterogeneity in income, tax rates, and tax rate changes. I assume $\tau^m = \tau^f$ because of tax system jointness. The pre-reform tax rates, $\tau$ and $a$, are given by the mean effective marginal and participation tax rates reported in the left panel of Figure \ref{fig:mt_at_tax}. The reform-induced tax changes, $d \tau / d \theta$ and $d a / d \theta$, are given by the mean changes in the tax rates reported in the right panel of Figure \ref{fig:mt_at_tax}. Overall, the expression for the effect of a small tax reform on economic efficiency is simplified to
\begin{equation}
\frac{d D/ d \theta}{W} = \frac{\tau}{1 - \tau} \cdot \frac{d \tau}{d \theta} \left[ \left( \varepsilon^m + \varepsilon^{mf} \right) s^m + \left( \varepsilon^f + \varepsilon^{fm} \right) s^f \right] + \frac{a}{1-a} \cdot \frac{d a}{d \theta} \eta s^f
\label{eq: representative_couple_welfare}
\end{equation}
where $s^m = w^m h^m / W$ denotes the labor income share of men, and $s^f = w^f h^f F \left( \tilde{q} \right) / W$ denotes the labor income share of women.

Table \ref{tab: results_main} reports the welfare gains for married couples resulting from each of four considered U.S. tax reforms. Total welfare gains (column 6) are decomposed into four parts: behavioral effects created along the intensive margin of men's labor supply (column 1), the intensive margin of women's labor supply (column 2), the participation margin of women (column 3), and, finally, the spousal cross-effects of working hours (column 4). To emphasize the quantitative importance of cross-effects, in column (5), I show total welfare gains if we abstract from them. Effectively, column (5) is the sum of own effects given in columns (1), (2), and (3). Column (7) displays the welfare gains calculated in a representative-couple economy, according to \eqref{eq: representative_couple_welfare}. Furthermore, to facilitate comparison across the reforms, in column (9), I report the welfare gain per dollar spent.

\begin{spacing}{1}
\begin{table}[t!]
\footnotesize
\centering
\caption{Welfare effects of labor income tax changes on married couples with working husbands}\label{tab: results_main}
\begin{tabular}{lccccccccc}
\hline \hline
& \multicolumn{7}{c}{Welfare gain, \% of aggregate labor income} && \\
\cline{2-8}
Reform & \begin{tabular}[t]{@{}c@{}}Intensive\\Males\end{tabular} & \begin{tabular}[t]{@{}c@{}}Intensive\\Females\end{tabular} & \begin{tabular}[t]{@{}c@{}}Extensive\\Females\end{tabular} & \begin{tabular}[t]{@{}c@{}}Cross-\\Effects\end{tabular} & \begin{tabular}[t]{@{}c@{}}Total\\w/o C.E.\end{tabular} & Total & RC & \begin{tabular}[t]{@{}c@{}}Tax Liab.\\Reduc., \%\end{tabular}& \begin{tabular}[t]{@{}c@{}}$\Delta$ Welfare/\\ \$ Spent\end{tabular} \\
& (1) & (2) & (3) & (4) & (5) & (6) & (7) & (8) & (9) \\
\hline
TRA86 & 0.19 & 0.18 & 0.45 & -0.27 & 0.82 & 0.55 & 0.44 & 7.20 & 1.08 \\
OBRA93 & -0.01 & -0.02 & -0.15 & 0.03 & -0.18 & -0.16 & -0.16 & 0.27 & 0.63 \\
EGTRRA01 & 0.09 & 0.12 & 0.40 & -0.17 & 0.61 & 0.44 & 0.42 & 7.19 & 1.07 \\
TCJA17 & 0.10 & 0.17 & 0.57 & -0.22 & 0.84 & 0.62 & 0.58 & 6.58 & 1.10 \\
\hline \hline
\end{tabular}
\justify\footnotesize{\textsc{Notes:} Welfare gains are calculated using \eqref{eq: main_formula} taken with the negative sign. I set $\varepsilon^{m} = 0.05$, $\varepsilon^{f} = 0.15$, $\varepsilon^{mf} = -0.05$, $\varepsilon^{fm} = -0.1$, and $\eta = 0.6$. The pre-reform tax rates and reform-induces changes in tax rates are calculated using NBER TAXSIM applied to the ASEC CPS data. Column (5) shows total welfare gains when the cross-effects are ignored, and calculated as $(1) + (2) + (3)$. Column (6) shows total welfare gains, and calculated as $(4) + (5)$. Column (7) shows the welfare gains in a representative-couple economy. Column (9) is calculated as $(8)/[(8) - (6)]$, where (8) is the decrease in tax liabilities as a share of labor income before behavioral responses.}
\end{table}
\end{spacing}
\smallskip

First, Table \ref{tab: results_main} shows that reform-induced changes in federal income tax rates result in the welfare gains that range from -0.16 to 0.62 percent of aggregate labor income. These numbers reflect the welfare effects that are driven by the labor supply behavioral responses. Three reforms---the TRA 1986, the EGTRRA 2001, and the TCJA 2017---created aggregate welfare gains, while the OBRA 1993 created welfare loss. Second, it follows from comparing columns (5) and (6) that the spousal working hours cross-effects are quantitatively important and therefore should not be ignored in the welfare analysis of policies. Otherwise, it may lead to overestimation of the welfare effects. For example, if I abstract from the cross-effects, I overestimate the welfare gains from the TCJA 2017 by 34.6\%. While this number seems to be high, the sensitivity analysis in Section \ref{Sensitivity Analysis} confirms the argument that the spousal cross-effects remain quantitatively important under any reasonable values of elasticities. The next conclusion from Table \ref{tab: results_main} is that the women's participation margin accounts for the bulk of total welfare gains. Again, ignoring this factor may lead to sizable bias in the estimates of policy welfare effects \citep{kleven2006marginal, eissa2008evaluation}. Another lesson from Table \ref{tab: results_main} is that a representative couple model, that uses the income-weighted mean tax rates and mean changes in tax rates, deliver the results that are close to ones reported in column (6). Hence, if we are primarily interested in assessing \textit{aggregate} welfare gains from tax reforms, a representative agent model may be a reasonable candidate device for this purpose. Finally, from the values in column (9), I conclude that the welfare gains vary between 0.63 and 1.10 USD per dollar spent.

The results reported in Table \ref{tab: results_main} provide a transparent decomposition of the aggregate welfare effects, thus highlighting one of the advantages of a sufficient statistics approach. Furthermore, when I construct the reform-induced changes in tax rates and then use them in formula \eqref{eq: main_formula}, I focus solely on the tax-driven labor supply behavioral responses. My results are not affected by the other incentives created by the reforms. For example, the TRA 1986 reform led to a shift of income that was previously labeled as corporate income to personal income \citep{guvenen2017top}. Despite these clear 
advantages, I also discuss the potential caveats and the ways to address them. First, my framework assumes that the reforms are small, and the measured efficiency gains represent a first-order approximation of the true effects. I use the pre-reform tax rates in \eqref{eq: main_formula}, however the reforms change the tax rates. The first-order approximation overstates the welfare gains of tax reductions and understates the welfare losses of tax increases \citep{kleven2021sufficient}. The only case that can be considered as a large reform is the reduction in the top tax rate during the TRA 1986. However, when I use the trapezoid approximation to evaluate the effects of this reform, the results do not dramatically change. Furthermore, since I consider each reform separately, I take into account the second-order effects across the reforms. Second, the elasticities may move in response to the reforms as well. \cite{blau2007changes} and \cite{heim2007incredible} report that between the 1970s and 2000s there was a dramatic reduction in own- and cross-elasticities of married women's labor supply. To address this caveat, I conduct sensitivity analysis using different combinations of elasticities. For example, moving from the 1980s to the 2010s may be viewed as moving from the ``high-elasticity'' to the ``low-elasticity'' parameterization described in Section \ref{Sensitivity Analysis}. Furthermore, there is substantial heterogeneity in labor supply elasticities, and the aggregate elasticities are not structural parameters \citep{attanasio2018aggregating}. To address this concern, I conduct sensitivity analysis and construct the lower and upper bounds for the welfare effects using reasonable ranges of elasticities. Finally, I derive \eqref{eq: main_formula} under the assumption that the tax and transfer function is linear. In Section \ref{Efficiency Loss and Nonlinear Taxation of Couples}, I show that the linearity assumption leads to overestimation of the welfare gains and characterize this linearization bias.

\begin{spacing}{1}
\begin{table}[t!]
\footnotesize
\centering
\caption{Welfare effects of labor income tax changes on married couples, sensitivity analysis}\label{tab: sensitivity_analysis}
\begin{tabular}{lccccccccc}
\hline \hline
& \multicolumn{7}{c}{Welfare gain, \% of aggregate labor income} & & \\
\cline{2-8}
Reform & \begin{tabular}[t]{@{}c@{}}Intensive\\Males\end{tabular} & \begin{tabular}[t]{@{}c@{}}Intensive\\Females\end{tabular} & \begin{tabular}[t]{@{}c@{}}Extensive\\Females\end{tabular} & \begin{tabular}[t]{@{}c@{}}Cross-\\Effects\end{tabular} & \begin{tabular}[t]{@{}c@{}}Total\\w/o C.E.\end{tabular} & Total & RC & \begin{tabular}[t]{@{}c@{}}Tax Liab.\\Reduc., \%\end{tabular} & \begin{tabular}[t]{@{}c@{}}$\Delta$ Welfare/\\ \$ Spent\end{tabular} \\
& (1) & (2) & (3) & (4) & (5) & (6) & (7) & (8) & (9) \\
\hline
\multicolumn{10}{c}{``Upper-Bound'' Parameterization: $\varepsilon^{m} = 0.1$, $\varepsilon^{f} = 0.2$, $\varepsilon^{mf} = 0$, $\varepsilon^{fm} = -0.05$, $\eta = 0.8$}\\
\hline
TRA86 & 0.39 & 0.24 & 0.60 & -0.08 & 1.23 & 1.15 & 1.03 & 7.20 & 1.19 \\
OBRA93$^*$ & 0.00 & -0.01 & -0.10 & 0.04 & -0.12 & -0.07 & -0.25 & 0.27 & 0.79 \\
EGTRRA01 & 0.18 & 0.16 & 0.54 & -0.04 & 0.88 & 0.84 & 0.77 & 7.19 & 1.13 \\
TCJA17 & 0.19 & 0.23 & 0.76 & -0.06 & 1.18 & 1.12 & 1.03 & 6.58 & 1.21 \\
\hline
\multicolumn{10}{c}{``Lower-Bound'' Parameterization: $\varepsilon^{m} = 0$, $\varepsilon^{f} = 0.1$, $\varepsilon^{mf} = -0.1$, $\varepsilon^{fm} = -0.15$, $\eta = 0.4$}\\
\hline
TRA86 & 0.00 & 0.12 & 0.30 & -0.47 & 0.42 & -0.05 & -0.14 & 7.20 & 0.99 \\
OBRA93$^*$ & -0.02 & -0.03 & -0.20 & 0.01 & -0.25 & -0.25 & -0.07 & 0.27 & 0.53 \\
EGTRRA01 & 0.00 & 0.08 & 0.27 & -0.30 & 0.35 & 0.05 & 0.06 & 7.19 & 1.01 \\
TCJA17 & 0.00 & 0.12 & 0.38 & -0.37 & 0.49 & 0.12 & 0.13 & 6.58 & 1.02 \\
\hline
\multicolumn{10}{c}{``High-Elasticity'' Parameterization: $\varepsilon^{m} = 0.1$, $\varepsilon^{f} = 0.2$, $\varepsilon^{mf} = -0.1$, $\varepsilon^{fm} = -0.15$, $\eta = 0.8$}\\
\hline
TRA86 & 0.39 & 0.24 & 0.60 & -0.47 & 1.23 & 0.75 & 0.57 & 7.20 & 1.12 \\
OBRA93 & -0.02 & -0.03 & -0.20 & 0.04 & -0.25 & -0.21 & -0.22 & 0.27 & 0.57 \\
EGTRRA01 & 0.18 & 0.16 & 0.54 & -0.30 & 0.88 & 0.57 & 0.54 & 7.19 & 1.09 \\
TCJA17 & 0.19 & 0.23 & 0.76 & -0.37 & 1.18 & 0.81 & 0.76 & 6.58 & 1.14 \\
\hline
\multicolumn{10}{c}{``Low-Elasticity'' Parameterization: $\varepsilon^{m} = 0$, $\varepsilon^{f} = 0.1$, $\varepsilon^{mf} = 0$, $\varepsilon^{fm} = -0.05$, $\eta = 0.4$}\\
\hline
TRA86 & 0.00 & 0.12 & 0.30 & -0.08 & 0.42 & 0.34 & 0.32 & 7.20 & 1.05 \\
OBRA93 & 0.00 & -0.01 & -0.10 & 0.01 & -0.12 & -0.11 & -0.11 & 0.27 & 0.72 \\
EGTRRA01 & 0.00 & 0.08 & 0.27 & -0.04 & 0.35 & 0.31 & 0.29 & 7.19 & 1.05 \\
TCJA17 & 0.00 & 0.12 & 0.38 & -0.06 & 0.49 & 0.44 & 0.40 & 6.58 & 1.07 \\
\hline
\multicolumn{10}{c}{Baseline Parameterization + Participation Elasticity Varies by Income Quintile}\\
\hline
TRA86 & 0.19 & 0.18 & 0.23 & -0.27 & 0.61 & 0.33 & - & 7.21 & 1.05 \\
OBRA93 & -0.01 & -0.02 & -0.21 & 0.03 & -0.24 & -0.21 & - & 0.27 & 0.56 \\
EGTRRA01 & 0.09 & 0.12 & 0.28 & -0.17 & 0.49 & 0.32 & - & 7.19 & 1.05 \\
TCJA17 & 0.10 & 0.17 & 0.34 & -0.22 & 0.61 & 0.39 & - & 6.58 & 1.06 \\
\hline \hline
\end{tabular}
\justify\footnotesize{\textsc{Notes:} Welfare gains are calculated using \eqref{eq: main_formula} taken with the negative sign. The pre-reform tax rates and reform-induces changes in tax rates are calculated using NBER TAXSIM applied to the ASEC CPS data. Column (5) shows total welfare gains when the cross-effects are ignored, and calculated as $(1) + (2) + (3)$. Column (6) shows total welfare gains, and calculated as $(4) + (5)$. Column (7) shows the welfare gains in a representative-couple economy. Column (9) is calculated as $(8)/[(8) - (6)]$, where (8) is the decrease in tax liabilities as a share of labor income before behavioral responses. In the last panel, the participation elasticity takes values 1/0.8/0.6/0.4/0.2 for the bottom/second/third/fourth/top couple's income quintiles, keeping the mean participation elasticity equal to 0.6.\\
$^*$ Since the OBRA 1993 increased or left unchanged the tax rates for most spouses in the sample, I use the parameters from the ``lower-bound'' scenario in the panel corresponding to the ``upper-bound'' scenario and vice-versa.}
\end{table}
\end{spacing}

\subsection{Sensitivity Analysis}\label{Sensitivity Analysis}

To explore the sensitivity of my results, I consider several alternative parameterizations of elasticities. The results are reported in Table \ref{tab: sensitivity_analysis}. First, I begin with the ``upper-bound'' scenario. Under this parameterization, own elasticities have reasonably high values ($\varepsilon^{m} = 0.1$, $\varepsilon^{f} = 0.2$, and $\eta = 0.8$) and cross-elasticities have reasonably low values ($\varepsilon^{mf} = 0$ and $\varepsilon^{fm} = -0.05$) for the TRA 1986, the EGTRRA 2001, and TCJA 2017 reforms, i.e. the reforms that feature reductions in the mean effective tax rates. On the contrary, for the OBRA 1993, I assume low own elasticities ($\varepsilon^{m} = 0$, $\varepsilon^{f} = 0.1$, and $\eta = 0.4$) and high cross-elasticities ($\varepsilon^{mf} = -0.1$ and $\varepsilon^{fm} = -0.15$). Under the ``upper-bound'' parameterization, the welfare gains range from -0.07 to 1.15 percent of aggregate labor income. Next, I consider the opposite scenario, namely, the ``lower-bound'' parameterization of elasticities. I flipped the values and assume low own elasticities and high cross-elasticities for the TRA 1986, the EGTRRA 2001, and TCJA 2017 reforms, and vice versa for the OBRA 1993. In this case, the welfare gains range from -0.25 to 0.12 percent of aggregate labor income. Overall, the first two panels of Table \ref{tab: sensitivity_analysis} can inform us about the bounds on efficiency gains resulting from the U.S. tax reforms.

Next, I consider two parameterizations labeled as ``high-elasticity'' and ``low-elasticity''. In particular, I set all the elasticities to high values in the former case, and low values in the latter case. The third and fourth panels of Table \ref{tab: sensitivity_analysis} may facilitate the comparison of reforms that were conducted in different time periods. If married women's elasticities shrank between the 1970s and 2000s \citep{blau2007changes, heim2007incredible}, then the conclusion from Table \ref{tab: results_main} that the TCJA 2017 reform created the largest efficiency gains among four reforms may be reconsidered.
\newpage

Finally, in the bottom panel of Table \ref{tab: sensitivity_analysis}, I report the results of an exercise where I allow the participation elasticity to decline in household income. In particular, I assign values 1/0.8/0.6/0.4/0.2 to the first/second/third/fourth/fifth income quintiles, keeping the mean participation elasticity equal to 0.6. Under this parameterization, total efficiency gains range from -0.21 to 0.39 percent of aggregate labor income. These numbers are lower than in the baseline scenario because high-income couples have smaller participation elasticities and hence benefit less from tax reductions. In turn, low-income couples benefit more, but the aggregate measure of efficiency gains masks heterogeneity in welfare effects. Beyond all these findings, the results from Table \ref{tab: sensitivity_analysis} reinforce my claim that cross-elasticities and participation elasticities quantitatively important and should be accounted for in the welfare analysis of economic policies.

\begin{figure}[t!]
\begin{subfigure}{.5\textwidth}
  \centering
  \includegraphics[width=\linewidth]{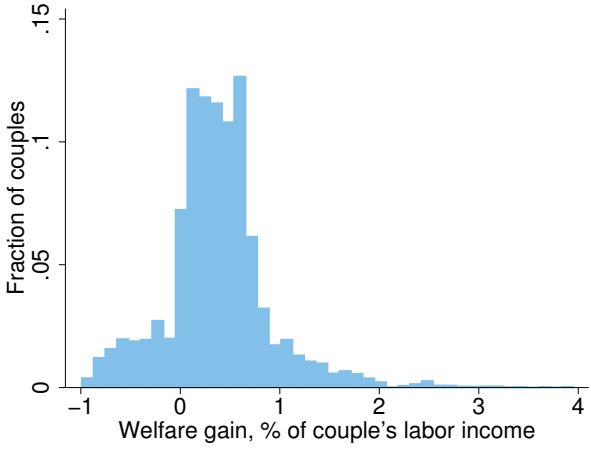}
  \subcaption{TRA 1986 reform}
  \label{fig: dwl_1986_distribution}
\end{subfigure}%
\begin{subfigure}{.5\textwidth}
  \centering
  \includegraphics[width=\linewidth]{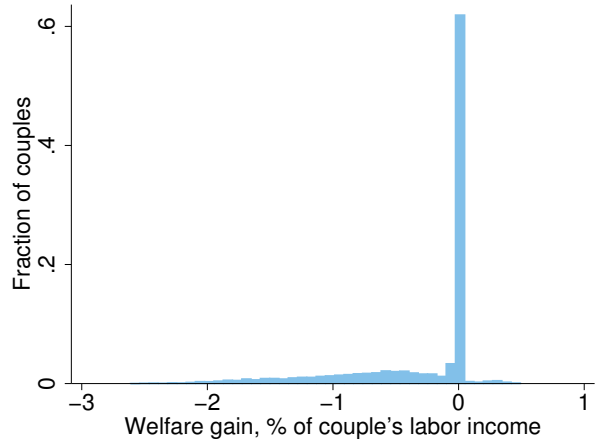}
  \subcaption{OBRA 1993 reform}
  \label{fig: dwl_1992_distribution}
\end{subfigure}
\bigskip

\begin{subfigure}{.5\textwidth}
  \centering
  \includegraphics[width=\linewidth]{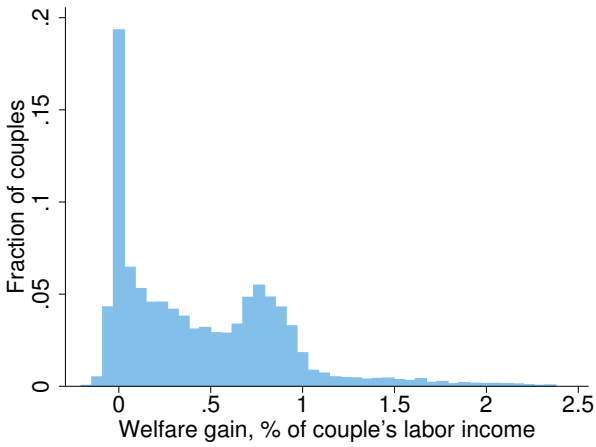}
  \subcaption{EGTRRA 2001 reform}
  \label{fig: dwl_2000_distribution}
\end{subfigure}%
\begin{subfigure}{.5\textwidth}
  \centering
  \includegraphics[width=\linewidth]{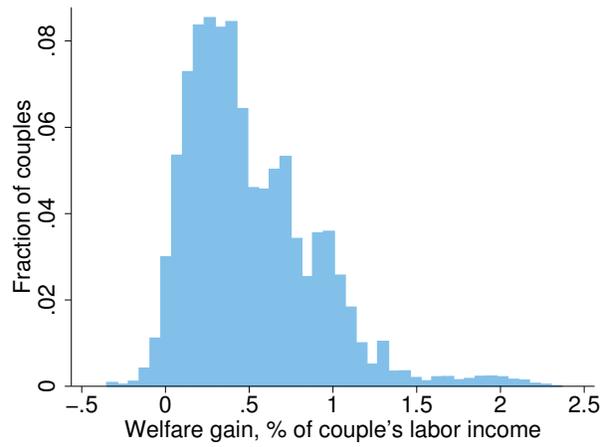}
  \subcaption{TCJA 2017 reform}
  \label{fig: dwl_2017_distribution}
\end{subfigure}
\caption{Distribution of reform-induced welfare gains among couples}
\label{fig: dwl_dist_couples}
\justify\footnotesize{\textsc{Notes:} Welfare gains are calculated under a baseline parameterization of elasticities.}
\end{figure}

\subsection{Welfare Gains Distribution}\label{Welfare Gains Distribution}

So far I consider the aggregate welfare effects of tax reforms, however the use of microdata combined with TAXSIM allows studying the distribution of welfare gains and losses. Indeed, the aggregate effects may mask significant heterogeneity across households.\footnote{~Using the U.S. data, \cite{zidar2019tax} finds that the positive relationship between tax cuts and employment growth is largely driven by tax cuts for low-income group, rather than high-income individuals.} In this section, I use the baseline parameterization of elasticities and answer the following questions. How are the efficiency gains from tax reforms distributed in the population of married couples? Furthermore, according to Table \ref{tab: results_main}, three reforms created aggregate welfare gains. Are there any losers? Finally, how do the welfare gains vary by income?

\begin{spacing}{1}
\begin{table}[t!]
\centering
\caption{Distribution of welfare gains for couples, \% of couple's labor income}\label{tab: gains_percentiles}
\begin{tabular}{lccccc}
\hline \hline
Reform & P10 & P25 & P50 & P75 & P90 \\
\hline
TRA86 & -0.21 & 0.13 & 0.37 & 0.61 & 0.94 \\
OBRA93 & -1.09 & -0.40 & 0.00 & 0.00 & 0.00 \\
EGTRRA01 & 0.00 & 0.06 & 0.36 & 0.76 & 0.95 \\
TCJA17 & 0.10 & 0.23 & 0.42 & 0.74 & 1.02 \\
\hline \hline
\end{tabular}
\justify\footnotesize{\textsc{Notes:} Welfare gains are calculated under a baseline parameterization of elasticities, and are measured as a share of couple's labor income.}
\end{table}
\end{spacing}

Figure \ref{fig: dwl_dist_couples} documents the distribution of welfare gains among couples for each reform. A simple visual inspection confirms the argument that income tax changes create heterogeneous welfare effects, and there are both winners and losers from each reform. In Table \ref{tab: gains_percentiles}, I report the percentiles of efficiency gains distribution. For the TRA 1986, the median welfare gain is equal to 0.37 percent of couple's labor income. While a substantial fraction of couples win from the reform, those at the 10th percentile experience welfare loss equal to 0.21 percent of the labor income. On the contrary, couples at the 90th percentile receive welfare gain of 0.94 percent of labor income. Next, following the OBRA 1993, the median couple stay welfare neutral. However, couples at the 10th percentile experience welfare loss of 1.09 percent of labor income. In the case of the other two reforms, the EGTRRA 2001 and the TCJA 2017, the values corresponding to the 10th percentile are non-negative. One more observation that follows from Figure \ref{fig: dwl_dist_couples} and Table \ref{tab: gains_percentiles} is that the dispersion of the efficiency gains significantly differs across the reforms. Apart from the OBRA 1993, where most of couples are welfare neutral, the P75-P25 ratio for the TRA 1986, the EGTRRA 2001, and the TCJA 2017 is equal to 4.7, 12.7, and 3.2 correspondingly.

Next, in Table \ref{tab: winners_losers}, I report the fractions of winners, losers, and welfare-neutral couples. I define winners as those with welfare gains above 0.1 percent of couple's labor income. Losers are defined as those with welfare losses greater than 0.1 percent of labor income. Finally, welfare-neutral couples are those whose absolute values of welfare gains or losses do not exceed 0.1 percent of labor income. It follows that, despite the TRA 1986 created aggregate welfare gains, it left 12.3 percent of couples with welfare losses. In turn, while the OBRA 1993 created aggregate welfare losses, for about two-thirds of the married couples this reform was welfare-neutral. In the case of EGTRRA 2001 and the TCJA 2017 the share of winners is equal to 69.6 and 90.3 percent correspondingly.

\begin{spacing}{1}
\begin{table}[t!]
\centering
\caption{Fractions of winners, losers, and welfare-neutral couples}\label{tab: winners_losers}
\begin{tabular}{lccc}
\hline \hline
Reform & Winners, \% & Losers, \% & Neutral, \% \\
\hline
TRA86 & 78.7 & 12.3 & 9.1 \\
OBRA93 & 1.4 & 31.2 & 67.4 \\
EGTRRA01 & 69.6 & 0.3 & 30.1 \\
TCJA17 & 90.3 & 0.6 & 9.0 \\
\hline \hline
\end{tabular}
\justify\footnotesize{\textsc{Notes:} Welfare gains are calculated under a baseline parameterization of elasticities. Winners are defined as couples with welfare gains above 0.1 percent of couple's labor income. Losers are defined as couples with welfare losses greater than 0.1 percent of labor income. Welfare-neutral couples are defined as those whose absolute values of welfare gains or losses do not exceed 0.1 percent of labor income.}
\end{table}
\end{spacing}

\begin{figure}[b!]
\centering
\begin{subfigure}{.5\textwidth}
  \centering
  \includegraphics[width=\linewidth]{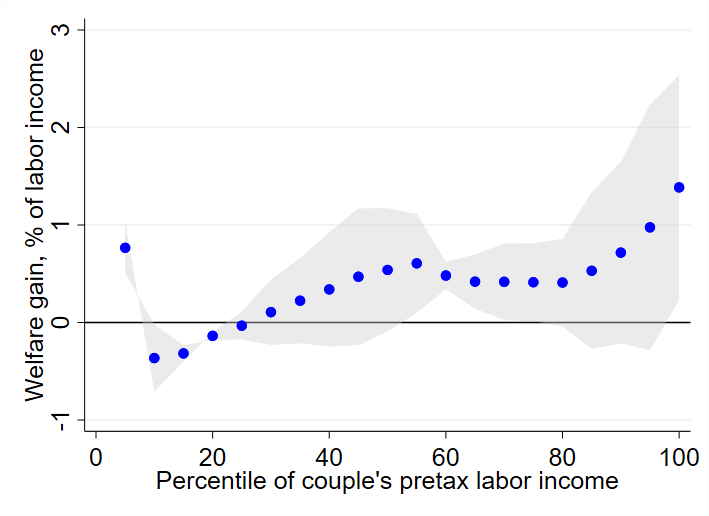}
  \subcaption{TRA 1986 reform}
  \label{fig: dwl_1986}
\end{subfigure}%
\begin{subfigure}{.5\textwidth}
  \centering
  \includegraphics[width=\linewidth]{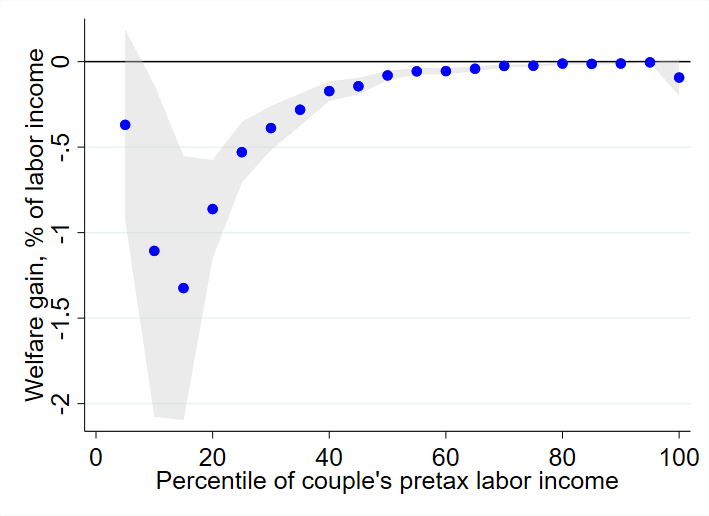}
  \subcaption{OBRA 1993 reform}
  \label{fig: dwl_1992}
\end{subfigure}
\medskip

\begin{subfigure}{.5\textwidth}
  \centering
  \includegraphics[width=\linewidth]{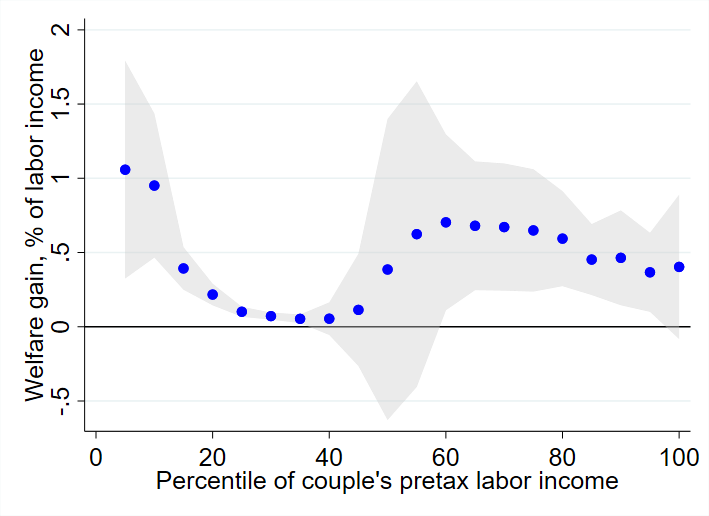}
  \subcaption{EGTRRA 2001 reform}
  \label{fig: dwl_2000}
\end{subfigure}%
\begin{subfigure}{.5\textwidth}
  \centering
  \includegraphics[width=\linewidth]{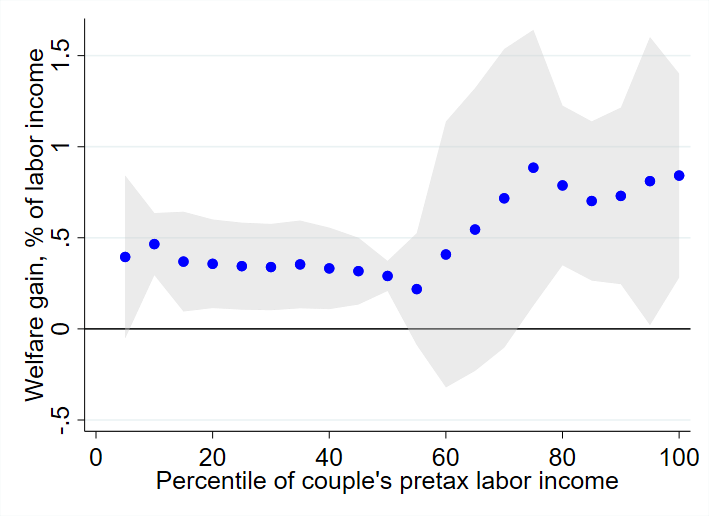}
  \subcaption{TCJA 2017 reform}
  \label{fig: dwl_2017}
\end{subfigure}
\caption{Welfare gains and income distribution, couples}
\label{fig: dwl_couples}
\justify\footnotesize{\textsc{Notes:} Welfare gains are measured as a percentage of the labor income. Each dot represents 5 percent of the sample. The grey shaded area represents the interval between ``lower-bound'' and ``upper-bound'' elasticity parameterizations.}
\end{figure}
\smallskip

Finally, in Figure \ref{fig: dwl_couples}, I explore how the welfare gains vary by income. Each dot represents 5 percent of the sample, and the grey shaded areas represent the interval between ``lower-bound'' and ``upper-bound'' elasticity parameterizations. Figure \ref{fig: dwl_couples} clearly shows that efficiency gains change nonlinearly with income. There are two general patterns. First, the TRA 1986, the OBRA 1993 (excluding the bottom 10 percent), and the TCJA 2017 can be characterized as monotonic tax reforms \citep{bierbrauer2021politically} and they resulted in monotonic relationships between welfare gains and income. Overall, richer taxpayers benefited from these reforms more than poorer taxpayers. Second, the OBRA 1993 and the EGTRRA 2001 reforms demonstrate a U-shaped pattern in the welfare gains. In this case, the main winners of the reforms are located at the lower and upper ends of the income distribution. Interestingly, \cite{hotchkiss2012assessing} and \cite{hotchkiss2021impact} discover similar patterns of the welfare gains despite using very different methodology. Overall, the results from this section reassure that despite a representative couple model can be a reasonable candidate for assessing aggregate efficiency gains, it does not capture rich heterogeneity and misses important distributional aspects of tax reforms.

\subsection{Counterfactual Tax Reforms}\label{Counterfactual Tax Reforms}

In this section, I conduct two sets of counterfactual tax reforms aimed at addressing the following questions. First, how does the pre-reform income distribution matter for my results? Second, how do the initial conditions---pre-reform income distribution and tax law---jointly matter for the estimates of welfare gains?

In Table \ref{tab: counterfactual_reform_1}, I report the results for the first set of counterfactual reforms. In this exercise, I take the couples' income distribution in pre-reform year $t$ (for example, in 1986), and apply the pre- and post-reform $X$'s (for example, the TCJA 2017) tax laws. Table \ref{tab: counterfactual_reform_1} consists of four panels, where each panel represents the income distribution that I use. For example, Panel A shows the results for four reforms applied to the 1986 income distribution. The first column displays the reforms. Column (9) reports the percentage difference between counterfactual and actual welfare gains per dollar spent, shown in column (8). By construction, the results in the first line of Panel A coincide with one from Table \ref{tab: results_main}, and hence there is zero in column (9). The results from the bottom line of Panel A should be interpreted in the following way: if the TCJA 2017 were to be applied to the 1986 income distribution, total welfare gains would be 0.40 percent of aggregate labor income. It follows from column (9) that this counterfactual 2017 TCJA reform would deliver 2.68\% less welfare gain per dollar spent than the actual TCJA 2017 (reported in the bottom line of Panel D). In turn, the first line of Panel D shows that if the TRA 1986 reform were to be applied to the 2017 income distribution, then the welfare gain per dollar spent would be 5.48\% higher than from the actual TRA 1986. Overall, the results from Table \ref{tab: counterfactual_reform_1} suggest that despite the initial income distribution matters, it has limited quantitative importance. Counterfactual welfare gains per dollar spent do not differ by more than 7.54\% from the actual ones.

\begin{spacing}{1}
\begin{table}[t!]
\footnotesize
\centering
\caption{Welfare effects of tax reforms applied to counterfactual income distribution}\label{tab: counterfactual_reform_1}
\begin{tabular}{lccccccccc}
\hline \hline
& \multicolumn{7}{c}{Welfare gain, \% of aggregate labor income} & & \\
\cline{2-7}
Reform & \begin{tabular}[t]{@{}c@{}}Intensive\\Males\end{tabular} & \begin{tabular}[t]{@{}c@{}}Intensive\\Females\end{tabular} & \begin{tabular}[t]{@{}c@{}}Extensive\\Females\end{tabular} & \begin{tabular}[t]{@{}c@{}}Cross-\\Effects\end{tabular} & Total & RC & \begin{tabular}[t]{@{}c@{}}Tax Liab.\\Reduc., \%\end{tabular} & \begin{tabular}[t]{@{}c@{}}$\Delta$ Welfare/\\ \$ Spent\end{tabular} & Diff., \% \\
& (1) & (2) & (3) & (4) & (5) & (6) & (7) & (8) & (9) \\
\hline
\multicolumn{10}{c}{Panel A: Tax Reforms Applied to Pre-TRA86 Distribution of Couples}\\
\hline
TRA86 & 0.19 & 0.18 & 0.45 & -0.27 & 0.55 & 0.44 & 7.20 & 1.08 & 0.00 \\
OBRA93 & -0.01 & -0.02 & -0.13 & 0.02 & -0.14 & -0.14 & 0.29 & 0.68 & +7.54 \\
EGTRRA01 & 0.09 & 0.11 & 0.36 & -0.16 & 0.40 & 0.37 & 7.46 & 1.06 & -0.80 \\
TCJA17 & 0.09 & 0.12 & 0.36 & -0.18 & 0.40 & 0.37 & 5.76 & 1.07 & -2.68 \\
\hline
\multicolumn{10}{c}{Panel B: Tax Reforms Applied to Pre-OBRA93 Distribution of Couples}\\
\hline
TRA86 & 0.19 & 0.22 & 0.53 & -0.30 & 0.63 & 0.51 & 7.38 & 1.09 & +1.09 \\
OBRA93 & -0.01 & -0.02 & -0.15 & 0.03 & -0.16 & -0.16 & 0.27 & 0.63 & 0.00 \\
EGTRRA01 & 0.08 & 0.12 & 0.39 & -0.16 & 0.43 & 0.40 & 7.38 & 1.06 & -0.32 \\
TCJA17 & 0.09 & 0.14 & 0.41 & -0.18 & 0.45 & 0.42 & 5.87 & 1.08 & -1.88 \\
\hline
\multicolumn{10}{c}{Panel C: Tax Reforms Applied to Pre-EGTRRA01 Distribution of Couples}\\
\hline
TRA86 & 0.33 & 0.31 & 0.82 & -0.48 & 0.97 & 0.76 & 10.23 & 1.11 & +2.11 \\
OBRA93 & -0.04 & -0.04 & -0.18 & 0.07 & -0.19 & -0.20 & -0.97 &  &  \\
EGTRRA01 & 0.09 & 0.12 & 0.40 & -0.17 & 0.44 & 0.42 & 7.19 & 1.07 & 0.00 \\
TCJA17 & 0.10 & 0.14 & 0.44 & -0.20 & 0.48 & 0.45 & 6.19 & 1.08 & -1.80 \\
\hline
\multicolumn{10}{c}{Panel D: Tax Reforms Applied to Pre-TCJA17 Distribution of Couples}\\
\hline
TRA86 & 0.29 & 0.42 & 1.13 & -0.52 & 1.32 & 1.05 & 10.62 & 1.14 & +5.48 \\
OBRA93 & -0.03 & -0.05 & -0.22 & 0.07 & -0.24 & -0.25 & -0.96 &  &  \\
EGTRRA01 & 0.08 & 0.13 & 0.48 & -0.17 & 0.52 & 0.49 & 7.15 & 1.08 & +1.18 \\
TCJA17 & 0.10 & 0.17 & 0.57 & -0.22 & 0.62 & 0.58 & 6.58 & 1.10 & 0.00 \\
\hline \hline
\end{tabular}
\justify\footnotesize{\textsc{Notes:} Welfare gains are calculated using \eqref{eq: main_formula} taken with the negative sign and under a baseline parameterization of elasticities. In each exercise, I take the distribution of couples corresponding to some pre-reform year, as indicated in four panels, and calculate the welfare effects from applying the reform that is shown in the left-most column. Column (5) shows total welfare gains, and calculated as $(1) + (2) + (3) + (4)$. Column (6) shows the welfare gains in a representative-couple economy. Column (8) is calculated as $(7)/[(7) - (5)]$, where (7) is the decrease in tax liabilities as a share of labor income before behavioral responses. Column (9) shows the percentage difference relative to the actual welfare gains from the reform.}
\end{table}
\end{spacing}

Next, I conduct a set of counterfactual reforms, where I take the distribution of couples and the tax law corresponding to some pre-reform year $t$ (for example, in 1986) and apply the post-reform $X$'s (for example, the TCJA 2017) tax law. In other words, in this example, I study the welfare consequences of moving from the pre-TRA 1986 economy to the post-TCJA 2017 economy. Table \ref{tab: counterfactual_reform_2} reports the results. I do not conduct the backward reforms between two consecutive reforms (e.g., I do not consider the welfare consequences of moving from the pre-OBRA 1993 economy to the post-TRA 1986 economy) because the welfare effects are likely to be negligible in these cases. Panel A suggests two interesting findings. First, moving from the pre-TRA 1986 economy to the post-EGTRRA 2001 and post-TCJA 2017 economies leads to substantial reduction in tax liabilities, 17.85 and 22.28 percent, relative to the actual TRA 1986 reform (7.20 percent), and higher efficiency gains, 0.88 and 1.19 percent of aggregate labor income (the actual TRA 1986 results in 0.55 percent of aggregate labor income). However, when I make the efficiency gains comparable, column (8) shows that the actual TRA 1986 generated more welfare gain per dollar spent than the alternative considered counterfactual reforms.

\begin{spacing}{1}
\begin{table}[t!]
\footnotesize
\centering
\caption{Welfare effects of counterfactual tax reforms}\label{tab: counterfactual_reform_2}
\begin{tabular}{lcccccccc}
\hline \hline
& \multicolumn{7}{c}{Welfare gain, \% of aggregate labor income} & \\
\cline{2-7}
Reform & \begin{tabular}[t]{@{}c@{}}Intensive\\Males\end{tabular} & \begin{tabular}[t]{@{}c@{}}Intensive\\Females\end{tabular} & \begin{tabular}[t]{@{}c@{}}Extensive\\Females\end{tabular} & \begin{tabular}[t]{@{}c@{}}Cross-\\Effects\end{tabular} & Total & RC & \begin{tabular}[t]{@{}c@{}}Tax Liab.\\Reduc., \%\end{tabular} & \begin{tabular}[t]{@{}c@{}}$\Delta$ Welfare/\\ \$ Spent\end{tabular} \\
& (1) & (2) & (3) & (4) & (5) & (6) & (7) & (8) \\
\hline
\multicolumn{9}{c}{Panel A: Tax Reforms Applied to Pre-TRA86 Distribution of Couples and Tax Law}\\
\hline
TRA86 & 0.19 & 0.18 & 0.45 & -0.27 & 0.55 & 0.44 & 7.20 & 1.08 \\
OBRA93 & 0.19 & 0.17 & 0.35 & -0.27 & 0.44 & 0.29 & 7.73 & 1.06 \\
EGTRRA01 & 0.27 & 0.27 & 0.75 & -0.41 & 0.88 & 0.74 & 17.85 & 1.05 \\
TCJA17 & 0.36 & 0.38 & 1.02 & -0.58 & 1.19 & 0.96 & 22.28 & 1.06 \\
\hline
\multicolumn{9}{c}{Panel B: Tax Reforms Applied to Pre-OBRA93 Distribution of Couples and Tax Law}\\
\hline
TRA86 & --- & --- & --- & --- & --- & --- & --- & --- \\
OBRA93 & -0.01 & -0.02 & -0.15 & 0.03 & -0.16 & -0.16 & 0.27 & 0.63 \\
EGTRRA01 & 0.06 & 0.09 & 0.26 & -0.12 & 0.29 & 0.27 & 10.09 & 1.03 \\
TCJA17 & 0.13 & 0.19 & 0.51 & -0.25 & 0.57 & 0.52 & 14.69 & 1.04 \\
\hline
\multicolumn{9}{c}{Panel C: Tax Reforms Applied to Pre-EGTRRA01 Distribution of Couples and Tax Law}\\
\hline
TRA86 & 0.09 & 0.08 & 0.25 & -0.15 & 0.27 & 0.22 & -0.74 &  \\
OBRA93 & --- & --- & --- & --- & --- & --- & --- & --- \\
EGTRRA01 & 0.09 & 0.12 & 0.40 & -0.17 & 0.44 & 0.42 & 7.19 & 1.07 \\
TCJA17 & 0.15 & 0.23 & 0.69 & -0.31 & 0.76 & 0.70 & 12.16 & 1.07 \\
\hline
\multicolumn{9}{c}{Panel D: Tax Reforms Applied to Pre-TCJA17 Distribution of Couples and Tax Law}\\
\hline
TRA86 & 0.03 & 0.02 & 0.05 & -0.05 & 0.05 & -0.02 & -6.40 &  \\
OBRA93 & -0.03 & -0.06 & -0.26 & 0.07 & -0.27 & -0.29 & -7.38 &  \\
EGTRRA01 & --- & --- & --- & --- & --- & --- & --- & --- \\
TCJA17 & 0.10 & 0.17 & 0.57 & -0.22 & 0.62 & 0.58 & 6.58 & 1.10 \\
\hline \hline
\end{tabular}
\justify\footnotesize{\textsc{Notes:} Welfare gains are calculated using \eqref{eq: main_formula} taken with the negative sign and under a baseline parameterization of elasticities. In each exercise, I take the distribution of couples and the tax law corresponding to some pre-reform year, as indicated in four panels, and calculate the welfare effects from applying the reform that is shown in the left-most column. Column (5) shows total welfare gains, and calculated as $(1) + (2) + (3) + (4)$. Column (6) shows the welfare gains in a representative-couple economy. Column (8) is calculated as $(7)/[(7) - (5)]$, where (7) is the decrease in tax liabilities as a share of labor income before behavioral responses.}
\end{table}
\end{spacing}

\section{Efficiency Loss and Nonlinear Taxation of Couples}\label{Efficiency Loss and Nonlinear Taxation of Couples}

\subsection{Linearization Bias}\label{Linearization Bias}

The framework in Section \ref{Model} is elaborated under linear tax and transfer function. Despite the real tax and transfer schedules often feature nonlinearities, this assumption is widely used in the literature studying the efficiency gains of tax reforms \citep{chetty2009sufficient}. How sizable is the bias in the estimates of welfare gains resulting from the linearity assumption? In this section, I address this question by extending the framework of \cite{blomquist2019marginal}, who study linearization bias through the lens of the model with singles, to the setting with couples.

I use a version of the model from Section \ref{Model}. In particular, consider an economy populated by couples with preferences $v \left( c, y_m, y_f, \upsilon_m, \upsilon_f \right)$ where $y_m$ and $y_f$ are taxable incomes of a male and a female, $\upsilon_m$ and $\upsilon_f$ are idiosyncratic preference parameters jointly drawn from continuous distribution $\varGamma$. To focus on the bias that comes from tax function linearization, I abstract from the labor force participation margin and consider an economy populated by dual-earner couples, i.e. in all couples both spouses are employed. Furthermore, I state the problem in terms of taxable rather than labor income.\footnote{~\cite{feldstein1999tax} suggests that the relevant statistic for calculating efficiency loss is the elasticity of taxable income because people can adjust their behavior along different margins, not only labor supply.} The couple's budget constraint is given by $c = y_m + y_f - T \left( y_m, y_f, \theta \right) + I$ where $I$ is lump-sum non-taxable income. Following the similar steps as in Section \ref{Model}, I obtain the expression for marginal deadweight loss, $d D / d \theta$.

Now suppose that the original tax and transfer function $T$ is replaced by a linearized function $T^L$ that delivers the same solution as the original problem, $\left( c^*, y^*_m, y^*_f \right)$. In particular, the latter is described by proportional tax rates $\tau_m = \partial T \left( y_m, y_f, \theta \right) / \partial y_m$ and $\tau_f = \partial T \left( y_m, y_f, \theta \right) / \partial y_f$ and a lump-sum component. Namely,
\begin{equation}
T^L \left( y_m, y_f, \tau_m, \tau_f \right) = \tau_m (\theta) y_m + \tau_f (\theta) y_f + T^*
\label{eq: linearized_tax}
\end{equation}
where I set $T^* = y_m^* + y_f^* - c^* - \tau_m y^*_m + \tau_f y^*_f + I$. Again, following the steps from Section \ref{Model} and using the linearized budget constraint \eqref{eq: linearized_tax}, I obtain the expression for marginal deadweight loss, $d D^L / d \theta$.

How does the original reform-induced efficiency loss, $d D / d \theta$, differ from the change in efficiency loss under a linearized tax and transfer function, $d D^L / d \theta$? Proposition 2 characterizes these objects and reveals that they depend on two sets of terms, utility curvature and tax function curvature. In particular, from binding $v \left( c, y_m, y_f, \upsilon_m, \upsilon_f \right) = \bar{U}$, obtain the inverse, $c = \psi \left( y_m, y_f, \upsilon_m, \upsilon_f, \bar{U} \right)$. Next, denote $\psi''_{ij} \equiv \partial^2 \psi (\cdot) / \partial \tilde{y}_i \partial \tilde{y}_j$, $T''_{ij} \equiv \partial^2 T (\cdot) / \partial \tilde{y}_i \partial \tilde{y}_j$, $T''_{i \theta} \equiv \partial^2 T (\cdot) / \partial \tilde{y}_i \partial \theta$, and, finally, $T'_i \equiv \partial T (\cdot) / \partial y_i$. Then, $\psi$-terms account for utility curvature and $T$-terms account for tax function curvature.
\bigskip

\noindent\textbf{Proposition 2 (Efficiency Loss under Nonlinear Taxation of Couples).} \textit{
Under nonlinear tax function $T$, efficiency loss from any arbitrary small tax reform $d \theta \approx 0$ is given by}
\begin{multline}
\frac{d D}{d \theta} = - \int \Bigg[ \frac{T'_m \left[ \left( \psi''_{mf} + T''_{mf} \right) T''_{f \theta} - \left( \psi''_{ff} + T''_{ff} \right) T''_{m \theta} \right]}{\left( \psi''_{mm} + T''_{mm} \right) \left( \psi''_{ff} + T''_{ff} \right) - \left( \psi''_{mf} + T''_{mf} \right)^2} +\\
\frac{T'_f \left[ \left( \psi''_{mf} + T''_{mf} \right) T''_{m \theta} - \left( \psi''_{mm} + T''_{mm} \right) T''_{f \theta} \right]}{\left( \psi''_{mm} + T''_{mm} \right) \left( \psi''_{ff} + T''_{ff} \right) - \left( \psi''_{mf} + T''_{mf} \right)^2} \Bigg] d\varGamma \left( \upsilon_m, \upsilon_f \right)
\label{eq: proposition_2_nonlinear}
\end{multline}

\noindent\textit{Under linearized tax function $T^L$, efficiency loss from any arbitrary small tax reform $d \theta \approx 0$ is given by}
\begin{equation}
\frac{d D^L}{d \theta} = - \int \Bigg[ \frac{T'_m \left( \psi''_{mf} T''_{f \theta} - \psi''_{ff} T''_{m \theta} \right)}{\psi''_{mm} \psi''_{ff} - \left( \psi''_{mf} \right)^2} + \frac{T'_f \left( \psi''_{mf} T''_{m \theta} - \psi''_{mm} T''_{f \theta} \right)}{\psi''_{mm} \psi''_{ff} - \left( \psi''_{mf} \right)^2} \Bigg] d\varGamma \left( \upsilon_m, \upsilon_f \right)
\label{eq: proposition_2_linear}
\end{equation}

\noindent\textbf{Proof.} \textit{See Appendix.}
\bigskip
\newpage

To obtain \eqref{eq: proposition_2_linear} from \eqref{eq: proposition_2_nonlinear}, one should simply set $T''_{mm} = T''_{ff} = T''_{mf} = 0$. Expressions \eqref{eq: proposition_2_nonlinear} and \eqref{eq: proposition_2_linear} are generalized versions of equations (5) and (8) from \cite{blomquist2019marginal} who derive them for the economy populated by singles. To demonstrate that their result is a special case of Proposition 2, set to zero all joint terms, $\psi''_{mf} = T''_{mf} = 0$, in \eqref{eq: proposition_2_nonlinear}, and then marginal deadweight loss for individual of gender $j$ is given by $T'_j T''_{j \theta} / \left( \psi''_{jj} + T''_{jj} \right)$, and this is (5) in \cite{blomquist2019marginal}. The linearized version \eqref{eq: proposition_2_linear} is given by $T'_j T''_{j \theta} / \psi''_{jj}$, and this is (8) in their paper.

Define the \textit{linearization bias} as the percentage difference between reform-induced efficiency loss under linearized and original tax and transfer functions:
\begin{equation}
\Delta = \frac{\frac{d D^L}{d \theta} - \frac{d D}{d \theta}}{\frac{d D}{d \theta}}
\label{eq: dwl_lin_bias_definition}
\end{equation}

While Proposition 2 suggests that the size of this bias is affected by the curvatures of utility and tax function, it is instructive to assume the functional forms for utility $v$ and tax and transfer function $T$ and obtain the expression for $\Delta$ in terms of model parameters.

\subsection{Efficiency Loss and HSV Tax Function}\label{Efficiency Loss and HSV Tax Function}

I assume that the couple's preferences are given by
\begin{equation}
v \left( c, y_m, y_f, \upsilon_m, \upsilon_f \right) = c - \frac{\upsilon_m}{\sigma+1} \left( \frac{y_m}{\upsilon_m} \right)^{\sigma+1} - \frac{\upsilon_f}{\sigma+1} \left( \frac{y_f}{\upsilon_f} \right)^{\sigma+1}
\label{eq: preferences_quasilinear}
\end{equation}
where parameter $\sigma$ is the inverse elasticity of taxable income. The quasilinearity assumption implies that there is no income effect, and hence the compensated and uncompensated taxable income functions coincide. Furthermore, $\upsilon_m$ and $\upsilon_f$ may be interpreted as wages.

Next, I choose the functional form for $T$ to summarize the tax and transfer system in a simplified way. \cite{heathcote2017optimal} show that the log-linear function $T(y) = y - \lambda y^{1-\theta}$ (henceforth, HSV tax function) yields a remarkably good approximation of the actual tax and transfer system in the United States. In this specification, parameter $\theta$ is interpreted as a measure of tax progressivity, and parameter $\lambda$ determines the level of tax rates. To capture joint and separate taxation of spousal incomes, I consider the following tax and transfer functions \citep{bick2017taxation, borella2021marriage}:
\begin{equation}
T \left( y_m, y_f \right) = \lambda \left( y_m + y_f \right)^{1 - \theta}
\label{eq: joint_hsv}
\end{equation}
\begin{equation}
T \left( y_m, y_f \right) = \tilde{\lambda} y_m^{1 - \theta} + \tilde{\lambda} y_f^{1 - \theta}
\label{eq: separate_hsv}
\end{equation}
Function \eqref{eq: joint_hsv} describes joint taxation, while function \eqref{eq: separate_hsv} describes separate taxation. To close the model, I assume that the government uses all tax revenue to fund the government expenditures, and runs a balanced budget. Denote by $g$ the share of government consumption in aggregate income. Finally, I set lump-sum non-taxable income of couples to zero, $I = 0$.

I consider a small reform that changes progressivity of the tax and transfer system, $d \theta \approx 0$. How sizable is the linearization bias? Proposition 3 states that it is given by the ratio between the tax progressivity parameter $\theta$ and the inverse elasticity of taxable income $\sigma$. Hence, the magnitude of the bias is jointly determined by a policy parameter (tax function curvature) and a preference parameter (utility curvature).
\bigskip

\noindent\textbf{Proposition 3 (Linearization Bias with HSV Tax Function).} \textit{Consider a small reform that changes tax progressivity, $d \theta \approx 0$. Under both joint and separate taxation of spousal incomes, the linearization bias is given by
\begin{equation}
\Delta = \frac{\theta}{\sigma}
\label{eq: proposition_hsv_bias}
\end{equation}}

\noindent\textbf{Proof.} \textit{See Appendix.}
\bigskip

Higher initial progressivity of the tax system and higher elasticity of taxable income result in the greater magnitude of the linearization bias or, alternatively, greater overestimation of aggregate efficiency gain. Using \eqref{eq: proposition_hsv_bias} and estimates of $\theta$ and $\sigma$ from the literature, I can quantify $\Delta$. Using the information on 1720 estimates of the elasticity of taxable income from 61 papers, \cite{neisser2021elasticity} report that the majority lies between 0 and 1 with a peak around 0.3 and an excess mass between 0.7 and 1. Next, \cite{heathcote2017optimal}, using the sample of the U.S. households aged 25-60, where at least one adult has strong labor market attachment, and the time period between 2000 and 2006, estimate $\theta = 0.181$. Hence, for the range of taxable income elasticities between 0.2 and 1 (hence, between $\sigma = 5$ and $\sigma = 1$), the size of the linearization bias varies between 3.6\% (0.181/5 $\times$ 100\%) and 18.1\% (0.181/1 $\times$ 100\%). In other words, aggregate efficiency gain from a tax reform is overestimated by 3.6-18.1\%.

\section{Conclusion}\label{Conclusion}

This paper develops a framework to study the welfare effects of income tax changes on married couples. I build a static model of couples' labor supply that accounts for the presence of both intensive and extensive margins. My main result is an expression for efficiency gains, resulting from any arbitrary small tax reform, as a function of (i) labor supply elasticities capturing the behavioral responses to the tax policy reforms, (ii) pre-reform tax rates and reform-induced changes in the tax rates, and (iii) labor income shares. This formula allows to transparently decompose welfare gains into the effects that operate through the spousal own working hours, spousal cross-effects of working hours, and the women's participation margin.

At the next step, I use this expression to quantify the welfare effects of the labor income tax changes induced by four tax reforms implemented in the United States: the Tax Reform Act of 1986, the Omnibus Budget Reconciliation Act of 1993, the Economic Growth and Tax Relief Reconciliation Act of 2001, and the Tax Cuts and Jobs Act of 2017. To parameterize the model, I use the CPS ASEC data combined with the NBER TAXSIM calculator. My baseline estimates suggest that these reforms created welfare gains ranging from -0.16 to 0.62 percent of aggregate labor income. Looking at the forces that shape the efficiency gains, I find that, first, the bulk of the gains is generated by the women's labor force participation responses, and, second, the spousal cross-effects of working hours are quantitatively important, and abstracting from them leads to an overestimation of the welfare effects. Although three out of four considered U.S. tax reforms delivered aggregate welfare gains, I show that this result masks significant heterogeneity. In fact, each reform created both winners and losers. Furthermore, the welfare gains are unequally distributed among the couples with different incomes.

Finally, I show the robustness of my findings to several possible caveats. In the first set of exercises, I consider alternative parameterizations of elasticities. It partially allows addressing the concerns about the sensitivity of the results to the choice of values of elasticities. Next, I conduct a set of counterfactual reforms aimed at addressing the concerns about the sensitivity of results to the initial income distribution and the levels of pre-reform tax rates. Furthermore, I address the concern about the biased estimates of welfare effects resulting from the assumption about linearity of the tax and transfer function. To do so, I characterize the linearization bias under the log-linear tax function that approximates the tax and transfer system in the United States. Assuming a small reform that translates into a change in tax progressivity, I show that the linearization bias is given by the ratio between the tax progressivity parameter and the inverse elasticity of taxable income. Existing estimates of these objects suggest that the size of this upward bias lies within the range of 3.6-18.1\%.

\begin{spacing}{1}
\bibliographystyle{ecta}
\bibliography{dp_malkov}

\begin{thebibliography}{55}
\newcommand{\enquote}[1]{``#1''}
\expandafter\ifx\csname natexlab\endcsname\relax\def\natexlab#1{#1}\fi

\bibitem[\protect\citeauthoryear{Attanasio, Levell, Low, and
  S{\'a}nchez-Marcos}{Attanasio et~al.}{2018}]{attanasio2018aggregating}
\textsc{Attanasio, O., P.~Levell, H.~Low, and V.~S{\'a}nchez-Marcos} (2018):
  \enquote{Aggregating Elasticities: Intensive and Extensive Margins of Women's
  Labor Supply,} \emph{Econometrica}, 86, 2049--2082.

\bibitem[\protect\citeauthoryear{Auerbach}{Auerbach}{1985}]{auerbach1985theory}
\textsc{Auerbach, A.~J.} (1985): \enquote{The Theory of Excess Burden and
  Optimal Taxation,} in \emph{Handbook of Public Economics}, Elsevier, vol.~1,
  61--127.

\bibitem[\protect\citeauthoryear{Auerbach and Hines}{Auerbach and
  Hines}{2002}]{auerbach2002taxation}
\textsc{Auerbach, A.~J. and J.~R. Hines} (2002): \enquote{Taxation and Economic
  Efficiency,} in \emph{Handbook of Public Economics}, Elsevier, vol.~3,
  1347--1421.

\bibitem[\protect\citeauthoryear{Bar and Leukhina}{Bar and
  Leukhina}{2009}]{bar2009work}
\textsc{Bar, M. and O.~Leukhina} (2009): \enquote{To Work or Not to Work: Did
  Tax Reforms Affect Labor Force Participation of Married Couples?} \emph{The
  BE Journal of Macroeconomics}, 9, 1--30.

\bibitem[\protect\citeauthoryear{Bargain, Orsini, and Peichl}{Bargain
  et~al.}{2014}]{bargain2014comparing}
\textsc{Bargain, O., K.~Orsini, and A.~Peichl} (2014): \enquote{Comparing Labor
  Supply Elasticities in Europe and the United States: New Results,}
  \emph{Journal of Human Resources}, 49, 723--838.

\bibitem[\protect\citeauthoryear{Barro and Furman}{Barro and
  Furman}{2018}]{barro2018macroeconomic}
\textsc{Barro, R.~J. and J.~Furman} (2018): \enquote{Macroeconomic Effects of
  the 2017 Tax Reform,} \emph{Brookings Papers on Economic Activity}, 2018,
  257--345.

\bibitem[\protect\citeauthoryear{Barro and Redlick}{Barro and
  Redlick}{2011}]{barro2011macroeconomic}
\textsc{Barro, R.~J. and C.~J. Redlick} (2011): \enquote{Macroeconomic Effects
  from Government Purchases and Taxes,} \emph{Quarterly Journal of Economics},
  126, 51--102.

\bibitem[\protect\citeauthoryear{Bick, Br{\"u}ggemann, Fuchs-Sch{\"u}ndeln, and
  Paule-Paludkiewicz}{Bick et~al.}{2019}]{bick2019long}
\textsc{Bick, A., B.~Br{\"u}ggemann, N.~Fuchs-Sch{\"u}ndeln, and
  H.~Paule-Paludkiewicz} (2019): \enquote{Long-Term Changes in Married Couples'
  Labor Supply and Taxes: Evidence from the US and Europe since the 1980s,}
  \emph{Journal of International Economics}, 118, 44--62.

\bibitem[\protect\citeauthoryear{Bick and Fuchs-Sch{\"u}ndeln}{Bick and
  Fuchs-Sch{\"u}ndeln}{2017{\natexlab{a}}}]{bick2017quantifying}
\textsc{Bick, A. and N.~Fuchs-Sch{\"u}ndeln} (2017{\natexlab{a}}):
  \enquote{Quantifying the Disincentive Effects of Joint Taxation on Married
  Women's Labor Supply,} \emph{American Economic Review: Papers \&
  Proceedings}, 107, 100--104.

\bibitem[\protect\citeauthoryear{Bick and Fuchs-Sch{\"u}ndeln}{Bick and
  Fuchs-Sch{\"u}ndeln}{2017{\natexlab{b}}}]{bick2017taxation}
---\hspace{-.1pt}---\hspace{-.1pt}--- (2017{\natexlab{b}}): \enquote{Taxation
  and Labour Supply of Married Couples Across Countries: A Macroeconomic
  Analysis,} \emph{Review of Economic Studies}, 85, 1543--1576.

\bibitem[\protect\citeauthoryear{Bierbrauer, Boyer, and Peichl}{Bierbrauer
  et~al.}{2021}]{bierbrauer2021politically}
\textsc{Bierbrauer, F.~J., P.~C. Boyer, and A.~Peichl} (2021):
  \enquote{Politically Feasible Reforms of Nonlinear Tax Systems,}
  \emph{American Economic Review}, 111, 153--91.

\bibitem[\protect\citeauthoryear{Bitler, Gelbach, and Hoynes}{Bitler
  et~al.}{2006}]{bitler2006mean}
\textsc{Bitler, M.~P., J.~B. Gelbach, and H.~W. Hoynes} (2006): \enquote{What
  Mean Impacts Miss: Distributional Effects of Welfare Reform Experiments,}
  \emph{American Economic Review}, 96, 988--1012.

\bibitem[\protect\citeauthoryear{Blau and Kahn}{Blau and
  Kahn}{2007}]{blau2007changes}
\textsc{Blau, F.~D. and L.~M. Kahn} (2007): \enquote{Changes in the Labor
  Supply Behavior of Married Women: 1980--2000,} \emph{Journal of Labor
  Economics}, 25, 393--438.

\bibitem[\protect\citeauthoryear{Blau and Kahn}{Blau and
  Kahn}{2013}]{blau2013female}
---\hspace{-.1pt}---\hspace{-.1pt}--- (2013): \enquote{Female Labor Supply: Why
  is the US Falling Behind,} \emph{American Economic Review: Papers \&
  Proceedings}, 103, 251--256.

\bibitem[\protect\citeauthoryear{Blomquist and Simula}{Blomquist and
  Simula}{2019}]{blomquist2019marginal}
\textsc{Blomquist, S. and L.~Simula} (2019): \enquote{Marginal Deadweight Loss
  When the Income Tax is Nonlinear,} \emph{Journal of Econometrics}, 211,
  47--60.

\bibitem[\protect\citeauthoryear{Blundell, Pistaferri, and
  Saporta-Eksten}{Blundell et~al.}{2018}]{blundell2018children}
\textsc{Blundell, R., L.~Pistaferri, and I.~Saporta-Eksten} (2018):
  \enquote{Children, Time Allocation, and Consumption Insurance,} \emph{Journal
  of Political Economy}, 126, S73--S115.

\bibitem[\protect\citeauthoryear{Borella, De~Nardi, and Yang}{Borella
  et~al.}{2021}]{borella2021marriage}
\textsc{Borella, M., M.~De~Nardi, and F.~Yang} (2021): \enquote{Are
  Marriage-Related Taxes and Social Security Benefits Holding Back Female Labor
  Supply?} \emph{NBER Working Paper No. 26097}.

\bibitem[\protect\citeauthoryear{Chetty}{Chetty}{2009}]{chetty2009sufficient}
\textsc{Chetty, R.} (2009): \enquote{Sufficient Statistics for Welfare
  Analysis: A Bridge between Structural and Reduced-Form Methods,} \emph{Annual
  Review of Economics}, 1, 451--488.

\bibitem[\protect\citeauthoryear{Cho and Rogerson}{Cho and
  Rogerson}{1988}]{cho1988family}
\textsc{Cho, J.-O. and R.~Rogerson} (1988): \enquote{Family Labor Supply and
  Aggregate Fluctuations,} \emph{Journal of Monetary Economics}, 21, 233--245.

\bibitem[\protect\citeauthoryear{Dahlby}{Dahlby}{1998}]{dahlby1998progressive}
\textsc{Dahlby, B.} (1998): \enquote{Progressive Taxation and the Social
  Marginal Cost of Public Funds,} \emph{Journal of Public Economics}, 67,
  105--122.

\bibitem[\protect\citeauthoryear{Domeij and Heathcote}{Domeij and
  Heathcote}{2004}]{domeij2004distributional}
\textsc{Domeij, D. and J.~Heathcote} (2004): \enquote{On the Distributional
  Effects of Reducing Capital Taxes,} \emph{International Economic Review}, 45,
  523--554.

\bibitem[\protect\citeauthoryear{Dupuy and Weber}{Dupuy and
  Weber}{2021}]{dupuy2021marriage}
\textsc{Dupuy, A. and S.~Weber} (2021): \enquote{Marriage Market
  Counterfactuals Using Matching Models,} \emph{Economica}.

\bibitem[\protect\citeauthoryear{Eissa and Hoynes}{Eissa and
  Hoynes}{2004}]{eissa2004taxes}
\textsc{Eissa, N. and H.~W. Hoynes} (2004): \enquote{Taxes and the Labor Market
  Participation of Married Couples: The Earned Income Tax Credit,}
  \emph{Journal of Public Economics}, 88, 1931--1958.

\bibitem[\protect\citeauthoryear{Eissa, Kleven, and Kreiner}{Eissa
  et~al.}{2008}]{eissa2008evaluation}
\textsc{Eissa, N., H.~J. Kleven, and C.~T. Kreiner} (2008): \enquote{Evaluation
  of Four Tax Reforms in the United States: Labor Supply and Welfare Effects
  for Single Mothers,} \emph{Journal of Public Economics}, 92, 795--816.

\bibitem[\protect\citeauthoryear{Eissa and Liebman}{Eissa and
  Liebman}{1996}]{eissa1996labor}
\textsc{Eissa, N. and J.~B. Liebman} (1996): \enquote{Labor Supply Response to
  the Earned Income Tax Credit,} \emph{Quarterly Journal of Economics}, 111,
  605--637.

\bibitem[\protect\citeauthoryear{Feenberg and Coutts}{Feenberg and
  Coutts}{1993}]{feenberg1993introduction}
\textsc{Feenberg, D. and E.~Coutts} (1993): \enquote{An Introduction to the
  TAXSIM Model,} \emph{Journal of Policy Analysis and Management}, 12,
  189--194.

\bibitem[\protect\citeauthoryear{Feldstein}{Feldstein}{1999}]{feldstein1999tax}
\textsc{Feldstein, M.} (1999): \enquote{Tax Avoidance and the Deadweight Loss
  of the Income Tax,} \emph{Review of Economics and Statistics}, 81, 674--680.

\bibitem[\protect\citeauthoryear{Finkelstein and Hendren}{Finkelstein and
  Hendren}{2020}]{finkelstein2020welfare}
\textsc{Finkelstein, A. and N.~Hendren} (2020): \enquote{Welfare Analysis Meets
  Causal Inference,} \emph{Journal of Economic Perspectives}, 34, 146--167.

\bibitem[\protect\citeauthoryear{Flood, King, Rodgers, Ruggles, and
  Warren}{Flood et~al.}{2020}]{flood2020cps}
\textsc{Flood, S., M.~King, R.~Rodgers, S.~Ruggles, and J.~R. Warren} (2020):
  \enquote{Integrated Public Use Microdata Series, Current Population Survey:
  Version 8.0 [dataset],} \emph{Minneapolis, MN: IPUMS, 2020.
  \href{https://doi.org/10.18128/D030.V8.0}{https://doi.org/10.18128/D030.V8.0}}.

\bibitem[\protect\citeauthoryear{Gayle and Shephard}{Gayle and
  Shephard}{2019}]{gayle2019optimal}
\textsc{Gayle, G.-L. and A.~Shephard} (2019): \enquote{Optimal Taxation,
  Marriage, Home Production, and Family Labor Supply,} \emph{Econometrica}, 87,
  291--326.

\bibitem[\protect\citeauthoryear{Goldin}{Goldin}{2014}]{goldin2014grand}
\textsc{Goldin, C.} (2014): \enquote{A Grand Gender Convergence: Its Last
  Chapter,} \emph{American Economic Review}, 104, 1091--1119.

\bibitem[\protect\citeauthoryear{Guner, Kaygusuz, and Ventura}{Guner
  et~al.}{2012}]{guner2012taxation}
\textsc{Guner, N., R.~Kaygusuz, and G.~Ventura} (2012): \enquote{Taxation and
  Household Labour Supply,} \emph{Review of Economic Studies}, 79, 1113--1149.

\bibitem[\protect\citeauthoryear{Guvenen and Kaplan}{Guvenen and
  Kaplan}{2017}]{guvenen2017top}
\textsc{Guvenen, F. and G.~Kaplan} (2017): \enquote{Top Income Inequality in
  the 21st Century: Some Cautionary Notes,} \emph{FRB Minneapolis Quarterly
  Review}, 38.

\bibitem[\protect\citeauthoryear{Guvenen, Ozkan, and Song}{Guvenen
  et~al.}{2014}]{guvenen2014nature}
\textsc{Guvenen, F., S.~Ozkan, and J.~Song} (2014): \enquote{The Nature of
  Countercyclical Income Risk,} \emph{Journal of Political Economy}, 122,
  621--660.

\bibitem[\protect\citeauthoryear{Harberger}{Harberger}{1964}]{harberger1964taxation}
\textsc{Harberger, A.} (1964): \enquote{Taxation, Resource Allocation, and
  Welfare,} in \emph{The Role of Direct and Indirect Taxes in the Federal
  Reserve System}, Princeton University Press, 25--80.

\bibitem[\protect\citeauthoryear{Heathcote, Storesletten, and
  Violante}{Heathcote et~al.}{2017}]{heathcote2017optimal}
\textsc{Heathcote, J., K.~Storesletten, and G.~L. Violante} (2017):
  \enquote{Optimal Tax Progressivity: An Analytical Framework,} \emph{Quarterly
  Journal of Economics}, 132, 1693--1754.

\bibitem[\protect\citeauthoryear{Heim}{Heim}{2007}]{heim2007incredible}
\textsc{Heim, B.~T.} (2007): \enquote{The Incredible Shrinking Elasticities:
  Married Female Labor Supply, 1978--2002,} \emph{Journal of Human Resources},
  42, 881--918.

\bibitem[\protect\citeauthoryear{Hendren and Sprung-Keyser}{Hendren and
  Sprung-Keyser}{2020}]{hendren2020unified}
\textsc{Hendren, N. and B.~Sprung-Keyser} (2020): \enquote{A Unified Welfare
  Analysis of Government Policies,} \emph{Quarterly Journal of Economics}, 135,
  1209--1318.

\bibitem[\protect\citeauthoryear{Holter, Krueger, and Stepanchuk}{Holter
  et~al.}{2019}]{holter2019tax}
\textsc{Holter, H.~A., D.~Krueger, and S.~Stepanchuk} (2019): \enquote{How Do
  Tax Progressivity and Household Heterogeneity Affect Laffer Curves?}
  \emph{Quantitative Economics}, 10, 1317--1356.

\bibitem[\protect\citeauthoryear{Hotchkiss, Moore, and Rios-Avila}{Hotchkiss
  et~al.}{2012}]{hotchkiss2012assessing}
\textsc{Hotchkiss, J.~L., R.~E. Moore, and F.~Rios-Avila} (2012):
  \enquote{Assessing the Welfare Impact of Tax Reform: A Case Study of the 2001
  US Tax Cut,} \emph{Review of Income and Wealth}, 58, 233--256.

\bibitem[\protect\citeauthoryear{Hotchkiss, Moore, and Rios-Avila}{Hotchkiss
  et~al.}{2021}]{hotchkiss2021impact}
---\hspace{-.1pt}---\hspace{-.1pt}--- (2021): \enquote{Impact of the 2017 Tax
  Cuts and Jobs Act on Labor Supply and Welfare of Married Households,}
  \emph{Federal Reserve Bank of Atlanta Working Paper 2021-18}.

\bibitem[\protect\citeauthoryear{Immervoll, Kleven, Kreiner, and
  Verdelin}{Immervoll et~al.}{2009}]{immervoll2009evaluation}
\textsc{Immervoll, H., H.~J. Kleven, C.~T. Kreiner, and N.~Verdelin} (2009):
  \enquote{An Evaluation of the Tax-Transfer Treatment of Married Couples in
  European Countries,} \emph{IZA Discussion Paper No. 3965}.

\bibitem[\protect\citeauthoryear{Kaygusuz}{Kaygusuz}{2010}]{kaygusuz2010taxes}
\textsc{Kaygusuz, R.} (2010): \enquote{Taxes and Female Labor Supply,}
  \emph{Review of Economic Dynamics}, 13, 725--741.

\bibitem[\protect\citeauthoryear{Kleven}{Kleven}{2020}]{kleven2020eitc}
\textsc{Kleven, H.~J.} (2020): \enquote{The EITC and the Extensive Margin: A
  Reappraisal,} \emph{NBER Working Paper No. 26405}.

\bibitem[\protect\citeauthoryear{Kleven}{Kleven}{2021}]{kleven2021sufficient}
---\hspace{-.1pt}---\hspace{-.1pt}--- (2021): \enquote{Sufficient Statistics
  Revisited,} \emph{Annual Review of Economics}, 13, 515--538.

\bibitem[\protect\citeauthoryear{Kleven and Kreiner}{Kleven and
  Kreiner}{2006}]{kleven2006marginal}
\textsc{Kleven, H.~J. and C.~T. Kreiner} (2006): \enquote{The Marginal Cost of
  Public Funds: Hours of Work versus Labor Force Participation,} \emph{Journal
  of Public Economics}, 90, 1955--1973.

\bibitem[\protect\citeauthoryear{Meghir and Phillips}{Meghir and
  Phillips}{2010}]{meghir2010labour}
\textsc{Meghir, C. and D.~Phillips} (2010): \enquote{Labour Supply and Taxes,}
  \emph{Dimensions of Tax Design: The Mirrlees Review}, 202--74.

\bibitem[\protect\citeauthoryear{Mertens and Ravn}{Mertens and
  Ravn}{2012}]{mertens2012empirical}
\textsc{Mertens, K. and M.~O. Ravn} (2012): \enquote{Empirical Evidence on the
  Aggregate Effects of Anticipated and Unanticipated US Tax Policy Shocks,}
  \emph{American Economic Journal: Economic Policy}, 4, 145--81.

\bibitem[\protect\citeauthoryear{Mertens and Ravn}{Mertens and
  Ravn}{2013}]{mertens2013dynamic}
---\hspace{-.1pt}---\hspace{-.1pt}--- (2013): \enquote{The Dynamic Effects of
  Personal and Corporate Income Tax Changes in the United States,}
  \emph{American Economic Review}, 103, 1212--47.

\bibitem[\protect\citeauthoryear{Meyer, Mok, and Sullivan}{Meyer
  et~al.}{2015}]{meyer2015household}
\textsc{Meyer, B.~D., W.~K. Mok, and J.~X. Sullivan} (2015): \enquote{Household
  Surveys in Crisis,} \emph{Journal of Economic Perspectives}, 29, 199--226.

\bibitem[\protect\citeauthoryear{Mulligan and Rubinstein}{Mulligan and
  Rubinstein}{2008}]{mulligan2008selection}
\textsc{Mulligan, C.~B. and Y.~Rubinstein} (2008): \enquote{Selection,
  Investment, and Women's Relative Wages over Time,} \emph{Quarterly Journal of
  Economics}, 123, 1061--1110.

\bibitem[\protect\citeauthoryear{Neisser}{Neisser}{2021}]{neisser2021elasticity}
\textsc{Neisser, C.} (2021): \enquote{The Elasticity of Taxable Income: A
  Meta-Regression Analysis,} \emph{The Economic Journal}.

\bibitem[\protect\citeauthoryear{Prescott}{Prescott}{2004}]{prescott2004americans}
\textsc{Prescott, E.~C.} (2004): \enquote{Why Do Americans Work So Much More
  than Europeans?} \emph{Federal Reserve Bank of Minneapolis Quarterly Review},
  28, 2--12.

\bibitem[\protect\citeauthoryear{Zidar}{Zidar}{2019}]{zidar2019tax}
\textsc{Zidar, O.} (2019): \enquote{Tax Cuts for Whom? Heterogeneous Effects of
  Income Tax Changes on Growth and Employment,} \emph{Journal of Political
  Economy}, 127, 1437--1472.

\bibitem[\protect\citeauthoryear{Ziliak, Hardy, and Bollinger}{Ziliak
  et~al.}{2011}]{ziliak2011earnings}
\textsc{Ziliak, J.~P., B.~Hardy, and C.~Bollinger} (2011): \enquote{Earnings
  Volatility in America: Evidence from Matched CPS,} \emph{Labour Economics},
  18, 742--754.

\end{thebibliography}
\end{spacing}

\newpage

\section*{Appendix}
\setcounter{equation}{0}
\setcounter{figure}{0}
\setcounter{table}{0}
\renewcommand\theequation{A.\arabic{equation}}
\renewcommand\thefigure{A.\arabic{figure}}
\renewcommand\thetable{A.\arabic{table}}

\subsection*{Proof of Proposition 1 (Sufficient Statistics Formula)}\label{Proof of Proposition 1 (Sufficient Statistics Formula)}

This proof extends \cite{eissa2008evaluation} to the framework with couples. First, differentiate the compensated labor supply functions for dual-earner and single-earner couples with respect to $\theta$:
\begin{multline}
\frac{d \tilde{h}^{m,2}_i}{d \theta} = - \sum_{j = m,f} \frac{\partial \tilde{h}^{m,2}_i \left( (1 - \tau^m_i) w^m_i, (1 - \tau^f_i) w^f_i, v_i \right)}{\partial \left( \left( 1 - \tau^j_i \right) w^j_i \right)} w^j_i \frac{d \tau^j_i}{d \theta} =\\
- \sum_{j = m,f} \left( \frac{\partial \tilde{h}^{m,2}_i}{\partial \left( 1 - \tau^j_i \right)} \cdot \frac{1 - \tau^j_i}{\tilde{h}^{m,2}_i} \right) \cdot \frac{\tilde{h}^{m,2}_i}{1 - \tau^j_i} \cdot \frac{d \tau^j_i}{d \theta} = - \left( \varepsilon^{m,2}_i \frac{\tilde{h}^{m,2}_i}{1 - \tau^m_i} \cdot \frac{d \tau^m_i}{d \theta} + \varepsilon^{mf}_i \frac{\tilde{h}^{m,2}_i}{1 - \tau^f_i} \cdot \frac{d \tau^f_i}{d \theta} \right)
\label{eq: d_hm2/d_theta}
\end{multline}
and, following similar arguments,
\begin{equation}
\frac{d \tilde{h}^f_i}{d \theta} = - \left( \varepsilon^f_i \frac{\tilde{h}^f_i}{1 - \tau^f_i} \cdot \frac{d \tau^f_i}{d \theta} + \varepsilon^{fm}_i \frac{\tilde{h}^f_i}{1 - \tau^m_i} \cdot \frac{d \tau^m_i}{d \theta} \right)
\label{eq: d_hf/d_theta}
\end{equation}
\begin{equation}
\frac{d \tilde{h}^{m,1}_i}{d \theta} = - \varepsilon^{m,1}_i \frac{\tilde{h}^{m,1}_i}{1 - \tau^m_i} \cdot \frac{d \tau^m_i}{d \theta}
\label{eq: d_hm1/d_theta}
\end{equation}

Next, I derive the expression for $d F_i \left( \tilde{q}_i \right) / d \theta = \left( \partial F_i \left( \tilde{q}_i \right) / \partial \tilde{q}_i \right) \cdot \left( d \tilde{q}_i / d \theta \right)$. First, differentiate the expression for threshold $\tilde{q}_i$, \eqref{eq: q_tilde_main}, with respect to $\theta$:
\begin{multline}
\frac{d \tilde{q}_i}{d \theta} = \frac{\partial v_i \left( \tilde{c}^2_i, \tilde{h}^{m,2}_i, \tilde{h}^f_i \right)}{\partial \tilde{c}^2_i} \cdot \frac{d \tilde{c}^2_i}{d \theta} + \frac{\partial v_i \left( \tilde{c}^2_i, \tilde{h}^{m,2}_i, \tilde{h}^f_i \right)}{\partial \tilde{h}^{m,2}_i} \cdot \frac{d \tilde{h}^{m,2}_i}{d \theta} + \frac{\partial v_i \left( \tilde{c}^2_i, \tilde{h}^{m,2}_i, \tilde{h}^f_i \right)}{\partial \tilde{h}^f_i} \cdot \frac{d \tilde{h}^f_i}{d \theta} -\\
\frac{\partial v_i \left( \tilde{c}^1_i, \tilde{h}^{m,1}_i, 0 \right)}{\partial \tilde{c}^1_i} \cdot \frac{d \tilde{c}^1_i}{d \theta} - \frac{\partial v_i \left( \tilde{c}^1_i, \tilde{h}^{m,1}_i, 0 \right)}{\partial \tilde{h}^{m,1}_i} \cdot \frac{d \tilde{h}^{m,1}_i}{d \theta}
\label{eq: d_tilde_q/d_theta}
\end{multline}

\newpage
Next, differentiate the equation that connects compensated consumption in dual-earner and single-earner couples, \eqref{eq: c_tilde}, with respect to $\theta$:
\begin{multline}
\frac{d \tilde{c}^2_i}{d \theta} = \frac{d \tilde{c}^1_i}{d \theta} + w^m_i \left( \frac{d \tilde{h}^{m,2}_i}{d \theta} - \frac{d \tilde{h}^{m,1}_i}{d \theta} \right) + w^f_i \frac{d \tilde{h}^f_i}{d \theta} - \Bigg[ \frac{\partial T \left( w^m_i \tilde{h}^{m,2}_i, w^f_i \tilde{h}^f_i, \theta \right)}{\partial \left( w^m_i \tilde{h}^{m,2}_i \right)} \cdot \frac{d \tilde{h}^{m,2}_i}{d \theta} w^m_i +\\
\frac{\partial T \left( w^m_i \tilde{h}^{m,2}_i, w^f_i \tilde{h}^f_i, \theta \right)}{\partial \left( w^f_i \tilde{h}^f_i \right)} \cdot \frac{d \tilde{h}^f_i}{d \theta} w^f_i + \frac{d T \left( w^m_i \tilde{h}^{m,2}_i, w^f_i \tilde{h}^f_i, \theta \right)}{d \theta} -\\
\frac{\partial T \left( w^m_i \tilde{h}^{m,1}_i, 0, \theta \right)}{\partial \left( w^m_i \tilde{h}^{m,1}_i \right)} \cdot \frac{d \tilde{h}^{m,1}_i}{d \theta} w^m_i - \frac{d T \left( w^m_i \tilde{h}^{m,1}_i, 0, \theta \right)}{d \theta} \Bigg] =\\
\frac{d \tilde{c}^1_i}{d \theta} + \left( 1 - \tau^m_i \right) w^m_i \left( \frac{d \tilde{h}^{m,2}_i}{d \theta} - \frac{d \tilde{h}^{m,1}_i}{d \theta} \right) + \left( 1 - \tau^f_i \right) w^f_i \frac{d \tilde{h}^f_i}{d \theta} - \frac{d a_i}{d \theta} \left[ w^m_i \left( \tilde{h}^{m,2}_i - \tilde{h}^{m,1}_i \right) + w^f_i \tilde{h}^f_i \right]
\label{eq: d_tilde_c/d_theta}
\end{multline}
where I denote the reform-induced change in the effective participation tax rate as
$$\frac{d a_i}{d \theta} \equiv \frac{d T \left( w^m_i \tilde{h}^{m,2}_i, w^f_i \tilde{h}^f_i, \theta \right) / d \theta - d T \left( w^m_i \tilde{h}^{m,1}_i, 0, \theta \right) / d \theta}{w^m_i \left( \tilde{h}^{m,2}_i - \tilde{h}^{m,1}_i \right) + w^f_i \tilde{h}^f_i}$$

Next, I plug \eqref{eq: d_tilde_c/d_theta} into \eqref{eq: d_tilde_q/d_theta} and use the first-order conditions from the expenditure minimization problem for dual-earner and single-earner couples to get
\begin{multline*}
\frac{d \tilde{q}_i}{d \theta} = \left( \frac{d \tilde{c}^1_i}{d \theta} - \left( 1 - \tau^m_i \right) w^m_i \frac{d \tilde{h}^{m,1}_i}{d \theta} \right) \left( \frac{\partial v_i \left( \tilde{c}^2_i, \tilde{h}^{m,2}_i, \tilde{h}^f_i \right)}{\partial \tilde{c}^2_i} - \frac{\partial v_i \left( \tilde{c}^1_i, \tilde{h}^{m,1}_i, 0 \right)}{\partial \tilde{c}^1_i} \right) -\\
\frac{\partial v_i \left( \tilde{c}^2_i, \tilde{h}^{m,2}_i, \tilde{h}^f_i \right)}{\partial \tilde{c}^2_i} \cdot \frac{d a_i}{d \theta} \left[ w^m_i \left( \tilde{h}^{m,2}_i - \tilde{h}^{m,1}_i \right) + w^f_i \tilde{h}^f_i \right]
\end{multline*}

Notice that the first multiplier in the first term is equal to zero, and therefore I obtain
\begin{equation}
\frac{d \tilde{q}_i}{d \theta} = - \frac{\partial v_i \left( \tilde{c}^2_i, \tilde{h}^{m,2}_i, \tilde{h}^f_i \right)}{\partial \tilde{c}^2_i} \cdot \frac{d a_i}{d \theta} \left[ w^m_i \left( \tilde{h}^{m,2}_i - \tilde{h}^{m,1}_i \right) + w^f_i \tilde{h}^f_i \right]
\label{eq: d_tilde_q/d_theta2}
\end{equation}

Plugging \eqref{eq: d_tilde_q/d_theta2} into $d F_i \left( \tilde{q}_i \right) / d \theta = \left( \partial F_i \left( \tilde{q}_i \right) / \partial \tilde{q}_i \right) \cdot \left( d \tilde{q}_i / d \theta \right)$, I get
\begin{equation}
\frac{d F_i \left( \tilde{q}_i \right)}{d \theta} = - \frac{\partial F_i \left( \tilde{q}_i \right)}{\partial \tilde{q}_i} \cdot \frac{\partial v_i \left( \tilde{c}^2_i, \tilde{h}^{m,2}_i, \tilde{h}^f_i \right)}{\partial \tilde{c}^2_i} \cdot \frac{d a_i}{d \theta} \left[ w^m_i \left( \tilde{h}^{m,2}_i - \tilde{h}^{m,1}_i \right) + w^f_i \tilde{h}^f_i \right]
\label{eq: dF/d_theta2}
\end{equation}

From the definition of the compensated participation elasticity, \eqref{eq: eta_definition},
\begin{equation}
\eta_i = \frac{\partial F_i \left( \tilde{q}_i \right)}{\partial \tilde{q}_i} \cdot \frac{d \tilde{q}_i}{d (1 - a_i)} \cdot \frac{1 - a_i}{F_i \left( \tilde{q}_i \right)} = - \frac{\partial F_i \left( \tilde{q}_i \right)}{\partial \tilde{q}_i} \cdot \frac{d \tilde{q}_i}{d a_i} \cdot \frac{1 - a_i}{F_i \left( \tilde{q}_i \right)}
\label{eq: eta_equation}
\end{equation}

To get the expression for $d \tilde{q}_i / d a_i$, I use \eqref{eq: d_tilde_q/d_theta2}:
\begin{equation}
\frac{d \tilde{q}_i}{d a_i} = \frac{d \tilde{q}_i / d \theta}{d a_i / d \theta} = - \frac{\partial v_i \left( \tilde{c}^2_i, \tilde{h}^{m,2}_i, \tilde{h}^f_i \right)}{\partial \tilde{c}^2_i} \left[ w^m_i \left( \tilde{h}^{m,2}_i - \tilde{h}^{m,1}_i \right) + w^f_i \tilde{h}^f_i \right]
\label{eq: dtilde_q/da}
\end{equation}

Plugging \eqref{eq: dtilde_q/da} into \eqref{eq: eta_equation}, we obtain:
\begin{equation}
\eta_i = \frac{\partial F_i \left( \tilde{q}_i \right)}{\partial \tilde{q}_i} \cdot \frac{1 - a_i}{F_i \left( \tilde{q}_i \right)} \cdot \frac{\partial v_i \left( \tilde{c}^2_i, \tilde{h}^{m,2}_i, \tilde{h}^f_i \right)}{\partial \tilde{c}^2_i} \left[ w^m_i \left( \tilde{h}^{m,2}_i - \tilde{h}^{m,1}_i \right) + w^f_i \tilde{h}^f_i \right]
\label{eq: eta_equation2}
\end{equation}

Next, I plug \eqref{eq: eta_equation2} into \eqref{eq: dF/d_theta2}, and obtain
\begin{equation}
\frac{d F_i \left( \tilde{q}_i \right)}{d \theta} = - \frac{F_i \left( \tilde{q}_i \right)}{1 - a_i} \cdot \frac{d a_i}{d \theta} \eta_i
\label{eq: dF/d_theta_main}
\end{equation}

Finally, I plug \eqref{eq: d_hm2/d_theta}-\eqref{eq: d_hm1/d_theta} and \eqref{eq: dF/d_theta_main} into \eqref{eq: dD/d_theta}, and obtain
\begin{multline}
\frac{d D}{d \theta} = \sum_{i = 1}^N \Bigg[ \left( \varepsilon^{m,2}_i \frac{\tau^m_i}{1 - \tau^m_i} \cdot \frac{d \tau^m_i}{d \theta} + \varepsilon^{mf}_i \frac{\tau^m_i}{1 - \tau^f_i} \cdot \frac{d \tau^f_i}{d \theta} \right) F_i \left( \tilde{q}_i \right) w^m_i \tilde{h}^{m,2}_i +\\
\varepsilon^{m,1}_i \frac{\tau^m_i}{1 - \tau^m_i} \cdot \frac{d \tau^m_i}{d \theta} \left( 1 - F_i \left( \tilde{q}_i \right) \right) w^m_i \tilde{h}^{m,1}_i + \left( \varepsilon^f_i \frac{\tau^f_i}{1 - \tau^f_i} \cdot \frac{d \tau^f_i}{d \theta} + \varepsilon^{fm}_i \frac{\tau^f_i}{1 - \tau^m_i} \cdot \frac{d \tau^m_i}{d \theta} \right) F_i \left( \tilde{q}_i \right) w^f_i \tilde{h}^f_i +\\
\eta_i \frac{a_i}{1 - a_i} \cdot \frac{d a_i}{d \theta} F_i \left( \tilde{q}_i \right) \left[ w^m_i \left( \tilde{h}^{m,2}_i - \tilde{h}^{m,1}_i \right) + w^f_i \tilde{h}^f_i \right] \Bigg]
\label{eq: dD/dtheta_app}
\end{multline}

Dividing this expression by aggregate labor income, \eqref{eq: agg_income}, I obtain equation \eqref{eq: main_formula}. This completes the proof of Proposition 1. $\blacksquare$

\newpage

\setcounter{equation}{0}
\renewcommand\theequation{B.\arabic{equation}}
\subsection*{Proof of Proposition 2 (Efficiency Loss under Nonlinear Taxation of Couples)}\label{Proof of Proposition 2 (Efficiency Loss under Nonlinear Taxation of Couples)}

This proof extends \cite{blomquist2019marginal} to the framework with couples. The utility maximization problem of couple is given by
\begin{equation}
\max_{c, y_m, y_f} v \left( c, y_m, y_f, \upsilon_m, \upsilon_f \right)
\label{eq: OA_utility}
\end{equation}
\begin{equation}
\text{s.t.}~~~~~c = y_m + y_f - T \left( y_m, y_f, \theta \right) + I
\label{eq: OA_budget_constraint}
\end{equation}
where $y_m$ and $y_f$ are taxable incomes of a male and a female, $\upsilon_m$ and $\upsilon_f$ are individual specific preference parameters of a male and a female, and $I$ is lump-sum non-taxable income. As before, $\theta$ is a treatment parameter that captures the tax policy reforms. Preference parameters $\boldsymbol{\upsilon} = \left( \upsilon_m, \upsilon_f \right)$ are jointly drawn from continuous distribution $\varGamma$. Denote the solution to this problem by $c \left( \upsilon_m, \upsilon_f, \theta, I \right)$, $y_m \left( \upsilon_m, \upsilon_f, \theta, I \right)$, and $y_f \left( \upsilon_m, \upsilon_f, \theta, I \right)$.

Next, to get the compensated functions, I turn to the expenditure minimization problem that is given by
\begin{equation}
\min_{c, y_m, y_f} c - y_m - y_f + T \left( y_m, y_f, \theta \right) - I
\label{eq: OA_minE}
\end{equation}
\begin{equation}
\text{s.t.}~~~~~v \left( c, y_m, y_f, \upsilon_m, \upsilon_f \right) \geq \bar{U}
\label{eq: OA_minE_constraint}
\end{equation}
where $\bar{U}$ is a fixed level of utility. The solution delivers compensated functions $\tilde{c} \left( \upsilon_m, \upsilon_f, \theta, \bar{U} \right)$, $\tilde{y}_m \left( \upsilon_m, \upsilon_f, \theta, \bar{U} \right)$, and $\tilde{y}_f \left( \upsilon_m, \upsilon_f, \theta, \bar{U} \right)$.

Note that, at the optimum, the following equality holds:
\begin{equation}
v \left( \tilde{c}, \tilde{y}_m, \tilde{y}_f, \upsilon_m, \upsilon_f \right) = \bar{U}
\label{eq: OA_utility_2}
\end{equation}

Using the compensated functions, I write the expenditure function as
\begin{equation}
E \left( \upsilon_m, \upsilon_f, \theta, \bar{U} \right) = \tilde{c} - \tilde{y}_m - \tilde{y}_f + T \left( \tilde{y}_m, \tilde{y}_f, \theta \right) - I
\label{eq: OA_E2}
\end{equation}

Next, set $\bar{U}$ to be the indirect utility delivered by the utility maximization problem \eqref{eq: OA_utility}-\eqref{eq: OA_budget_constraint}. Then the solutions to the expenditure minimization problem and utility maximization problem coincide.

Consistent with \eqref{eq: excess_burden_i}, I use the measure of excess burden based on the equivalent variation:
\begin{equation}
D \left( \upsilon_m, \upsilon_f, \theta, \bar{U} \right) = E \left( \upsilon_m, \upsilon_f, \theta, \bar{U} \right) - E \left( \upsilon_m, \upsilon_f, 0, \bar{U} \right) - R \left( \upsilon_m, \upsilon_f, \theta, \bar{U} \right)
\label{eq: OA_D_i_definition}
\end{equation}

The tax revenue function $R \left( \upsilon_m, \upsilon_f, \theta, \bar{U} \right)$ is given by
\begin{equation}
R \left( \upsilon_m, \upsilon_f, \theta, \bar{U} \right) = T \left( \tilde{y}_m, \tilde{y}_f, \theta \right)
\label{eq: OA_R_definition}
\end{equation}

Aggregate deadweight loss from a tax and transfer system $\theta$ is obtained by integrating excess burdens over all couples:
\begin{equation}
D = \int D \left( \upsilon_m, \upsilon_f, \theta, \bar{U} \right) d\varGamma \left( \upsilon_m, \upsilon_f \right)
\label{eq: OA_D_definition}
\end{equation}

Next, plugging \eqref{eq: OA_R_definition} into \eqref{eq: OA_D_i_definition}, I obtain
\begin{equation}
D = \int \left[ E \left( \upsilon_m, \upsilon_f, \theta, \bar{U} \right) - T \left( \tilde{y}_m, \tilde{y}_f, \theta \right) - E \left( \upsilon_m, \upsilon_f, 0, \bar{U} \right) \right] d\varGamma \left( \upsilon_m, \upsilon_f \right)
\label{eq: OA_D_expanded}
\end{equation}

Applying the envelope theorem to \eqref{eq: OA_E2}, I obtain $d E \left( \upsilon_m, \upsilon_f, \theta, \bar{U} \right) / d \theta = \partial T \left( \tilde{y}_m, \tilde{y}_f, \theta \right) / \partial \theta$, i.e. the small tax reform affects the expenditure function only through mechanical revenue effect. Differentiating aggregate excess burden \eqref{eq: OA_D_expanded} with respect to parameter $\theta$ and using the result from the envelope theorem, I get
\begin{equation}
\frac{d D}{d \theta} = - \int \Bigg[ \frac{\partial T \left( \tilde{y}_m, \tilde{y}_f, \theta \right)}{\partial \tilde{y}_m} \cdot \frac{d \tilde{y}_m}{d \theta} + \frac{\partial T \left( \tilde{y}_m, \tilde{y}_f, \theta \right)}{\partial \tilde{y}_f} \cdot \frac{d \tilde{y}_f}{d \theta} \Bigg] d\varGamma \left( \upsilon_m, \upsilon_f \right)
\label{eq: OA_dD/d_theta}
\end{equation}

Next, I rewrite the expression for marginal deadweight loss in terms of the curvatures of the indifference curve and the tax function. Consider a small reform captured by $d \theta \approx 0$. In the expenditure minimization problem, assume that constraint \eqref{eq: OA_minE_constraint} is binding:
\begin{equation}
v \left( c, y_m, y_f, \upsilon_m, \upsilon_f \right) = \bar{U}
\label{eq: OA_2ecase1_constraint}
\end{equation}

Define function $c = \psi \left( y_m, y_f, \upsilon_m, \upsilon_f, \bar{U} \right)$, and plug it into the objective \eqref{eq: OA_minE}:
\begin{equation}
\min_{y_m, y_f} \psi \left( y_m, y_f, \upsilon_m, \upsilon_f, \bar{U} \right) - y_m - y_f + T \left( y_m, y_f, \theta \right) - I
\label{eq: OA_2ecase1_minE}
\end{equation}

From \eqref{eq: OA_2ecase1_minE}, I get the following first-order conditions:
\begin{equation}
\frac{\partial \psi \left( \tilde{y}_m, \tilde{y}_f, \upsilon_m, \upsilon_f, \bar{U} \right)}{\partial \tilde{y}_j} - 1 + \frac{\partial T \left( \tilde{y}_m, \tilde{y}_f, \theta \right)}{\partial \tilde{y}_j} = 0,~~~~~~~j = m, f
\label{eq: OA_2ecase1_foc}
\end{equation}

Differentiating \eqref{eq: OA_2ecase1_foc} with respect to policy parameter $\theta$, I obtain
\begin{multline}
\frac{\partial^2 \psi \left( \tilde{y}_m, \tilde{y}_f, \upsilon_m, \upsilon_f, \bar{U} \right)}{\partial \left( \tilde{y}_m \right)^2} \frac{d \tilde{y}_m}{d \theta} + \frac{\partial^2 \psi \left( \tilde{y}_m, \tilde{y}_f, \upsilon_m, \upsilon_f, \bar{U} \right)}{\partial \tilde{y}_m \partial \tilde{y}_f} \frac{d \tilde{y}_f}{d \theta} + \frac{\partial^2 T \left( \tilde{y}_m, \tilde{y}_f, \theta \right)}{\partial \tilde{y}_m \partial \theta} +\\
\frac{\partial^2 T \left( \tilde{y}_m, \tilde{y}_f, \theta \right)}{\partial \left( \tilde{y}_m \right)^2} \frac{d \tilde{y}_m}{d \theta} + \frac{\partial^2 T \left( \tilde{y}_m, \tilde{y}_f, \theta \right)}{\partial \tilde{y}_m \partial \tilde{y}_f} \frac{d \tilde{y}_f}{d \theta} = 0
\label{eq: OA_2ecase1_foc_m_theta}
\end{multline}
\begin{multline}
\frac{\partial^2 \psi \left( \tilde{y}_m, \tilde{y}_f, \upsilon_m, \upsilon_f, \bar{U} \right)}{\partial \left( \tilde{y}_f \right)^2} \frac{d \tilde{y}_f}{d \theta} + \frac{\partial^2 \psi \left( \tilde{y}_m, \tilde{y}_f, \upsilon_m, \upsilon_f, \bar{U} \right)}{\partial \tilde{y}_m \partial \tilde{y}_f} \frac{d \tilde{y}_m}{d \theta} + \frac{\partial^2 T \left( \tilde{y}_m, \tilde{y}_f, \theta \right)}{\partial \tilde{y}_f \partial \theta} +\\
\frac{\partial^2 T \left( \tilde{y}_m, \tilde{y}_f, \theta \right)}{\partial \left( \tilde{y}_f \right)^2} \frac{d \tilde{y}_f}{d \theta} + \frac{\partial^2 T \left( \tilde{y}_m, \tilde{y}_f, \theta \right)}{\partial \tilde{y}_m \partial \tilde{y}_f} \frac{d \tilde{y}_m}{d \theta} = 0
\label{eq: OA_2ecase1_foc_f_theta}
\end{multline}

Denote $\psi''_{ij} \equiv \partial^2 \psi (\cdot) / \partial \tilde{y}_i \partial \tilde{y}_j$, $T''_{ij} \equiv \partial^2 T (\cdot) / \partial \tilde{y}_i \partial \tilde{y}_j$, and $T''_{i \theta} \equiv \partial^2 T (\cdot) / \partial \tilde{y}_i \partial \theta$. Solving for $d \tilde{y}_m / d \theta$ and $d \tilde{y}_f / d \theta$ from equations \eqref{eq: OA_2ecase1_foc_m_theta} and \eqref{eq: OA_2ecase1_foc_f_theta}, I get the expressions in terms of curvatures of the indifference curve and tax function:
\begin{equation}
\frac{d \tilde{y}_m}{d \theta} = \frac{\left( \psi''_{mf} + T''_{mf} \right) T''_{f \theta} - \left( \psi''_{ff} + T''_{ff} \right) T''_{m \theta}}{\left( \psi''_{mm} + T''_{mm} \right) \left( \psi''_{ff} + T''_{ff} \right) - \left( \psi''_{mf} + T''_{mf} \right)^2}
\label{eq: OA_2ecase1_dym_dtheta}
\end{equation}
\begin{equation}
\frac{d \tilde{y}_f}{d \theta} = \frac{\left( \psi''_{mf} + T''_{mf} \right) T''_{m \theta} - \left( \psi''_{mm} + T''_{mm} \right) T''_{f \theta}}{\left( \psi''_{mm} + T''_{mm} \right) \left( \psi''_{ff} + T''_{ff} \right) - \left( \psi''_{mf} + T''_{mf} \right)^2}
\label{eq: OA_2ecase1_dyf_dtheta}
\end{equation}

Denoting $T'_i \equiv \partial T (\cdot) / \partial y_i$ and plugging \eqref{eq: OA_2ecase1_dym_dtheta} and \eqref{eq: OA_2ecase1_dyf_dtheta} into \eqref{eq: OA_dD/d_theta}, I obtain the expression for reform-induced efficiency loss under nonlinear taxation of couples:
\begin{multline}
\frac{d D}{d \theta} = - \int \Bigg[ \frac{T'_m \left[ \left( \psi''_{mf} + T''_{mf} \right) T''_{f \theta} - \left( \psi''_{ff} + T''_{ff} \right) T''_{m \theta} \right]}{\left( \psi''_{mm} + T''_{mm} \right) \left( \psi''_{ff} + T''_{ff} \right) - \left( \psi''_{mf} + T''_{mf} \right)^2} +\\
\frac{T'_f \left[ \left( \psi''_{mf} + T''_{mf} \right) T''_{m \theta} - \left( \psi''_{mm} + T''_{mm} \right) T''_{f \theta} \right]}{\left( \psi''_{mm} + T''_{mm} \right) \left( \psi''_{ff} + T''_{ff} \right) - \left( \psi''_{mf} + T''_{mf} \right)^2} \Bigg] d\varGamma \left( \upsilon_m, \upsilon_f \right)
\label{eq: OA_case1_mdwl}
\end{multline}
\bigskip

\noindent\textbf{Linearized Tax Function}
\bigskip

Given $\left( \upsilon_m, \upsilon_f, \theta, I \right)$, a couple solve the problem \eqref{eq: OA_utility}-\eqref{eq: OA_budget_constraint}, and I denote the solution by $c^* = c \left( \upsilon_m, \upsilon_f, \theta, I \right)$, $y_m^* = y_m \left( \upsilon_m, \upsilon_f, \theta, I \right)$, and $y_f^* = y_f \left( \upsilon_m, \upsilon_f, \theta, I \right)$. I linearize the tax system, so that now it is described by proportional tax rates $\tau_m = \partial T \left( y_m, y_f, \theta \right) / \partial y_m$ and $\tau_f = \partial T \left( y_m, y_f, \theta \right) / \partial y_f$ and a lump-sum component. Namely,
\begin{equation}
T^L \left( y_m, y_f, \tau_m, \tau_f \right) = \tau_m (\theta) y_m + \tau_f (\theta) y_f + T^*
\label{eq: appendix_linearized_tax}
\end{equation}

This linearized tax system delivers $\left(c^*, y_m^*, y_f^* \right)$ as a solution if I set $T^* = y_m^* + y_f^* - c^* - \tau_m y^*_m + \tau_f y^*_f + I$. To simplify notation, I omit explicit dependence of $\tau_m$ and $\tau_f$ on parameter $\theta$. Under linearized tax system, the couple's budget constraint is given by
\begin{equation}
c = \left( 1 - \tau_m \right) y_m + \left( 1 - \tau_f \right) y_f + I - T^*
\label{eq: appendix_linearized_budget_constraint}
\end{equation}

Denote the solution to the problem with a linearized tax function, given by \eqref{eq: OA_utility} and \eqref{eq: appendix_linearized_budget_constraint}, by $c^L \left( \upsilon_m, \upsilon_f, \tau_m, \tau_f, I-T^* \right)$, $y_m^L \left( \upsilon_m, \upsilon_f, \tau_m, \tau_f, I-T^* \right)$, and $y_f^L \left( \upsilon_m, \upsilon_f, \tau_m, \tau_f, I-T^* \right)$. By construction, it coincides with $\left(c^*, y_m^*, y_f^* \right)$.

Next, solving the expenditure minimization problem with a linearized tax function, I get the compensated functions $\tilde{c}^L \left( \upsilon_m, \upsilon_f, \tau_m, \tau_f, \bar{U} \right)$, $\tilde{y}_m^L \left( \upsilon_m, \upsilon_f, \tau_m, \tau_f, \bar{U} \right)$, and $\tilde{y}_f^L \left( \upsilon_m, \upsilon_f, \tau_m, \tau_f, \bar{U} \right)$. Using the expression for marginal deadweight loss \eqref{eq: OA_dD/d_theta}, I obtain
\begin{equation}
\frac{d D^L}{d \theta} = - \int \Bigg[ \frac{\partial T^L \left( \tilde{y}^L_m, \tilde{y}^L_f, \tau_m, \tau_f \right)}{\partial \tilde{y}^L_m} \cdot \frac{d \tilde{y}^L_m}{d \theta} + \frac{\partial T^L \left( \tilde{y}^L_m, \tilde{y}^L_f, \tau_m, \tau_f \right)}{\partial \tilde{y}^L_f} \cdot \frac{d \tilde{y}^L_f}{d \theta} \Bigg] d\varGamma \left( \upsilon_m, \upsilon_f \right)
\label{eq: OA_dD/d_theta_linearization}
\end{equation}

Now, I rewrite this expression for reform-induced efficiency loss in terms of the curvatures of the indifference curve and tax function. Under a linearized tax function, it is true that $\left( T^L \right)''_{ij} \equiv \partial^2 T^L (\cdot) / \partial \tilde{y}_i^L \partial \tilde{y}_j^L = 0$. Furthermore, since, by construction, $\left( T^L \right)'_i = T'_i$, then it is also true that $\left( T^L \right)''_{i\theta} = T''_{i\theta}$. Therefore, using these results in \eqref{eq: OA_2ecase1_dym_dtheta} and \eqref{eq: OA_2ecase1_dyf_dtheta}, I obtain
\begin{equation}
\frac{d \tilde{y}_m^L}{d \theta} = \frac{\psi''_{mf} T''_{f \theta} - \psi''_{ff} T''_{m \theta}}{\psi''_{mm} \psi''_{ff} - \left( \psi''_{mf} \right)^2}
\label{eq: OA_2ecase1_dym_dtheta_linearization}
\end{equation}
\begin{equation}
\frac{d \tilde{y}_f^L}{d \theta} = \frac{\psi''_{mf} T''_{m \theta} - \psi''_{mm} T''_{f \theta}}{\psi''_{mm} \psi''_{ff} - \left( \psi''_{mf} \right)^2}
\label{eq: OA_2ecase1_dyf_dtheta_linearization}
\end{equation}

Plugging \eqref{eq: OA_2ecase1_dym_dtheta_linearization} and \eqref{eq: OA_2ecase1_dyf_dtheta_linearization} into \eqref{eq: OA_dD/d_theta_linearization}, I obtain the expression for marginal deadweight loss under linearized tax function:
\begin{equation}
\frac{d D^L}{d \theta} = - \int \Bigg[ \frac{T'_m \left( \psi''_{mf} T''_{f \theta} - \psi''_{ff} T''_{m \theta} \right)}{\psi''_{mm} \psi''_{ff} - \left( \psi''_{mf} \right)^2} + \frac{T'_f \left( \psi''_{mf} T''_{m \theta} - \psi''_{mm} T''_{f \theta} \right)}{\psi''_{mm} \psi''_{ff} - \left( \psi''_{mf} \right)^2} \Bigg] d\varGamma \left( \upsilon_m, \upsilon_f \right)
\label{eq: OA_case1_mdwl_linearization}
\end{equation}

This completes the proof of Proposition 2. $\blacksquare$

\newpage
\setcounter{equation}{0}
\renewcommand\theequation{C.\arabic{equation}}
\subsection*{Proof of Proposition C.1 (Efficiency Loss with HSV Tax Function)}\label{Proof of Proposition C.1 (Efficiency Loss with HSV Tax Function)}

Before proving Proposition 3, first, I state and prove Proposition C.1 that characterizes the expressions for marginal deadweight losses under joint and separate taxation of spousal incomes.
\bigskip

\noindent\textbf{Proposition C.1 (Efficiency Loss with HSV Tax Function).} \textit{Under joint taxation of spousal incomes, described by \eqref{eq: joint_hsv}, efficiency loss for $\left( \upsilon_m, \upsilon_f \right)$-couple from a small change in tax progressivity $d \theta \approx 0$ is given by}
{\small\begin{multline*}
\frac{d D_{joint} \left( \upsilon_m, \upsilon_f \right)}{d \theta} = \left[ 1 - \lambda^{\frac{\sigma}{\sigma+\theta}} (1 - \theta)^{\frac{\sigma}{\sigma+\theta}} \left( \upsilon_m + \upsilon_f \right)^{- \frac{\sigma \theta}{\sigma+\theta}} \right] \times\\
\frac{\left[ \lambda (1 - \theta)^{1-\sigma-\theta} \left( \upsilon_m + \upsilon_f \right)^\sigma \right]^{\frac{1}{\sigma+\theta}}}{\sigma+\theta} \left[ 1 + \frac{(1-\theta) \log \left( \lambda (1-\theta) \left( \upsilon_m + \upsilon_f \right)^\sigma \right)}{\sigma+\theta} \right]
\end{multline*}}

\noindent\textit{Under joint taxation of spousal incomes and linearized tax function, efficiency loss for $\left( \upsilon_m, \upsilon_f \right)$-couple from a small reform $d \theta \approx 0$ is given by}
{\small\begin{multline*}
\frac{d D^L_{joint} \left( \upsilon_m, \upsilon_f \right)}{d \theta} = \left[ 1 - \lambda^{\frac{\sigma}{\sigma+\theta}} (1 - \theta)^{\frac{\sigma}{\sigma+\theta}} \left( \upsilon_m + \upsilon_f \right)^{- \frac{\sigma \theta}{\sigma+\theta}} \right] \times\\
\frac{\left[ \lambda (1-\theta)^{1-\sigma-\theta} \left( \upsilon_m + \upsilon_f \right)^\sigma \right]^{\frac{1}{\sigma+\theta}}}{\sigma} \left[ 1 + \frac{(1-\theta) \log \left( \lambda (1-\theta) \left( \upsilon_m + \upsilon_f \right)^\sigma \right)}{\sigma+\theta} \right]
\end{multline*}}
\textit{where $\lambda$ in both expressions above is given by}
{\small\begin{equation*}
\lambda = (1-g)^{\frac{\sigma+\theta}{\sigma}} (1-\theta)^{\frac{\theta}{\sigma}} \left[ \frac{\int \left( \upsilon_m + \upsilon_f \right)^{\frac{\sigma}{\sigma+\theta}} d\varGamma \left( \upsilon_m, \upsilon_f \right)}{\int \left( \upsilon_m + \upsilon_f \right)^{\frac{\sigma (1-\theta)}{\sigma+\theta}} d\varGamma \left( \upsilon_m, \upsilon_f \right)} \right]^{\frac{\sigma+\theta}{\sigma}}
\end{equation*}}

\noindent\textit{Under separate taxation of spousal incomes, described by \eqref{eq: separate_hsv}, efficiency loss for $\left( \upsilon_m, \upsilon_f \right)$-couple from a small change in tax progressivity $d \theta \approx 0$ is given by}
{\small\begin{multline*}
\frac{d D_{sep} \left( \upsilon_m, \upsilon_f \right)}{d \theta} =\\
\left[ 1 - \tilde{\lambda}^{\frac{\sigma}{\sigma+\theta}} (1 - \theta)^{\frac{\sigma}{\sigma+\theta}} \upsilon_m^{-\frac{\sigma \theta}{\sigma+\theta}} \right] \frac{\left[ \tilde{\lambda} (1 - \theta)^{1-\sigma-\theta} \upsilon_m^\sigma \right]^{\frac{1}{\sigma+\theta}}}{\sigma+\theta} \left[ 1 + \frac{(1-\theta) \log \left( \tilde{\lambda} (1-\theta) \upsilon_m^\sigma \right)}{\sigma+\theta} \right] +\\
\left[ 1 - \tilde{\lambda}^{\frac{\sigma}{\sigma+\theta}} (1 - \theta)^{\frac{\sigma}{\sigma+\theta}} \upsilon_f^{-\frac{\sigma \theta}{\sigma+\theta}} \right] \frac{\left[ \tilde{\lambda} (1 - \theta)^{1-\sigma-\theta} \upsilon_f^\sigma \right]^{\frac{1}{\sigma+\theta}}}{\sigma+\theta} \left[ 1 + \frac{(1-\theta) \log \left( \tilde{\lambda} (1-\theta) \upsilon_f^\sigma \right)}{\sigma+\theta} \right]
\end{multline*}}

\newpage
\noindent\textit{Under separate taxation of spousal incomes and linearized tax function, efficiency loss for $\left( \upsilon_m, \upsilon_f \right)$-couple from a small reform $d \theta \approx 0$ is given by}
{\small\begin{multline*}
\frac{d D^L_{sep} \left( \upsilon_m, \upsilon_f \right)}{d \theta} =\\
\left[ 1 - \tilde{\lambda}^{\frac{\sigma}{\sigma+\theta}} (1 - \theta)^{\frac{\sigma}{\sigma+\theta}} \upsilon_m^{-\frac{\sigma \theta}{\sigma+\theta}} \right] \frac{\left[ \tilde{\lambda} (1-\theta)^{1-\sigma-\theta} \upsilon_m^\sigma \right]^{\frac{1}{\sigma+\theta}}}{\sigma} \left[ 1 + \frac{(1 - \theta) \log \left( \tilde{\lambda} (1-\theta) \upsilon_m^\sigma \right)}{\sigma+\theta} \right] +\\
\left[ 1 - \tilde{\lambda}^{\frac{\sigma}{\sigma+\theta}} (1 - \theta)^{\frac{\sigma}{\sigma+\theta}} \upsilon_f^{-\frac{\sigma \theta}{\sigma+\theta}} \right] \frac{\left[ \tilde{\lambda} (1-\theta)^{1-\sigma-\theta} \upsilon_f^\sigma \right]^{\frac{1}{\sigma+\theta}}}{\sigma} \left[ 1 + \frac{(1 - \theta) \log \left( \tilde{\lambda} (1-\theta) \upsilon_f^\sigma \right)}{\sigma+\theta} \right]
\end{multline*}}
\textit{where $\tilde{\lambda}$ in both expressions above is given by}
{\small\begin{equation*}
\tilde{\lambda} = (1-g)^{\frac{\sigma+\theta}{\sigma}} (1-\theta)^{\frac{\theta}{\sigma}} \left[ \frac{\int \left( \upsilon_m^{\frac{\sigma}{\sigma+\theta}} + \upsilon_f^{\frac{\sigma}{\sigma+\theta}} \right) d\varGamma \left( \upsilon_m, \upsilon_f \right)}{\int \left( \upsilon_m^{\frac{\sigma (1-\theta)}{\sigma+\theta}} + \upsilon_f^{\frac{\sigma (1-\theta)}{\sigma+\theta}} \right) d\varGamma \left( \upsilon_m, \upsilon_f \right)} \right]^{\frac{\sigma+\theta}{\sigma}}
\end{equation*}}

\noindent\textbf{Proof.}

\noindent\textbf{Joint Taxation of Spouses}
\bigskip

The problem of couple characterized by preference parameters $\left( \upsilon_m, \upsilon_f \right)$ is given by
\begin{equation}
\max_{c, y_m, y_f} c - \frac{\upsilon_m}{\sigma+1} \left( \frac{y_m}{\upsilon_m} \right)^{\sigma+1} - \frac{\upsilon_f}{\sigma+1} \left( \frac{y_f}{\upsilon_f} \right)^{\sigma+1}
\label{eq: hsv_joint_utility_appendix}
\end{equation}
\begin{equation}
\text{s.t.}~~~~~c = \lambda \left( y_m + y_f \right)^{1 - \theta}
\label{eq: hsv_joint_bc_appendix}
\end{equation}

Plugging \eqref{eq: hsv_joint_bc_appendix} into \eqref{eq: hsv_joint_utility_appendix} and maximizing with respect to $y_m$ and $y_f$, I obtain the following first-order conditions:
\begin{equation}
\left( \frac{y_m}{\upsilon_m} \right)^{\sigma} = \lambda (1 - \theta) \left( y_m + y_f \right)^{-\theta}
\label{eq: hsv_joint_ym_foc}
\end{equation}
\begin{equation}
\left( \frac{y_f}{\upsilon_f} \right)^{\sigma} = \lambda (1 - \theta) \left( y_m + y_f \right)^{-\theta}
\label{eq: eq: hsv_joint_yf_foc}
\end{equation}

Then, using $y_m / \upsilon_m = y_f / \upsilon_f$, derive the expressions for the optimal taxable income:
\begin{equation}
\tilde{y}_m = \lambda^{\frac{1}{\sigma+\theta}} (1 - \theta)^{\frac{1}{\sigma+\theta}} \left( \upsilon_m + \upsilon_f \right)^{-\frac{\theta}{\sigma+\theta}} \upsilon_m
\label{eq: hsv_joint_ym}
\end{equation}
\begin{equation}
\tilde{y}_f = \lambda^{\frac{1}{\sigma+\theta}} (1 - \theta)^{\frac{1}{\sigma+\theta}} \left( \upsilon_m + \upsilon_f \right)^{-\frac{\theta}{\sigma+\theta}} \upsilon_f
\label{eq: hsv_joint_yf}
\end{equation}

\begin{spacing}{1.14}
Plugging these expressions into \eqref{eq: hsv_joint_bc_appendix}, I obtain the optimal consumption:
\begin{equation}
\tilde{c} = \lambda^{\frac{\sigma+1}{\sigma+\theta}} (1 - \theta)^{\frac{1-\theta}{\sigma+\theta}} \left( \upsilon_m + \upsilon_f \right)^{\frac{\sigma(1-\theta)}{\sigma+\theta}}
\label{eq: hsv_joint_c}
\end{equation}

Given the quasilinear preferences, the income effect on taxable income is zero, and hence the Marshallian and Hicksian functions coincide. Next, using \eqref{eq: hsv_joint_ym} and \eqref{eq: hsv_joint_yf}, I obtain the compensated tax revenue function:
\begin{multline}
T \left( \tilde{y}_m, \tilde{y}_f, \theta \right) = \tilde{y}_m + \tilde{y}_f - \lambda \left( \tilde{y}_m + \tilde{y}_f \right)^{1 - \theta} =\\
\lambda^{\frac{1}{\sigma+\theta}} (1 - \theta)^{\frac{1}{\sigma+\theta}} \left( \upsilon_m + \upsilon_f \right)^{\frac{\sigma}{\sigma+\theta}} - \lambda^{\frac{\sigma+1}{\sigma+\theta}} (1 - \theta)^{\frac{1-\theta}{\sigma+\theta}} \left( \upsilon_m + \upsilon_f \right)^{\frac{\sigma(1-\theta)}{\sigma+\theta}}
\label{eq: hsv_joint_T}
\end{multline}

Differentiating the compensated taxable income functions $\tilde{y}_m$ and $\tilde{y}_f$ with respect to the tax progressivity parameter $\theta$, I get
\begin{equation}
\frac{d \tilde{y}_m}{d \theta} = - \upsilon_m \left[ \frac{\lambda (1-\theta)}{\left( \upsilon_m + \upsilon_f \right)^\theta} \right]^{\frac{1}{\sigma+\theta}} \left[ \frac{1}{(1-\theta)(\sigma+\theta)} + \frac{\log \left( \lambda (1-\theta) \left( \upsilon_m + \upsilon_f \right)^\sigma \right)}{(\sigma+\theta)^2} \right]
\label{eq: hsv_joint_diff_ym}
\end{equation}
\begin{equation}
\frac{d \tilde{y}_f}{d \theta} = - \upsilon_f \left[ \frac{\lambda (1-\theta)}{\left( \upsilon_m + \upsilon_f \right)^\theta} \right]^{\frac{1}{\sigma+\theta}} \left[ \frac{1}{(1-\theta)(\sigma+\theta)} + \frac{\log \left( \lambda (1-\theta) \left( \upsilon_m + \upsilon_f \right)^\sigma \right)}{(\sigma+\theta)^2} \right]
\label{eq: hsv_joint_diff_yf}
\end{equation}

Next, differentiating the compensated tax revenue function \eqref{eq: hsv_joint_T} with respect to taxable income, obtain
\begin{equation}
\frac{\partial T}{\partial \tilde{y}_m} = \frac{\partial T}{\partial \tilde{y}_f} = 1 - \lambda (1 - \theta) \left( \tilde{y}_m + \tilde{y}_f \right)^{-\theta} = 1 - \lambda^{\frac{\sigma}{\sigma+\theta}} (1 - \theta)^{\frac{\sigma}{\sigma+\theta}} \left( \upsilon_m + \upsilon_f \right)^{- \frac{\sigma \theta}{\sigma+\theta}}
\label{eq: hsv_joint_diff_T_ym_yf}
\end{equation}

Using \eqref{eq: hsv_joint_diff_ym}-\eqref{eq: hsv_joint_diff_yf} and \eqref{eq: hsv_joint_diff_T_ym_yf}, I obtain the expression for marginal efficiency loss for $\left( \upsilon_m, \upsilon_f \right)$-couple under HSV tax and transfer function:
\begin{multline}
\frac{d D_{joint} \left( \upsilon_m, \upsilon_f \right)}{d \theta} = - \left[ \frac{\partial T}{\partial \tilde{y}_m} \cdot \frac{d \tilde{y}_m}{d \theta} + \frac{\partial T}{\partial \tilde{y}_f} \cdot \frac{d \tilde{y}_f}{d \theta} \right] = \left[ 1 - \lambda^{\frac{\sigma}{\sigma+\theta}} (1 - \theta)^{\frac{\sigma}{\sigma+\theta}} \left( \upsilon_m + \upsilon_f \right)^{- \frac{\sigma \theta}{\sigma+\theta}} \right] \times\\
\frac{\left[ \lambda (1 - \theta)^{1-\sigma-\theta} \left( \upsilon_m + \upsilon_f \right)^\sigma \right]^{\frac{1}{\sigma+\theta}}}{\sigma+\theta} \left[ 1 + \frac{(1-\theta) \log \left( \lambda (1-\theta) \left( \upsilon_m + \upsilon_f \right)^\sigma \right)}{\sigma+\theta} \right]
\label{eq: hsv_joint_dwl}
\end{multline}

Now I turn to the linearized program. Given the values of parameters $\left( \theta, \lambda, \sigma, \upsilon_m, \upsilon_f \right)$, denote the solution to the couple's problem by $\left( c^*, y^*_m, y^*_f \right)$. Linearizing the budget constraint \eqref{eq: hsv_joint_bc_appendix} around this point, I obtain
\begin{equation}
c = \lambda (1 - \theta) \left( y_m^* + y_f^* \right)^{-\theta} \left( y_m + y_f \right) + T^*
\label{eq: hsv_joint_lin_bc}
\end{equation}
where $T^* = \lambda \theta \left( y_m^* + y_f^* \right)^{1-\theta} + c^*$.\end{spacing}

\newpage
Next, I plug the linearized budget constraint \eqref{eq: hsv_joint_lin_bc} into the objective function \eqref{eq: hsv_joint_utility_appendix} and obtain the following first-order conditions:
\begin{equation}
\left( \frac{y_m}{\upsilon_m} \right)^{\sigma} = \lambda (1 - \theta) \left( y_m^* + y_f^* \right)^{-\theta}
\label{eq: hsv_joint_ym_foc_lin}
\end{equation}
\begin{equation}
\left( \frac{y_f}{\upsilon_f} \right)^{\sigma} = \lambda (1 - \theta) \left( y_m^* + y_f^* \right)^{-\theta}
\label{eq: hsv_joint_yf_foc_lin}
\end{equation}

Optimal taxable incomes in the problem with a linearized budget constraint are given by
\begin{equation}
\tilde{y}^L_m = \lambda^{\frac{1}{\sigma}} (1-\theta)^{\frac{1}{\sigma}} \left( y_m^* + y_f^* \right)^{-\frac{\theta}{\sigma}} \upsilon_m
\label{eq: hsv_joint_ym_lin}
\end{equation}
\begin{equation}
\tilde{y}^L_f = \lambda^{\frac{1}{\sigma}} (1-\theta)^{\frac{1}{\sigma}} \left( y_m^* + y_f^* \right)^{-\frac{\theta}{\sigma}} \upsilon_f
\label{eq: hsv_joint_yf_lin}
\end{equation}

Differentiating the compensated taxable income functions $\tilde{y}^L_m$ and $\tilde{y}^L_f$ with respect to the tax progressivity parameter $\theta$, I obtain
\begin{equation}
\frac{d \tilde{y}^L_m}{d \theta} = - \left[ \lambda (1-\theta)^{1-\sigma} \left( y_m^* + y_f^* \right)^{-\theta} \right]^{\frac{1}{\sigma}} \left[ \frac{1}{\sigma} + \frac{(1 - \theta) \log \left( y_m^* + y_f^* \right)}{\sigma} \right] \upsilon_m
\label{eq: hsv_joint_diff_ym_lin_auxiliary}
\end{equation}
\begin{equation}
\frac{d \tilde{y}^L_f}{d \theta} = - \left[ \lambda (1-\theta)^{1-\sigma} \left( y_m^* + y_f^* \right)^{-\theta} \right]^{\frac{1}{\sigma}} \left[ \frac{1}{\sigma} + \frac{(1 - \theta) \log \left( y_m^* + y_f^* \right)}{\sigma} \right] \upsilon_f
\label{eq: hsv_joint_diff_yf_lin_auxiliary}
\end{equation}

Plugging the optimal taxable income, \eqref{eq: hsv_joint_ym_lin} and \eqref{eq: hsv_joint_yf_lin}, into these expressions, I obtain
\begin{equation}
\frac{d \tilde{y}^L_m}{d \theta} = - \upsilon_m \left[ \frac{\lambda (1-\theta)^{1-\sigma-\theta}}{\left( \upsilon_m + \upsilon_f \right)^\theta} \right]^{\frac{1}{\sigma+\theta}} \left[ \frac{1}{\sigma} + \frac{(1-\theta) \log \left( \lambda (1-\theta) \left( \upsilon_m + \upsilon_f \right)^\sigma \right)}{\sigma (\sigma+\theta)} \right]
\label{eq: hsv_joint_diff_ym_lin}
\end{equation}
\begin{equation}
\frac{d \tilde{y}^L_f}{d \theta} = - \upsilon_f \left[ \frac{\lambda (1-\theta)^{1-\sigma-\theta}}{\left( \upsilon_m + \upsilon_f \right)^\theta} \right]^{\frac{1}{\sigma+\theta}} \left[ \frac{1}{\sigma} + \frac{(1-\theta) \log \left( \lambda (1-\theta) \left( \upsilon_m + \upsilon_f \right)^\sigma \right)}{\sigma (\sigma+\theta)} \right]
\label{eq: hsv_joint_diff_yf_lin}
\end{equation}

By construction, optimal taxable income in the problems with nonlinear and linearized tax functions coincide, i.e. $\tilde{y}_m = \tilde{y}^L_m$ and $\tilde{y}_f = \tilde{y}^L_f$. Therefore, $\partial T / \partial \tilde{y}_m = \partial T / \partial \tilde{y}^L_m$ and $\partial T / \partial \tilde{y}_f = \partial T / \partial \tilde{y}^L_f$. Using \eqref{eq: hsv_joint_diff_T_ym_yf} and \eqref{eq: hsv_joint_diff_ym_lin}-\eqref{eq: hsv_joint_diff_yf_lin}, I obtain marginal deadweight loss for $\left( \upsilon_m, \upsilon_f \right)$-couple in the problem with a linearized HSV tax function:
\begin{multline}
\frac{d D^L_{joint} \left( \upsilon_m, \upsilon_f \right)}{d \theta} = - \left[ \frac{\partial T}{\partial \tilde{y}^L_m} \cdot \frac{d \tilde{y}^L_m}{d \theta} + \frac{\partial T}{\partial \tilde{y}^L_f} \cdot \frac{d \tilde{y}^L_f}{d \theta} \right] = \left[ 1 - \lambda^{\frac{\sigma}{\sigma+\theta}} (1 - \theta)^{\frac{\sigma}{\sigma+\theta}} \left( \upsilon_m + \upsilon_f \right)^{- \frac{\sigma \theta}{\sigma+\theta}} \right] \times\\
\frac{\left[ \lambda (1-\theta)^{1-\sigma-\theta} \left( \upsilon_m + \upsilon_f \right)^\sigma \right]^{\frac{1}{\sigma+\theta}}}{\sigma} \left[ 1 + \frac{(1-\theta) \log \left( \lambda (1-\theta) \left( \upsilon_m + \upsilon_f \right)^\sigma \right)}{\sigma+\theta} \right]
\label{eq: hsv_joint_dwl_lin}
\end{multline}

Finally, using the government budget constraint, I solve for $\lambda$ as a function of policy parameters $(\theta, g)$ and primitives of the economy. The government budget constraint under joint taxation of spouses is given by
\begin{equation}
g \int \left( y_m + y_f \right) d\varGamma \left( \upsilon_m, \upsilon_f \right) = \int \left( y_m + y_f \right) d\varGamma \left( \upsilon_m, \upsilon_f \right) - \lambda \int \left( y_m + y_f \right)^{1-\theta} d\varGamma \left( \upsilon_m, \upsilon_f \right)
\label{eq: hsv_joint_government_bc}
\end{equation}

Solving for $\lambda$, I obtain
\begin{equation}
\lambda = \frac{(1-g) \int \left( y_m + y_f \right) d\varGamma \left( \upsilon_m, \upsilon_f \right)}{\int \left( y_m + y_f \right)^{1-\theta} d\varGamma \left( \upsilon_m, \upsilon_f \right)}
\label{eq: hsv_joint_lambda_aux}
\end{equation}

Finally, plugging \eqref{eq: hsv_joint_ym} and \eqref{eq: hsv_joint_yf}, I derive the expression for the equilibrium value of $\lambda$ under joint taxation of spouses:
\begin{equation}
\lambda = (1-g)^{\frac{\sigma+\theta}{\sigma}} (1-\theta)^{\frac{\theta}{\sigma}} \left[ \frac{\int \left( \upsilon_m + \upsilon_f \right)^{\frac{\sigma}{\sigma+\theta}} d\varGamma \left( \upsilon_m, \upsilon_f \right)}{\int \left( \upsilon_m + \upsilon_f \right)^{\frac{\sigma (1-\theta)}{\sigma+\theta}} d\varGamma \left( \upsilon_m, \upsilon_f \right)} \right]^{\frac{\sigma+\theta}{\sigma}}
\label{eq: hsv_joint_lambda}
\end{equation}

Note that, by construction, $\tilde{y}_m = \tilde{y}^L_m$ and $\tilde{y}_f = \tilde{y}^L_f$, and hence the values of $\lambda$ in the original and linearized programs coincide. This completes the derivation of the expressions for reform-induced efficiency loss in the case of joint taxation of spousal incomes.
\bigskip
\bigskip
\bigskip

\noindent\textbf{Separate Taxation of Spouses}
\bigskip

The problem of couple characterized by preference parameters $\left( \upsilon_m, \upsilon_f \right)$ is given by
\begin{equation}
\max_{c, y_m, y_f} c - \frac{\upsilon_m}{\sigma+1} \left( \frac{y_m}{\upsilon_m} \right)^{\sigma+1} - \frac{\upsilon_f}{\sigma+1} \left( \frac{y_f}{\upsilon_f} \right)^{\sigma+1}
\label{eq: hsv_separate_utility_appendix}
\end{equation}
\begin{equation}
\text{s.t.}~~~~~c = \tilde{\lambda} y_m^{1 - \theta} + \tilde{\lambda} y_f^{1 - \theta}
\label{eq: hsv_separate_bc_appendix}
\end{equation}

Substituting \eqref{eq: hsv_separate_bc_appendix} into \eqref{eq: hsv_separate_utility_appendix} and maximizing with respect to $y_m$ and $y_f$, I obtain the following first-order conditions:
\begin{equation}
\left( \frac{y_m}{\upsilon_m} \right)^{\sigma} = \tilde{\lambda} (1 - \theta) y_m^{-\theta}
\label{eq: hsv_separate_ym_foc}
\end{equation}
\begin{equation}
\left( \frac{y_f}{\upsilon_f} \right)^{\sigma} = \tilde{\lambda} (1 - \theta) y_f^{-\theta}
\label{eq: eq: hsv_separate_yf_foc}
\end{equation}

Then, using $y_m / y_f = \left( \upsilon_m / \upsilon_f \right)^{\frac{\sigma}{\sigma+\theta}}$, I derive the expressions for the optimal taxable income:
\begin{equation}
\tilde{y}_m = \tilde{\lambda}^{\frac{1}{\sigma+\theta}} (1 - \theta)^{\frac{1}{\sigma+\theta}} \upsilon_m^{\frac{\sigma}{\sigma+\theta}}
\label{eq: hsv_separate_ym}
\end{equation}
\begin{equation}
\tilde{y}_f = \tilde{\lambda}^{\frac{1}{\sigma+\theta}} (1 - \theta)^{\frac{1}{\sigma+\theta}} \upsilon_f^{\frac{\sigma}{\sigma+\theta}}
\label{eq: hsv_separate_yf}
\end{equation}

Substituting these expressions into \eqref{eq: hsv_separate_bc_appendix}, obtain the optimal consumption:
\begin{equation}
\tilde{c} = \tilde{\lambda}^{\frac{\sigma+1}{\sigma+\theta}} (1 - \theta)^{\frac{1-\theta}{\sigma+\theta}} \left[ \upsilon_m^{\frac{\sigma (1-\theta)}{\sigma+\theta}} + \upsilon_f^{\frac{\sigma (1-\theta)}{\sigma+\theta}} \right]
\label{eq: hsv_separate_c}
\end{equation}

Given the quasilinear preferences, the income effect on taxable incomes is zero, and hence the Marshallian and Hicksian functions coincide. Next, using \eqref{eq: hsv_separate_ym} and \eqref{eq: hsv_separate_yf}, I obtain
\begin{multline}
T \left( \tilde{y}_m, \tilde{y}_f, \theta \right) = \tilde{y}_m + \tilde{y}_f - \tilde{\lambda} \tilde{y}_m^{1 - \theta} - \tilde{\lambda} \tilde{y}_f^{1 - \theta} =\\
\tilde{\lambda}^{\frac{1}{\sigma+\theta}} (1 - \theta)^{\frac{1}{\sigma+\theta}} \left[ \upsilon_m^{\frac{\sigma}{\sigma+\theta}} + \upsilon_f^{\frac{\sigma}{\sigma+\theta}} \right] - \tilde{\lambda}^{\frac{\sigma+1}{\sigma+\theta}} (1 - \theta)^{\frac{1-\theta}{\sigma+\theta}} \left[ \upsilon_m^{\frac{\sigma (1-\theta)}{\sigma+\theta}} + \upsilon_f^{\frac{\sigma (1-\theta)}{\sigma+\theta}} \right]
\label{eq: hsv_separate_T}
\end{multline}

Differentiating the compensated taxable income functions $\tilde{y}_m$ and $\tilde{y}_f$ by parameter $\theta$, I obtain
\begin{equation}
\frac{d \tilde{y}_m}{d \theta} = - \left[ \tilde{\lambda} (1 - \theta) \upsilon_m^\sigma \right]^{\frac{1}{\sigma+\theta}} \left[ \frac{1}{(1-\theta)(\sigma+\theta)} + \frac{\log \left( \tilde{\lambda} (1-\theta) \upsilon_m^\sigma \right)}{(\sigma+\theta)^2} \right]
\label{eq: hsv_separate_diff_ym}
\end{equation}
\begin{equation}
\frac{d \tilde{y}_f}{d \theta} = - \left[ \tilde{\lambda} (1 - \theta) \upsilon_f^\sigma \right]^{\frac{1}{\sigma+\theta}} \left[ \frac{1}{(1-\theta)(\sigma+\theta)} + \frac{\log \left( \tilde{\lambda} (1-\theta) \upsilon_f^\sigma \right)}{(\sigma+\theta)^2} \right]
\label{eq: hsv_separate_diff_yf}
\end{equation}

Next, differentiating the compensated tax revenue function \eqref{eq: hsv_separate_T} by taxable income,
\begin{equation}
\frac{\partial T}{\partial \tilde{y}_m} = 1 - \tilde{\lambda} (1 - \theta) \tilde{y}_m^{-\theta} = 1 - \tilde{\lambda}^{\frac{\sigma}{\sigma+\theta}} (1 - \theta)^{\frac{\sigma}{\sigma+\theta}} \upsilon_m^{-\frac{\sigma \theta}{\sigma+\theta}}
\label{eq: hsv_separate_diff_T_ym}
\end{equation}
\begin{equation}
\frac{\partial T}{\partial \tilde{y}_f} = 1 - \tilde{\lambda} (1 - \theta) \tilde{y}_f^{-\theta} = 1 - \tilde{\lambda}^{\frac{\sigma}{\sigma+\theta}} (1 - \theta)^{\frac{\sigma}{\sigma+\theta}} \upsilon_f^{-\frac{\sigma \theta}{\sigma+\theta}}
\label{eq: hsv_separate_diff_T_yf}
\end{equation}

Using \eqref{eq: hsv_separate_diff_ym}-\eqref{eq: hsv_separate_diff_yf} and \eqref{eq: hsv_separate_diff_T_ym}-\eqref{eq: hsv_separate_diff_T_yf}, I obtain the expression for marginal efficiency loss for $\left( \upsilon_m, \upsilon_f \right)$-couple under HSV tax and transfer function:
\begin{multline}
\frac{d D_{sep} \left( \upsilon_m, \upsilon_f \right)}{d \theta} = - \left[ \frac{\partial T}{\partial \tilde{y}_m} \cdot \frac{d \tilde{y}_m}{d \theta} + \frac{\partial T}{\partial \tilde{y}_f} \cdot \frac{d \tilde{y}_f}{d \theta} \right] =\\
\left[ 1 - \tilde{\lambda}^{\frac{\sigma}{\sigma+\theta}} (1 - \theta)^{\frac{\sigma}{\sigma+\theta}} \upsilon_m^{-\frac{\sigma \theta}{\sigma+\theta}} \right] \frac{\left[ \tilde{\lambda} (1 - \theta)^{1-\sigma-\theta} \upsilon_m^\sigma \right]^{\frac{1}{\sigma+\theta}}}{\sigma+\theta} \left[ 1 + \frac{(1-\theta) \log \left( \tilde{\lambda} (1-\theta) \upsilon_m^\sigma \right)}{\sigma+\theta} \right] +\\
\left[ 1 - \tilde{\lambda}^{\frac{\sigma}{\sigma+\theta}} (1 - \theta)^{\frac{\sigma}{\sigma+\theta}} \upsilon_f^{-\frac{\sigma \theta}{\sigma+\theta}} \right] \frac{\left[ \tilde{\lambda} (1 - \theta)^{1-\sigma-\theta} \upsilon_f^\sigma \right]^{\frac{1}{\sigma+\theta}}}{\sigma+\theta} \left[ 1 + \frac{(1-\theta) \log \left( \tilde{\lambda} (1-\theta) \upsilon_f^\sigma \right)}{\sigma+\theta} \right]
\label{eq: hsv_separate_dwl}
\end{multline}

\newpage
Now I turn to the linearized program. Given the values of parameters $\left( \theta, \tilde{\lambda}, \sigma, \upsilon_m, \upsilon_f \right)$, denote the solution to the couple's problem by $\left( c^*, y^*_m, y^*_f \right)$. Linearizing the budget constraint \eqref{eq: hsv_separate_bc_appendix} around this point, I obtain
\begin{equation}
c = \tilde{\lambda} (1 - \theta) \left( y_m^* \right)^{-\theta} y_m + \tilde{\lambda} (1 - \theta) \left( y_f^* \right)^{-\theta} y_f + T^*
\label{eq: hsv_separate_lin_bc}
\end{equation}
where $T^* = \tilde{\lambda} \theta \left( y_m^* \right)^{1-\theta} + \tilde{\lambda} \theta \left( y_f^* \right)^{1-\theta} + c^*$.

Next, I plug the linearized budget constraint \eqref{eq: hsv_separate_lin_bc} into the objective function \eqref{eq: hsv_separate_utility_appendix} and obtain the following first-order conditions:
\begin{equation}
\left( \frac{y_m}{\upsilon_m} \right)^{\sigma} = \tilde{\lambda} (1 - \theta) \left( y_m^* \right)^{-\theta}
\label{eq: hsv_separate_ym_foc_lin}
\end{equation}
\begin{equation}
\left( \frac{y_f}{\upsilon_f} \right)^{\sigma} = \tilde{\lambda} (1 - \theta) \left( y_f^* \right)^{-\theta}
\label{eq: hsv_separate_yf_foc_lin}
\end{equation}

Optimal taxable incomes in the problem with a linearized budget constraint are given by
\begin{equation}
\tilde{y}^L_m = \tilde{\lambda}^{\frac{1}{\sigma}} (1-\theta)^{\frac{1}{\sigma}} \left( y_m^* \right)^{-\frac{\theta}{\sigma}} \upsilon_m
\label{eq: hsv_separate_ym_lin}
\end{equation}
\begin{equation}
\tilde{y}^L_f = \tilde{\lambda}^{\frac{1}{\sigma}} (1-\theta)^{\frac{1}{\sigma}} \left( y_f^* \right)^{-\frac{\theta}{\sigma}} \upsilon_f
\label{eq: hsv_separate_yf_lin}
\end{equation}

Differentiating the compensated taxable income functions $\tilde{y}^L_m$ and $\tilde{y}^L_f$ with respect to the tax progressivity parameter $\theta$, I obtain
\begin{equation}
\frac{d \tilde{y}^L_m}{d \theta} = - \left[ \tilde{\lambda} (1-\theta)^{1-\sigma} \left( y_m^* \right)^{-\theta} \right]^{\frac{1}{\sigma}} \left[ \frac{1}{\sigma} + \frac{(1 - \theta) \log y_m^*}{\sigma} \right] \upsilon_m
\label{eq: hsv_separate_diff_ym_lin_auxiliary}
\end{equation}
\begin{equation}
\frac{d \tilde{y}^L_f}{d \theta} = - \left[ \tilde{\lambda} (1-\theta)^{1-\sigma} \left( y_f^* \right)^{-\theta} \right]^{\frac{1}{\sigma}} \left[ \frac{1}{\sigma} + \frac{(1 - \theta) \log y_f^*}{\sigma} \right] \upsilon_f
\label{eq: hsv_separate_diff_yf_lin_auxiliary}
\end{equation}

Plugging the optimal taxable income, \eqref{eq: hsv_separate_ym} and \eqref{eq: hsv_separate_yf}, into these expressions, I get
\begin{equation}
\frac{\partial \tilde{y}^L_m}{\partial \theta} = - \left[ \tilde{\lambda} (1-\theta)^{1-\sigma-\theta} \upsilon_m^\sigma \right]^{\frac{1}{\sigma+\theta}} \left[ \frac{1}{\sigma} + \frac{(1 - \theta) \log \left( \tilde{\lambda} (1-\theta) \upsilon_m^\sigma \right)}{\sigma (\sigma+\theta)} \right]
\label{eq: hsv_separate_diff_ym_lin}
\end{equation}
\begin{equation}
\frac{\partial \tilde{y}^L_f}{\partial \theta} = - \left[ \tilde{\lambda} (1-\theta)^{1-\sigma-\theta} \upsilon_f^\sigma \right]^{\frac{1}{\sigma+\theta}} \left[ \frac{1}{\sigma} + \frac{(1 - \theta) \log \left( \tilde{\lambda} (1-\theta) \upsilon_f^\sigma \right)}{\sigma (\sigma+\theta)} \right]
\label{eq: hsv_separate_diff_yf_lin}
\end{equation}

\newpage
By construction, optimal taxable income in the problems with nonlinear and linearized tax functions coincide, i.e. $\tilde{y}_m = \tilde{y}^L_m$ and $\tilde{y}_f = \tilde{y}^L_f$. Therefore, $\partial T / \partial \tilde{y}_m = \partial T / \partial \tilde{y}^L_m$ and $\partial T / \partial \tilde{y}_f = \partial T / \partial \tilde{y}^L_f$. Using \eqref{eq: hsv_separate_diff_T_ym}-\eqref{eq: hsv_separate_diff_T_yf} and \eqref{eq: hsv_separate_diff_ym_lin}-\eqref{eq: hsv_separate_diff_yf_lin}, I obtain marginal deadweight loss for $\left( \upsilon_m, \upsilon_f \right)$-couple in the problem with a linearized HSV tax function:
\begin{multline}
\frac{d D^L_{sep} \left( \upsilon_m, \upsilon_f \right)}{d \theta} = - \left[ \frac{\partial T}{\partial \tilde{y}^L_m} \cdot \frac{d \tilde{y}^L_m}{d \theta} + \frac{\partial T}{\partial \tilde{y}^L_f} \cdot \frac{d \tilde{y}^L_f}{d \theta} \right] =\\
\left[ 1 - \tilde{\lambda}^{\frac{\sigma}{\sigma+\theta}} (1 - \theta)^{\frac{\sigma}{\sigma+\theta}} \upsilon_m^{-\frac{\sigma \theta}{\sigma+\theta}} \right] \frac{\left[ \tilde{\lambda} (1-\theta)^{1-\sigma-\theta} \upsilon_m^\sigma \right]^{\frac{1}{\sigma+\theta}}}{\sigma} \left[ 1 + \frac{(1 - \theta) \log \left( \tilde{\lambda} (1-\theta) \upsilon_m^\sigma \right)}{\sigma+\theta} \right] +\\
\left[ 1 - \tilde{\lambda}^{\frac{\sigma}{\sigma+\theta}} (1 - \theta)^{\frac{\sigma}{\sigma+\theta}} \upsilon_f^{-\frac{\sigma \theta}{\sigma+\theta}} \right] \frac{\left[ \tilde{\lambda} (1-\theta)^{1-\sigma-\theta} \upsilon_f^\sigma \right]^{\frac{1}{\sigma+\theta}}}{\sigma} \left[ 1 + \frac{(1 - \theta) \log \left( \tilde{\lambda} (1-\theta) \upsilon_f^\sigma \right)}{\sigma+\theta} \right]
\label{eq: hsv_separate_dwl_lin}
\end{multline}

Finally, using the government budget constraint, I solve for $\tilde{\lambda}$ as a function of policy parameters $(\theta, g)$ and primitives of the economy. The government budget constraint under separate taxation of spousal incomes takes the following form:
\begin{equation}
g \int \left( y_m + y_f \right) d\varGamma \left( \upsilon_m, \upsilon_f \right) = \int \left( y_m + y_f \right) d\varGamma \left( \upsilon_m, \upsilon_f \right) - \tilde{\lambda} \int \left( y_m^{1-\theta} + y_f^{1-\theta} \right) d\varGamma \left( \upsilon_m, \upsilon_f \right)
\label{eq: hsv_separate_government_bc}
\end{equation}

Solving for $\tilde{\lambda}$, I obtain
\begin{equation}
\tilde{\lambda} = \frac{(1-g) \int \left( y_m + y_f \right) d\varGamma \left( \upsilon_m, \upsilon_f \right)}{\int \left( y_m^{1-\theta} + y_f^{1-\theta} \right) d\varGamma \left( \upsilon_m, \upsilon_f \right)}
\label{eq: hsv_separate_lambda_aux}
\end{equation}

Finally, plugging \eqref{eq: hsv_separate_ym} and \eqref{eq: hsv_separate_yf}, I get the expression for the equilibrium value of $\tilde{\lambda}$ under separate taxation of spouses:
\begin{equation}
\tilde{\lambda} = (1-g)^{\frac{\sigma+\theta}{\sigma}} (1-\theta)^{\frac{\theta}{\sigma}} \left[ \frac{\int \left( \upsilon_m^{\frac{\sigma}{\sigma+\theta}} + \upsilon_f^{\frac{\sigma}{\sigma+\theta}} \right) d\varGamma \left( \upsilon_m, \upsilon_f \right)}{\int \left( \upsilon_m^{\frac{\sigma (1-\theta)}{\sigma+\theta}} + \upsilon_f^{\frac{\sigma (1-\theta)}{\sigma+\theta}} \right) d\varGamma \left( \upsilon_m, \upsilon_f \right)} \right]^{\frac{\sigma+\theta}{\sigma}}
\label{eq: hsv_separate_lambda}
\end{equation}

Note that, by construction, $\tilde{y}_m = \tilde{y}^L_m$ and $\tilde{y}_f = \tilde{y}^L_f$, and hence the values of $\tilde{\lambda}$ in the original and linearized programs coincide. This completes the derivation of the expressions for reform-induced efficiency loss in the case of separate taxation of spousal incomes.

Overall, this completes the proof of Proposition C.1. $\blacksquare$

\newpage
\subsection*{Proof of Proposition 3 (Linearization Bias with HSV Tax Function)}\label{Proof of Proposition 3 (Linearization Bias with HSV Tax Function)}

\noindent\textbf{Joint Taxation of Spouses}
\bigskip

By definition, the linearization bias is given by
$$\Delta_{joint} = \frac{\frac{d D^L_{joint}}{d \theta} - \frac{d D_{joint}}{d \theta}}{\frac{d D_{joint}}{d \theta}} = \frac{\int \frac{d D^L_{joint} \left( \upsilon_m, \upsilon_f \right)}{d \theta} d\varGamma \left( \upsilon_m, \upsilon_f \right) - \int \frac{d D_{joint} \left( \upsilon_m, \upsilon_f \right)}{d \theta} d\varGamma \left( \upsilon_m, \upsilon_f \right)}{\int \frac{d D_{joint} \left( \upsilon_m, \upsilon_f \right)}{d \theta} d\varGamma \left( \upsilon_m, \upsilon_f \right)}$$

Plugging \eqref{eq: hsv_joint_dwl} and \eqref{eq: hsv_joint_dwl_lin} from the first part of Proposition C.1, I obtain
\begin{equation*}
\resizebox{\textwidth}{!}
    {%
$\Delta_{joint} = \frac{\left(\frac{1}{\sigma} - \frac{1}{\sigma+\theta} \right) \int \left[ 1 - \lambda^{\frac{\sigma}{\sigma+\theta}} (1 - \theta)^{\frac{\sigma}{\sigma+\theta}} \left( \upsilon_m + \upsilon_f \right)^{- \frac{\sigma \theta}{\sigma+\theta}} \right] \left[ \lambda (1-\theta)^{1-\sigma-\theta} \left( \upsilon_m + \upsilon_f \right)^\sigma \right]^{\frac{1}{\sigma+\theta}} \left[ 1 + \frac{(1-\theta) \log \left( \lambda (1-\theta) \left( \upsilon_m + \upsilon_f \right)^\sigma \right)}{\sigma+\theta} \right] d\varGamma \left( \upsilon_m, \upsilon_f \right)}{\frac{1}{\sigma+\theta} \int \left[ 1 - \lambda^{\frac{\sigma}{\sigma+\theta}} (1 - \theta)^{\frac{\sigma}{\sigma+\theta}} \left( \upsilon_m + \upsilon_f \right)^{- \frac{\sigma \theta}{\sigma+\theta}} \right] \left[ \lambda (1-\theta)^{1-\sigma-\theta} \left( \upsilon_m + \upsilon_f \right)^\sigma \right]^{\frac{1}{\sigma+\theta}} \left[ 1 + \frac{(1-\theta) \log \left( \lambda (1-\theta) \left( \upsilon_m + \upsilon_f \right)^\sigma \right)}{\sigma+\theta} \right] d\varGamma \left( \upsilon_m, \upsilon_f \right)}$%
     }
\end{equation*}

Finally, simplifying, I find that the linearization bias is given by the ratio between the tax progressivity parameter $\theta$ and the inverse elasticity of taxable income $\sigma$:
\begin{equation}
\Delta_{joint} = \frac{\frac{1}{\sigma} - \frac{1}{\sigma+\theta}}{\frac{1}{\sigma+\theta}} = \frac{\theta}{\sigma}
\label{eq: hsv_joint_linearization_bias}
\end{equation}

\noindent\textbf{Separate Taxation of Spouses}
\bigskip

The linearization bias is given by
$$\Delta_{sep} = \frac{\frac{d D^L_{sep}}{d \theta} - \frac{d D_{sep}}{d \theta}}{\frac{d D_{sep}}{d \theta}} = \frac{\int \frac{d D^L_{sep} \left( \upsilon_m, \upsilon_f \right)}{d \theta} d\varGamma \left( \upsilon_m, \upsilon_f \right) - \int \frac{d D_{sep} \left( \upsilon_m, \upsilon_f \right)}{d \theta} d\varGamma \left( \upsilon_m, \upsilon_f \right)}{\int \frac{d D_{sep} \left( \upsilon_m, \upsilon_f \right)}{d \theta} d\varGamma \left( \upsilon_m, \upsilon_f \right)}$$

Plugging \eqref{eq: hsv_separate_dwl} and \eqref{eq: hsv_separate_dwl_lin} from the second part of Proposition C.1, I obtain
\begin{equation*}
\resizebox{\textwidth}{!}
    {%
$\Delta_{sep} = \frac{\left(\frac{1}{\sigma} - \frac{1}{\sigma+\theta} \right) \int \sum_{j=m,f} \left[ 1 - \tilde{\lambda}^{\frac{\sigma}{\sigma+\theta}} (1 - \theta)^{\frac{\sigma}{\sigma+\theta}} \upsilon_j^{-\frac{\sigma \theta}{\sigma+\theta}} \right] \left[ \tilde{\lambda} (1-\theta)^{1-\sigma-\theta} \upsilon_j^\sigma \right]^{\frac{1}{\sigma+\theta}} \left[ 1 + \frac{(1 - \theta) \log \left( \tilde{\lambda} (1-\theta) \upsilon_j^\sigma \right)}{\sigma+\theta} \right] d\varGamma \left( \upsilon_m, \upsilon_f \right)}{\frac{1}{\sigma+\theta} \int \sum_{j=m,f} \left[ 1 - \tilde{\lambda}^{\frac{\sigma}{\sigma+\theta}} (1 - \theta)^{\frac{\sigma}{\sigma+\theta}} \upsilon_j^{-\frac{\sigma \theta}{\sigma+\theta}} \right] \left[ \tilde{\lambda} (1-\theta)^{1-\sigma-\theta} \upsilon_j^\sigma \right]^{\frac{1}{\sigma+\theta}} \left[ 1 + \frac{(1 - \theta) \log \left( \tilde{\lambda} (1-\theta) \upsilon_j^\sigma \right)}{\sigma+\theta} \right] d\varGamma \left( \upsilon_m, \upsilon_f \right)}$%
     }
\end{equation*}

Finally, simplifying, I find that the linearization bias coincides with one I obtain under joint taxation of couples:
\begin{equation}
\Delta_{sep} = \frac{\frac{1}{\sigma} - \frac{1}{\sigma+\theta}}{\frac{1}{\sigma+\theta}} = \frac{\theta}{\sigma}
\label{eq: hsv_separate_linearization_bias}
\end{equation}

This completes the proof of Proposition 3. $\blacksquare$

\newpage
\section*{Additional Figures and Tables}
\setcounter{figure}{0}
\setcounter{table}{0}
\renewcommand\thefigure{A.\arabic{figure}}
\renewcommand\thetable{A.\arabic{table}}

\begin{figure}[H]
\centering
\includegraphics[width=0.75\linewidth]{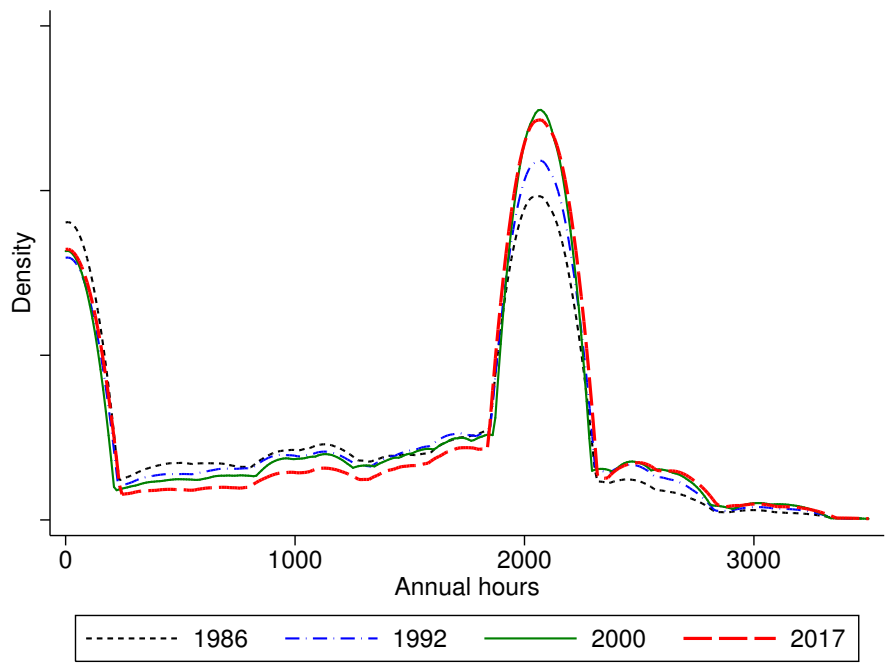}
\caption{Annual working hours of married women with working husbands in the United States}
\label{fig:hoursf}
\justify\footnotesize{\textsc{Notes:} Data is from the Annual Social and Economic Supplement of the Current Population Survey. The sample includes married women aged 25-54 with working husbands.}
\end{figure}

\begin{figure}[H]
\centering
\begin{subfigure}{.5\textwidth}
  \centering
  \includegraphics[width=\linewidth]{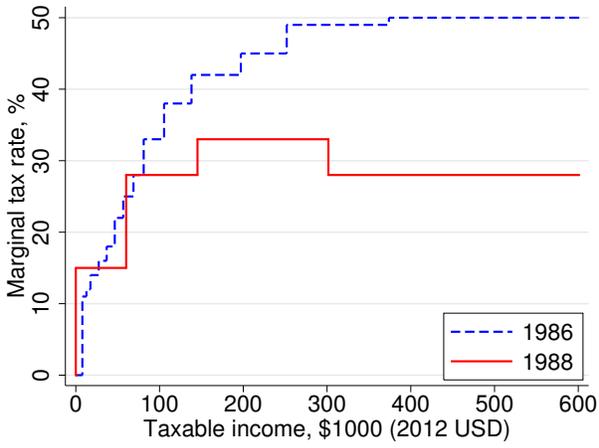}
  \subcaption{TRA 1986 reform}
  \label{fig: fed_1986_88m}
\end{subfigure}%
\begin{subfigure}{.5\textwidth}
  \centering
  \includegraphics[width=\linewidth]{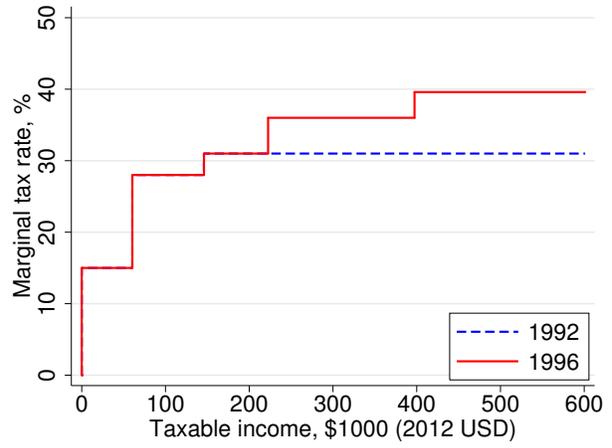}
  \subcaption{OBRA 1993 reform}
  \label{fig: fed_1992_96m}
\end{subfigure}
\bigskip

\begin{subfigure}{.5\textwidth}
  \centering
  \includegraphics[width=\linewidth]{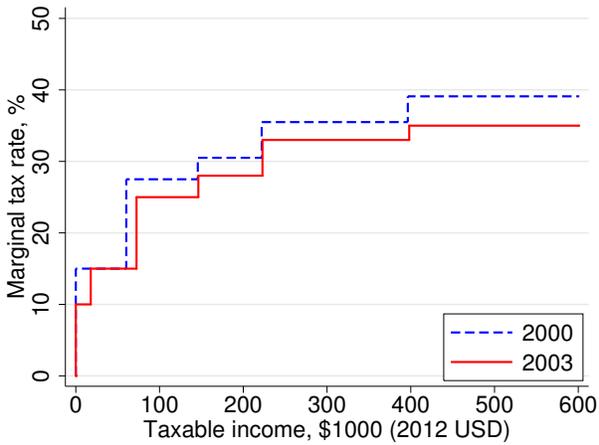}
  \subcaption{EGTRRA 2001 reform}
  \label{fig: fed_2000_03m}
\end{subfigure}%
\begin{subfigure}{.5\textwidth}
  \centering
  \includegraphics[width=\linewidth]{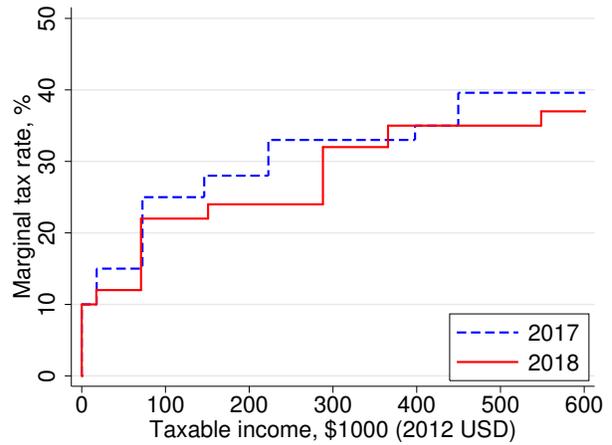}
  \subcaption{TCJA 2017 reform}
  \label{fig: fed_2017_18m}
\end{subfigure}
\caption{Pre-reform and post-reform U.S. federal income tax brackets and statutory marginal tax rates for married couples filing jointly}
\label{fig:tax_changes_m}
\end{figure}

\begin{spacing}{1}
\begin{table}[H]
\centering
\caption{Summary statistics, 1986 and 1992} \label{tab: cps_summary_1986_1992}
\begin{tabular}{lccc|ccc}
\hline \hline
& \multicolumn{3}{c}{1986} & \multicolumn{3}{c}{1992} \\
\cline{2-7}
& Mean & Median & St. Dev. & Mean & Median & St. Dev. \\
\hline
Males &&&&&\\
\hline
Age & 38.94 & 38 & 7.88 & 39.48 & 39 & 7.61 \\
White & 0.896 & 1 & 0.305 & 0.892 & 1 & 0.311 \\
College degree & 0.291 & 0 & 0.454 & 0.311 & 0 & 0.463 \\
Annual hours & 2201 & 2080 & 588 & 2217 & 2080 & 606 \\
Earnings (2012 USD) & 52873 & 47893 & 30218 & 53919 & 47610 & 33521 \\
\hline
Females &&&&&\\
\hline
Age & 36.66 & 36 & 7.55 & 37.47 & 37 & 7.37 \\
White & 0.895 & 1 & 0.306 & 0.891 & 1 & 0.312 \\
College degree & 0.211 & 0 & 0.408 & 0.259 & 0 & 0.438 \\
Employment & 0.732 & 1 & 0.443 & 0.764 & 1 & 0.425 \\
Annual hours & 1214 & 1400 & 940 & 1330 & 1664 & 939 \\
Earnings (2012 USD) & 25946 & 22104 & 19507 & 29740 & 25293 & 21906 \\
\hline
Number of children & 1.62 & 2 & 1.22 & 1.54 & 2 & 1.19 \\
Number of children under 6 & 0.50 & 0 & 0.78 & 0.49 & 0 & 0.77 \\
Female --- secondary earner & 0.834 & 1 & 0.372 & 0.788 & 1 & 0.409 \\
\hline
Number of observations & \multicolumn{3}{c|}{17127} & \multicolumn{3}{c}{18032} \\
\hline \hline
\end{tabular}
\end{table}
\justify\footnotesize{\textsc{Notes:} Data is from the Annual Social and Economic Supplement of the Current Population Survey. The sample includes married couples aged 25-54 with working husbands. Secondary earner is defined as the person with the lowest income among two spouses. Details about sample selection are given in the main text. To calculate the summary statistics, I use the CPS ASEC weights.}
\end{spacing}

\begin{spacing}{1}
\begin{table}[H]
\centering
\caption{Summary statistics, 2000 and 2017} \label{tab: cps_summary_2000_2017}
\begin{tabular}{lccc|ccc}
\hline \hline
& \multicolumn{3}{c}{2000} & \multicolumn{3}{c}{2017} \\
\cline{2-7}
& Mean & Median & St. Dev. & Mean & Median & St. Dev. \\
\hline
Males &&&&&\\
\hline
Age & 40.63 & 41 & 7.63 & 40.76 & 41 & 7.77 \\
White & 0.865 & 1 & 0.341 & 0.812 & 1 & 0.391 \\
College degree & 0.351 & 0 & 0.477 & 0.440 & 0 & 0.496 \\
Annual hours & 2294 & 2080 & 558 & 2229 & 2080 & 532 \\
Earnings (2012 USD) & 72918 & 53688 & 74811 & 76318 & 56644 & 81251 \\
\hline
Females &&&&&\\
\hline
Age & 38.78 & 39 & 7.57 & 38.96 & 39 & 7.78 \\
White & 0.862 & 1 & 0.345 & 0.804 & 1 & 0.397 \\
College degree & 0.324 & 0 & 0.468 & 0.493 & 0 & 0.500 \\
Employment rate & 0.777 & 1 & 0.416 & 0.747 & 1 & 0.435 \\
Annual hours & 1393 & 1820 & 947 & 1388 & 1872 & 971 \\
Earnings (2012 USD) & 37659 & 31332 & 37063 & 49817 & 37763 & 54504 \\
\hline
Number of children & 1.55 & 2 & 1.23 & 1.61 & 2 & 1.27 \\
Number of children under 6 & 0.46 & 0 & 0.76 & 0.51 & 0 & 0.79 \\
Female --- secondary earner & 0.775 & 1 & 0.417 & 0.720 & 1 & 0.449 \\
\hline
Number of observations & \multicolumn{3}{c|}{26883} & \multicolumn{3}{c}{17415} \\
\hline \hline
\end{tabular}
\end{table}
\justify\footnotesize{\textsc{Notes:} Data is from the Annual Social and Economic Supplement of the Current Population Survey. The sample includes married couples aged 25-54 with working husbands. Secondary earner is defined as the person with the lowest income among two spouses. Details about sample selection are given in the main text. To calculate the summary statistics, I use the CPS ASEC weights.}
\end{spacing}

\newpage
\appendix

\section*{Online Appendix (Not for Publication)}

\setcounter{equation}{0}
\setcounter{figure}{0}
\setcounter{table}{0}
\renewcommand\theequation{OA.\arabic{equation}}
\renewcommand\thefigure{OA.\arabic{figure}}
\renewcommand\thetable{OA.\arabic{table}}

\section{NBER TAXSIM Inputs}

To calculate the tax liabilities, the Internet NBER TAXSIM uses 32 input variables. As described in the text, I set most of them to zero because I do not model such things as childcare, capital income, housing, etc. Below I provide the full list of input variables with the details on each field.
\begin{enumerate}
\item \textit{taxsimid}: Individual ID.
\item \textit{year}: Tax year.
\item \textit{state} = 23 (Michigan): State.
\item \textit{mstat} = 2 (married filing jointly): Marital status.
\item \textit{page}: Age of primary taxpayer.
\item \textit{sage}: Age of spouse.
\item \textit{depx} = 2: Number of dependents.
\item \textit{dep13} = 2: Number of children under 13.
\item \textit{dep17} = 2: Number of children under 17.
\item \textit{dep18} = 2: Number of qualifying children for EITC.
\item \textit{pwages}: Wage and salary income of primary taxpayer (including self-employment).
\item \textit{swages}: Wage and salary income of spouse (including self-employment).
\item \textit{dividends} = 0: Dividend income.
\item \textit{intrec} = 0: Interest received.
\item \textit{stcg} = 0: Short term capital gains or losses.
\item \textit{ltcg} = 0: Long term capital gains or losses.
\item \textit{otherprop} = 0: Other property income subject to Net Investment Income Tax (NIIT).
\item \textit{nonprop} = 0: Other non-property income not subject to Medicare NIIT.
\item \textit{pensions} = 0: Taxable pensions and IRA distributions.
\item \textit{gssi} = 0: Gross Social Security benefits.
\item \textit{ui} = 0: Unemployment compensation received.
\item \textit{transfers} = 0: Other non-taxable transfer income.
\item \textit{rentpaid} = 0: Rent paid.
\item \textit{proptax} = 0: Real estate taxes paid.
\item \textit{otheritem} = 0: Other itemized deductions that are a preference for the Alternative Minimum Tax (AMT).
\item \textit{childcare} = 0: Child care expenses.
\item \textit{mortgage} = 0: Deductions not included in \textit{otheritem} and not a preference for the AMT.
\item \textit{scorp} = 0: Active S-Corp income.
\item \textit{pbusinc} = 0: Primary taxpayer's Qualified Business Income (QBI) subject to a preferential rate without phaseout.
\item \textit{pprofinc} = 0: Primary taxpayer's Specialized Service Trade or Business service (SSTB) with a preferential rate subject to claw-back.
\item \textit{sbusinc} = 0: Spouse's QBI.
\item \textit{sprofinc} = 0: Spouse's SSTB.
\end{enumerate}

\newpage

\section{Supplementary Tables}

\setcounter{table}{0}
\begin{spacing}{1}
\centering
\begin{longtable}{lccccccc}
\caption{EITC parameters for U.S. married couples filing jointly, 1986-2018} \label{tab:eitc_parameters}\\
\hline \hline
Year & \begin{tabular}[t]{@{}c@{}}Eligible\\Children\end{tabular} & \begin{tabular}[t]{@{}c@{}}Phase-In\\Rate, \%\end{tabular} & \begin{tabular}[t]{@{}c@{}}First\\Kink, \$\end{tabular} & \begin{tabular}[t]{@{}c@{}}Maximum\\Credit, \$\end{tabular} & \begin{tabular}[t]{@{}c@{}}Second\\Kink, \$\end{tabular} & \begin{tabular}[t]{@{}c@{}}Phase-Out\\Rate, \%\end{tabular} & \begin{tabular}[t]{@{}c@{}}Exhaustion\\Point, \$\end{tabular} \\ \hline 
\endfirsthead
\multicolumn{8}{c}%
{{\tablename\ \thetable{} -- continued from previous page}} \\
\hline \hline Year & \begin{tabular}[t]{@{}c@{}}Eligible\\Children\end{tabular} & \begin{tabular}[t]{@{}c@{}}Phase-In\\Rate, \%\end{tabular} & \begin{tabular}[t]{@{}c@{}}First\\Kink, \$\end{tabular} & \begin{tabular}[t]{@{}c@{}}Maximum\\Credit, \$\end{tabular} & \begin{tabular}[t]{@{}c@{}}Second\\Kink, \$\end{tabular} & \begin{tabular}[t]{@{}c@{}}Phase-Out\\Rate, \%\end{tabular} & \begin{tabular}[t]{@{}c@{}}Exhaustion\\Point, \$\end{tabular} \\ \hline 
\endhead

\hline \multicolumn{8}{r}{{Continued on next page}} \\
\endfoot

\hline \hline
\endlastfoot

1986 & any & 11 & 5000 & 550 & 6500 & 12.22 & 11000 \\
1987 & any & 14 & 6080 & 851 & 6920 & 10 & 15432 \\
1988 & any & 14 & 6240 & 874 & 9840 & 10 & 18576 \\
1989 & any & 14 & 6500 & 910 & 10240 & 10 & 19340 \\
1990 & any & 14 & 6810 & 953 & 10730 & 10 & 20264 \\
1991 & \begin{tabular}[t]{@{}c@{}}1\\2+\end{tabular} & \begin{tabular}[t]{@{}c@{}}16.7\\17.3\end{tabular} & \begin{tabular}[t]{@{}c@{}}7140\\7140\end{tabular} & \begin{tabular}[t]{@{}c@{}}1192\\1235\end{tabular} & \begin{tabular}[t]{@{}c@{}}11250\\11250\end{tabular} & \begin{tabular}[t]{@{}c@{}}11.93\\12.36\end{tabular} & \begin{tabular}[t]{@{}c@{}}21250\\21250\end{tabular} \\
1992 & \begin{tabular}[t]{@{}c@{}}1\\2+\end{tabular} & \begin{tabular}[t]{@{}c@{}}17.6\\18.4\end{tabular} & \begin{tabular}[t]{@{}c@{}}7520\\7520\end{tabular} & \begin{tabular}[t]{@{}c@{}}1324\\1384\end{tabular} & \begin{tabular}[t]{@{}c@{}}11840\\11840\end{tabular} & \begin{tabular}[t]{@{}c@{}}12.57\\13.14\end{tabular} & \begin{tabular}[t]{@{}c@{}}22370\\22370\end{tabular} \\
1993 & \begin{tabular}[t]{@{}c@{}}1\\2+\end{tabular} & \begin{tabular}[t]{@{}c@{}}18.5\\19.5\end{tabular} & \begin{tabular}[t]{@{}c@{}}7750\\7750\end{tabular} & \begin{tabular}[t]{@{}c@{}}1434\\1511\end{tabular} & \begin{tabular}[t]{@{}c@{}}12200\\12200\end{tabular} & \begin{tabular}[t]{@{}c@{}}13.21\\13.93\end{tabular} & \begin{tabular}[t]{@{}c@{}}23050\\23050\end{tabular} \\
1994 & \begin{tabular}[t]{@{}c@{}}0\\1\\2+\end{tabular} & \begin{tabular}[t]{@{}c@{}}7.65\\26.3\\30\end{tabular} & \begin{tabular}[t]{@{}c@{}}4000\\7750\\8425\end{tabular} & \begin{tabular}[t]{@{}c@{}}306\\2038\\2528\end{tabular} & \begin{tabular}[t]{@{}c@{}}5000\\11000\\11000\end{tabular} & \begin{tabular}[t]{@{}c@{}}7.65\\15.98\\17.68\end{tabular} & \begin{tabular}[t]{@{}c@{}}9000\\23755\\25296\end{tabular} \\
1995 & \begin{tabular}[t]{@{}c@{}}0\\1\\2+\end{tabular} & \begin{tabular}[t]{@{}c@{}}7.65\\34\\36\end{tabular} & \begin{tabular}[t]{@{}c@{}}4100\\6160\\8640\end{tabular} & \begin{tabular}[t]{@{}c@{}}314\\2094\\3110\end{tabular} & \begin{tabular}[t]{@{}c@{}}5130\\11290\\11290\end{tabular} & \begin{tabular}[t]{@{}c@{}}7.65\\15.98\\20.22\end{tabular} & \begin{tabular}[t]{@{}c@{}}9230\\24396\\26673\end{tabular} \\
1996 & \begin{tabular}[t]{@{}c@{}}0\\1\\2+\end{tabular} & \begin{tabular}[t]{@{}c@{}}7.65\\34\\40\end{tabular} & \begin{tabular}[t]{@{}c@{}}4220\\6330\\8890\end{tabular} & \begin{tabular}[t]{@{}c@{}}323\\2152\\3556\end{tabular} & \begin{tabular}[t]{@{}c@{}}5280\\11610\\11610\end{tabular} & \begin{tabular}[t]{@{}c@{}}7.65\\15.98\\21.06\end{tabular} & \begin{tabular}[t]{@{}c@{}}9500\\25078\\28495\end{tabular} \\
1997 & \begin{tabular}[t]{@{}c@{}}0\\1\\2+\end{tabular} & \begin{tabular}[t]{@{}c@{}}7.65\\34\\40\end{tabular} & \begin{tabular}[t]{@{}c@{}}4340\\6500\\9140\end{tabular} & \begin{tabular}[t]{@{}c@{}}332\\2210\\3656\end{tabular} & \begin{tabular}[t]{@{}c@{}}5430\\11930\\11930\end{tabular} & \begin{tabular}[t]{@{}c@{}}7.65\\15.98\\21.06\end{tabular} & \begin{tabular}[t]{@{}c@{}}9770\\25750\\29290\end{tabular} \\
1998 & \begin{tabular}[t]{@{}c@{}}0\\1\\2+\end{tabular} & \begin{tabular}[t]{@{}c@{}}7.65\\34\\40\end{tabular} & \begin{tabular}[t]{@{}c@{}}4460\\6680\\9390\end{tabular} & \begin{tabular}[t]{@{}c@{}}341\\2271\\3756\end{tabular} & \begin{tabular}[t]{@{}c@{}}5570\\12260\\12260\end{tabular} & \begin{tabular}[t]{@{}c@{}}7.65\\15.98\\21.06\end{tabular} & \begin{tabular}[t]{@{}c@{}}10030\\26473\\30095\end{tabular} \\
1999 & \begin{tabular}[t]{@{}c@{}}0\\1\\2+\end{tabular} & \begin{tabular}[t]{@{}c@{}}7.65\\34\\40\end{tabular} & \begin{tabular}[t]{@{}c@{}}4530\\6800\\9540\end{tabular} & \begin{tabular}[t]{@{}c@{}}347\\2312\\3816\end{tabular} & \begin{tabular}[t]{@{}c@{}}5670\\12460\\12460\end{tabular} & \begin{tabular}[t]{@{}c@{}}7.65\\15.98\\21.06\end{tabular} & \begin{tabular}[t]{@{}c@{}}10200\\26928\\30580\end{tabular} \\
2000 & \begin{tabular}[t]{@{}c@{}}0\\1\\2+\end{tabular} & \begin{tabular}[t]{@{}c@{}}7.65\\34\\40\end{tabular} & \begin{tabular}[t]{@{}c@{}}4610\\6920\\9720\end{tabular} & \begin{tabular}[t]{@{}c@{}}353\\2353\\3888\end{tabular} & \begin{tabular}[t]{@{}c@{}}5770\\12690\\12690\end{tabular} & \begin{tabular}[t]{@{}c@{}}7.65\\15.98\\21.06\end{tabular} & \begin{tabular}[t]{@{}c@{}}10380\\27413\\31152\end{tabular} \\
2001 & \begin{tabular}[t]{@{}c@{}}0\\1\\2+\end{tabular} & \begin{tabular}[t]{@{}c@{}}7.65\\34\\40\end{tabular} & \begin{tabular}[t]{@{}c@{}}4760\\7140\\10020\end{tabular} & \begin{tabular}[t]{@{}c@{}}364\\2428\\4008\end{tabular} & \begin{tabular}[t]{@{}c@{}}5950\\13090\\13090\end{tabular} & \begin{tabular}[t]{@{}c@{}}7.65\\15.98\\21.06\end{tabular} & \begin{tabular}[t]{@{}c@{}}10710\\28281\\32121\end{tabular} \\
2002 & \begin{tabular}[t]{@{}c@{}}0\\1\\2+\end{tabular} & \begin{tabular}[t]{@{}c@{}}7.65\\34\\40\end{tabular} & \begin{tabular}[t]{@{}c@{}}4910\\7370\\10350\end{tabular} & \begin{tabular}[t]{@{}c@{}}376\\2506\\4140\end{tabular} & \begin{tabular}[t]{@{}c@{}}7150\\14520\\14520\end{tabular} & \begin{tabular}[t]{@{}c@{}}7.65\\15.98\\21.06\end{tabular} & \begin{tabular}[t]{@{}c@{}}12060\\30201\\34178\end{tabular} \\
2003 & \begin{tabular}[t]{@{}c@{}}0\\1\\2+\end{tabular} & \begin{tabular}[t]{@{}c@{}}7.65\\34\\40\end{tabular} & \begin{tabular}[t]{@{}c@{}}4990\\7490\\10510\end{tabular} & \begin{tabular}[t]{@{}c@{}}382\\2547\\4204\end{tabular} & \begin{tabular}[t]{@{}c@{}}7240\\14730\\14730\end{tabular} & \begin{tabular}[t]{@{}c@{}}7.65\\15.98\\21.06\end{tabular} & \begin{tabular}[t]{@{}c@{}}12230\\30666\\34692\end{tabular} \\
2004 & \begin{tabular}[t]{@{}c@{}}0\\1\\2+\end{tabular} & \begin{tabular}[t]{@{}c@{}}7.65\\34\\40\end{tabular} & \begin{tabular}[t]{@{}c@{}}5100\\7660\\10750\end{tabular} & \begin{tabular}[t]{@{}c@{}}390\\2604\\4300\end{tabular} & \begin{tabular}[t]{@{}c@{}}7390\\15040\\15040\end{tabular} & \begin{tabular}[t]{@{}c@{}}7.65\\15.98\\21.06\end{tabular} & \begin{tabular}[t]{@{}c@{}}12490\\31338\\35458\end{tabular} \\
2005 & \begin{tabular}[t]{@{}c@{}}0\\1\\2+\end{tabular} & \begin{tabular}[t]{@{}c@{}}7.65\\34\\40\end{tabular} & \begin{tabular}[t]{@{}c@{}}5220\\7830\\11000\end{tabular} & \begin{tabular}[t]{@{}c@{}}399\\2662\\4400\end{tabular} & \begin{tabular}[t]{@{}c@{}}8530\\16370\\16370\end{tabular} & \begin{tabular}[t]{@{}c@{}}7.65\\15.98\\21.06\end{tabular} & \begin{tabular}[t]{@{}c@{}}13750\\33030\\37263\end{tabular} \\
2006 & \begin{tabular}[t]{@{}c@{}}0\\1\\2+\end{tabular} & \begin{tabular}[t]{@{}c@{}}7.65\\34\\40\end{tabular} & \begin{tabular}[t]{@{}c@{}}5380\\8080\\11340\end{tabular} & \begin{tabular}[t]{@{}c@{}}412\\2747\\4536\end{tabular} & \begin{tabular}[t]{@{}c@{}}8740\\16810\\16810\end{tabular} & \begin{tabular}[t]{@{}c@{}}7.65\\15.98\\21.06\end{tabular} & \begin{tabular}[t]{@{}c@{}}14120\\34001\\38348\end{tabular} \\
2007 & \begin{tabular}[t]{@{}c@{}}0\\1\\2+\end{tabular} & \begin{tabular}[t]{@{}c@{}}7.65\\34\\40\end{tabular} & \begin{tabular}[t]{@{}c@{}}5590\\8390\\11790\end{tabular} & \begin{tabular}[t]{@{}c@{}}428\\2853\\4716\end{tabular} & \begin{tabular}[t]{@{}c@{}}9000\\17390\\17390\end{tabular} & \begin{tabular}[t]{@{}c@{}}7.65\\15.98\\21.06\end{tabular} & \begin{tabular}[t]{@{}c@{}}14590\\35241\\39783\end{tabular} \\
2008 & \begin{tabular}[t]{@{}c@{}}0\\1\\2+\end{tabular} & \begin{tabular}[t]{@{}c@{}}7.65\\34\\40\end{tabular} & \begin{tabular}[t]{@{}c@{}}5720\\8580\\12060\end{tabular} & \begin{tabular}[t]{@{}c@{}}438\\2917\\4824\end{tabular} & \begin{tabular}[t]{@{}c@{}}10160\\18740\\18740\end{tabular} & \begin{tabular}[t]{@{}c@{}}7.65\\15.98\\21.06\end{tabular} & \begin{tabular}[t]{@{}c@{}}15880\\36995\\41646\end{tabular} \\
2009 & \begin{tabular}[t]{@{}c@{}}0\\1\\2\\3+\end{tabular} & \begin{tabular}[t]{@{}c@{}}7.65\\34\\40\\45\end{tabular} & \begin{tabular}[t]{@{}c@{}}5970\\8950\\12570\\12570\end{tabular} & \begin{tabular}[t]{@{}c@{}}457\\3043\\5028\\5657\end{tabular} & \begin{tabular}[t]{@{}c@{}}12470\\21420\\21420\\21420\end{tabular} & \begin{tabular}[t]{@{}c@{}}7.65\\15.98\\21.06\\21.06\end{tabular} & \begin{tabular}[t]{@{}c@{}}18440\\40463\\45295\\48279\end{tabular} \\
2010 & \begin{tabular}[t]{@{}c@{}}0\\1\\2\\3+\end{tabular} & \begin{tabular}[t]{@{}c@{}}7.65\\34\\40\\45\end{tabular} & \begin{tabular}[t]{@{}c@{}}5980\\8970\\12590\\12590\end{tabular} & \begin{tabular}[t]{@{}c@{}}457\\3050\\5036\\5666\end{tabular} & \begin{tabular}[t]{@{}c@{}}12490\\21460\\21460\\21460\end{tabular} & \begin{tabular}[t]{@{}c@{}}7.65\\15.98\\21.06\\21.06\end{tabular} & \begin{tabular}[t]{@{}c@{}}18470\\40545\\45373\\48362\end{tabular} \\
2011 & \begin{tabular}[t]{@{}c@{}}0\\1\\2\\3+\end{tabular} & \begin{tabular}[t]{@{}c@{}}7.65\\34\\40\\45\end{tabular} & \begin{tabular}[t]{@{}c@{}}6070\\9100\\12780\\12780\end{tabular} & \begin{tabular}[t]{@{}c@{}}464\\3094\\5112\\5751\end{tabular} & \begin{tabular}[t]{@{}c@{}}12670\\21770\\21770\\21770\end{tabular} & \begin{tabular}[t]{@{}c@{}}7.65\\15.98\\21.06\\21.06\end{tabular} & \begin{tabular}[t]{@{}c@{}}15740\\41132\\46044\\49078\end{tabular} \\
2012 & \begin{tabular}[t]{@{}c@{}}0\\1\\2\\3+\end{tabular} & \begin{tabular}[t]{@{}c@{}}7.65\\34\\40\\45\end{tabular} & \begin{tabular}[t]{@{}c@{}}6210\\9320\\13090\\13090\end{tabular} & \begin{tabular}[t]{@{}c@{}}475\\3169\\5236\\5891\end{tabular} & \begin{tabular}[t]{@{}c@{}}12980\\22300\\22300\\22300\end{tabular} & \begin{tabular}[t]{@{}c@{}}7.65\\15.98\\21.06\\21.06\end{tabular} & \begin{tabular}[t]{@{}c@{}}19190\\42130\\47162\\50270\end{tabular} \\
2013 & \begin{tabular}[t]{@{}c@{}}0\\1\\2\\3+\end{tabular} & \begin{tabular}[t]{@{}c@{}}7.65\\34\\40\\45\end{tabular} & \begin{tabular}[t]{@{}c@{}}6370\\9560\\13430\\13430\end{tabular} & \begin{tabular}[t]{@{}c@{}}487\\3250\\5372\\6044\end{tabular} & \begin{tabular}[t]{@{}c@{}}13310\\22870\\22870\\22870\end{tabular} & \begin{tabular}[t]{@{}c@{}}7.65\\15.98\\21.06\\21.06\end{tabular} & \begin{tabular}[t]{@{}c@{}}19680\\43210\\48378\\51567\end{tabular} \\
2014 & \begin{tabular}[t]{@{}c@{}}0\\1\\2\\3+\end{tabular} & \begin{tabular}[t]{@{}c@{}}7.65\\34\\40\\45\end{tabular} & \begin{tabular}[t]{@{}c@{}}6480\\9720\\13650\\13650\end{tabular} & \begin{tabular}[t]{@{}c@{}}496\\3305\\5460\\6143\end{tabular} & \begin{tabular}[t]{@{}c@{}}13540\\23260\\23260\\23260\end{tabular} & \begin{tabular}[t]{@{}c@{}}7.65\\15.98\\21.06\\21.06\end{tabular} & \begin{tabular}[t]{@{}c@{}}20020\\43941\\49186\\52427\end{tabular} \\
2015 & \begin{tabular}[t]{@{}c@{}}0\\1\\2\\3+\end{tabular} & \begin{tabular}[t]{@{}c@{}}7.65\\34\\40\\45\end{tabular} & \begin{tabular}[t]{@{}c@{}}6580\\9880\\13870\\13870\end{tabular} & \begin{tabular}[t]{@{}c@{}}503\\3359\\5548\\6242\end{tabular} & \begin{tabular}[t]{@{}c@{}}13760\\23630\\23630\\23630\end{tabular} & \begin{tabular}[t]{@{}c@{}}7.65\\15.98\\21.06\\21.06\end{tabular} & \begin{tabular}[t]{@{}c@{}}20340\\44651\\49974\\53267\end{tabular} \\
2016 & \begin{tabular}[t]{@{}c@{}}0\\1\\2\\3+\end{tabular} & \begin{tabular}[t]{@{}c@{}}7.65\\34\\40\\45\end{tabular} & \begin{tabular}[t]{@{}c@{}}6610\\9920\\13931\\13930\end{tabular} & \begin{tabular}[t]{@{}c@{}}506\\3373\\5572\\6269\end{tabular} & \begin{tabular}[t]{@{}c@{}}13820\\23740\\23740\\23740\end{tabular} & \begin{tabular}[t]{@{}c@{}}7.65\\15.98\\21.06\\21.06\end{tabular} & \begin{tabular}[t]{@{}c@{}}20430\\44846\\50198\\53505\end{tabular} \\
2017 & \begin{tabular}[t]{@{}c@{}}0\\1\\2\\3+\end{tabular} & \begin{tabular}[t]{@{}c@{}}7.65\\34\\40\\45\end{tabular} & \begin{tabular}[t]{@{}c@{}}6670\\10000\\14040\\14040\end{tabular} & \begin{tabular}[t]{@{}c@{}}510\\3400\\5616\\6318\end{tabular} & \begin{tabular}[t]{@{}c@{}}13930\\23930\\23930\\23930\end{tabular} & \begin{tabular}[t]{@{}c@{}}7.65\\15.98\\21.06\\21.06\end{tabular} & \begin{tabular}[t]{@{}c@{}}20600\\45207\\50597\\53930\end{tabular} \\
2018 & \begin{tabular}[t]{@{}c@{}}0\\1\\2\\3+\end{tabular} & \begin{tabular}[t]{@{}c@{}}7.65\\34\\40\\45\end{tabular} & \begin{tabular}[t]{@{}c@{}}6780\\10180\\14290\\14290\end{tabular} & \begin{tabular}[t]{@{}c@{}}519\\3461\\5716\\6431\end{tabular} & \begin{tabular}[t]{@{}c@{}}14170\\24350\\24350\\24350\end{tabular} & \begin{tabular}[t]{@{}c@{}}7.65\\15.98\\21.06\\21.06\end{tabular} & \begin{tabular}[t]{@{}c@{}}20950\\46010\\51492\\54884\end{tabular} \\
\hline \hline
\end{longtable}
\justify\footnotesize{\textsc{Notes:} This table shows the federal Earned Income Tax Credit parameters by family size for married couples filing jointly. Eligible children are under age 19 (under 24 if a full-time student) or permanently disabled and must reside with the taxpayer for more than half a year. Since 2002, the values of the second kink and exhaustion point were increased for married taxpayers filing jointly relative to taxpayers filing as single or the head of household. The phase-in rate is defined as the increase in the tax credit for an additional dollar of income. The first kink point corresponds to minimum income that is needed for maximizing the size of tax credit. The second kink point corresponds to maximum income allowed before the phasing-out region. The phase-out rate is defined as the reduction in the tax credit for an additional dollar of income above the second kink point. The exhaustion point corresponds to income at which the Earned Income Tax Credit is completely phased out.}
\end{spacing}

\newpage

\begin{spacing}{1}
\begin{table}[H]
\centering
\caption{Standard deductions and personal exemptions for U.S. married couples filing jointly} \label{tab: tax_parameters}
\begin{tabular}{lcc}
\hline \hline
Year & Standard Deduction & Personal Exemption \\
\hline
1986 & 3670 & 1080 \\
1987 & 3760 & 1900 \\
1988 & 5000 & 1950 \\
1989 & 5200 & 2000 \\
1990 & 5450 & 2050 \\
1991 & 5700 & 2150 \\
1992 & 6000 & 2300 \\
1993 & 6200 & 2350 \\
1994 & 6350 & 2450 \\
1995 & 6550 & 2500 \\
1996 & 6700 & 2550 \\
1997 & 6900 & 2650 \\
1998 & 7100 & 2700 \\
1999 & 7200 & 2750 \\
2000 & 7350 & 2800 \\
2001 & 7600 & 2900 \\
2002 & 7850 & 3000 \\
2003 & 9500 & 3050 \\
2004 & 9700 & 3100 \\
2005 & 10000 & 3200 \\
2006 & 10300 & 3300 \\
2007 & 10700 & 3400 \\
2008 & 10900 & 3500 \\
2009 & 11400 & 3650 \\
2010 & 11400 & 3650 \\
2011 & 11600 & 3700 \\
2012 & 11900 & 3800 \\
2013 & 12200 & 3900 \\
2014 & 12400 & 3950 \\
2015 & 12600 & 4000 \\
2016 & 12600 & 4050 \\
2017 & 12700 & 4050 \\
2018 & 24000 & 0 \\
\hline \hline
\end{tabular}
\end{table}
\justify\footnotesize{\textsc{Notes:} The Tax Cuts and Jobs Act of 2017 eliminated personal exemptions for tax years 2018-2025.}
\end{spacing}

\end{document}